\documentclass[aps,prx,reprint,showpacs,floatfix,nofootinbib,superscriptaddress]{revtex4-2}
\usepackage{amsmath,amssymb,bm,bbm,mathrsfs} 
\usepackage{graphicx}  
\usepackage{color} 
\usepackage[dvipsnames]{xcolor}
\usepackage[papersize={10in,11in}]{geometry}
\usepackage[colorlinks=true]{hyperref}  
\usepackage{ulem}

\usepackage{adjustbox}

\usepackage{graphicx}
\usepackage{relsize}
\usepackage{soul}
\usepackage{pgfplots}
\usepackage{ifthen}

\usepackage[mathscr]{euscript}
\usepackage[section]{placeins}
\hypersetup{
    bookmarks=true,         % show bookmarks bar?
    unicode=false,          % non-Latin characters 
    pdftoolbar=true,        % show Acrobat
    pdfmenubar=true,        % show Acrobat 
    pdffitwindow=false,     % window fit to page when opened
    pdfstartview={FitH},    % fits the width of the page to the window
    pdfsubject={},   % subject of the document
    pdfcreator={},   % creator of the document
    pdfproducer={}, % producer of the document
    pdfkeywords={} {} {}, % list of keywords
    pdfnewwindow=true,      % links in new window
    colorlinks=true,       % false: boxed links; true: colored links
    linkcolor=magenta, %red,          % color of internal links (change box color with linkbordercolor)
    citecolor=blue,        % color of links to bibliography
    filecolor=magenta,      % color of file links
    urlcolor=blue           % color of external links
} 

\usepackage{amsfonts}
\usepackage{upgreek}
\usepackage{slashed}
\usepackage{latexsym}
\usepackage{bbm}

\usepackage{multirow}

\newcommand{\beq}{\begin{equation}}
\newcommand{\eeq}{\end{equation}}
\def\bea{\begin{eqnarray}}
\def\eea{\end{eqnarray}}

\usepackage{braket}
\usepackage{comment}
\usepackage{slashed}
\usepackage[justification=raggedright]{caption}
\usepackage{subcaption}

\renewcommand{\vec}[1]{\boldsymbol{#1}}

\begin{document}

%\preprint{arXiv:xxxx}

\title{Intrinsic first and higher-order topological superconductivity in a doped topological insulator}

\author{Harley D. Scammell}
\email{h.scammell@unsw.edu.au}
\affiliation{School of Physics, University of New South Wales, Sydney 2052, Australia}
\affiliation{Australian Research Council Centre of Excellence in Future Low-Energy Electronics Technologies, University of New South Wales, Sydney 2052, Australia}

\author{Julian Ingham}
%\email{jingham@bu.edu}
\affiliation{Physics Department, Boston University, Commonwealth Avenue, Boston, MA 02215, USA}

\author{Max Geier}
\affiliation{Dahlem Center for Complex Quantum Systems and Fachbereich Physik, Freie Universit\"{a}t Berlin, Arnimallee 14, 14195 Berlin, Germany}
\affiliation{Center for Quantum Devices, Niels Bohr Institute, University of Copenhagen, DK-2100 Copenhagen, Denmark}

\author{Tommy Li}
\affiliation{Dahlem Center for Complex Quantum Systems and Fachbereich Physik, Freie Universit\"{a}t Berlin, Arnimallee 14, 14195 Berlin, Germany}

\date{\today}

\begin{abstract}

We explore higher order topological superconductivity in an artificial Dirac material with intrinsic spin-orbit coupling, which is a doped $\mathbb{Z}_2$ topological insulator in the normal state. A mechanism for superconductivity due to repulsive interactions -- \textit{pseudospin pairing} -- has recently been shown to naturally result in higher-order topology in Dirac systems past a minimum chemical potential \cite{Li2021}. Here we apply this theory through microscopic modeling of a superlattice potential imposed on an inversion symmetric hole-doped semiconductor heterostructure, known as hole-based semiconductor artificial graphene, and extend previous work to include the effects of spin-orbit coupling. We find that spin-orbit coupling enhances interaction effects, providing an experimental handle to increase the efficiency of the superconducting mechanism. We show that the phase diagram of these systems, as a function of chemical potential and interaction strength, contains three superconducting states -- a first-order topological $p+ip$ state, a second-order topological spatially modulated $p+i\tau p$ state, and a second-order topological extended $s$-wave state, $s_\tau$. We calculate the symmetry-based indicators for the $p+i\tau p$ and $s_\tau$ states, which prove these states possess second-order topology.  Exact diagonalisation results are presented which illustrate the interplay between the boundary physics and spin orbit interaction.  We argue that this class of systems offer an experimental platform to engineer and explore first and higher-order topological superconducting states.

\end{abstract}

\maketitle

\tableofcontents

\section{Introduction}

Higher-order topological superconductors are topological phases which exhibit gapless corner (hinge) modes in two (three) dimensions protected by spatial symmetries and the bulk gap, and have recently attracted immense interest \cite{Benalcazar2017,Benalcazar2017a,Peng2017,Langbehn2017,Geier2018,Geier2020,Trifunovic2019,Trifunovic2020,Shiozaki2019,Ono2020,Khalaf2018,Zhang2020,Zhang2019,Zhu2019,Franca2019,Wu2019,Roy2020,Anh2020,Zhang2021,Chew2021,Hsu2020,Wang2018,Gray2019,Choi2020,Schindler2018,Teo2013,Benalcazar2014,Zhu2018,Geier2021,Roy2021, OnoPoShiozaki,Slager2015,Kruthoff2017,Manna2021,Hua2022,Pan2021,MayMann2022,Wu2022,Wub2022,Lin2021secd,Lei2022,Zhangc2021secd,Roy2021b,Zhangc2022secd,Chen2022secd,Zhangb2022secd,Zhang2022secd,Luo2022secd,Li2021secd,Jahin2022}. It was recently proposed that Dirac materials, with purely repulsive interactions and sufficiently localised orbitals, intrinsically give rise to higher-order topological superconductivity \cite{Li2021}. We will refer to this as mechanism as {\it pseudospin pairing}.

Superlattices are a promising platform for this mechanism \cite{Li2020}, since they allow the experimental study of materials with tunable lattice constants, atomic orbitals and interactions \cite{Polini2013}, and have been extensively explored in the context of optical lattices \cite{Bloch2008,Zhang2018,Cooper2019,Browaeys2020} and van der Waals heterostructures \cite{Cao2018,Cao2018a,Yoo2019,Weston2020,Geim2013,Ajayan2016,Forsythe2018}.  Recently, significant experimental progress has also been made in forming honeycomb superlattices in patterned semiconductor heterostructures \cite{Park2009,Gibertini2009,Singha2011,AG,AG2,Du2021,FVergel2021,Gardenier2020,Chen2021ag,Freeney2022}.
Motivated by these developments, in this paper we discuss a $p$-type quantum well overlaid with a periodic potential with honeycomb symmetry (see e.g. Refs \cite{Sushkov2013,Tkachenko2015,Li2016,Li2017, Scammell2019,Krix2022}) as an explicit realisation of the pseudospin pairing mechanism. Here we extend the theory to include the influence of intrinsic spin-orbit coupling. The superlattice potential gives rise to Dirac band crossings at the $K,K'$ points; accounting for the intrinsic spin-orbit coupling gives rise to a spin-dependent mass for the Dirac fermions, opening up a $\mathbb{Z}_2$ topological bandgap. The low energy effective theory is equivalent to the Kane--Mele model for a topological insulator \cite{Kane2005, Sushkov2013, Scammell2019}, with an effective Dirac velocity controlled by the strength of spin-orbit coupling. We find that spin-orbit coupling enhances the superconducting instability and provides an additional handle to manipulate the topological superconducting phases.

 We present results specifically for a model of an artificial honeycomb lattice based on a nanopatterned hole-doped semiconductor quantum well, having in mind the fact that in this situation there is a high degree of experimental control over the electron-electron interaction as well as the band structure. However, our field theory treatment is generic and we anticipate our the results are relevant to a number of other Dirac materials, in which similar spin-orbit physics is present alongside localised orbitals. Unconventional superconductivity has recently been observed in twisted transition metal dichalogenides (TMDs) \cite{Wang2020}, which are Dirac systems where spin-orbit coupling plays an important role. Theoretical studies of twisted TMDs, e.g. Ref. \cite{Wu2019b}, have suggested effective models for the superlattice potential similar to the one we examine in the present paper. Superconductivity has also been seen in the intrinsic heterostructure Ba$_6$Nb$_{11}$S$_{28}$, a material which can be modelled as a stack of decoupled NbS$_2$ layers subjected to a superlattice potential arising from the Ba$_3$NbS$_5$ spacer layers \cite{Devarakonda2020}. Other than superlattice systems, superconductivity is seen in spin-orbit coupled topological materials including Pb$_{1/3}$TaS$_2$ \cite{Yang2021}, few-layer stanene \cite{Liao2018}, monolayer TMDs \cite{Barrera2018,Lu2018,Lu2015,Yang2018,Ye2012}, doped topological insulators \cite{Yonezawa2019,Kreiner2011,Sasaki2012,Liu2015,Sato2013,Novak2013,Fu2010,Fatemi2018,Sajadi2018}, and recently discovered vanadium-based kagome metals \cite{Ortiz2020,Zhu2021,Chen2021,Ortiz2021,Ni2021,Chenb2021,Liang2021,Zhao2021,Kang2021,Jiang2021,Li2021b,Ortiz2019,Zhao2021b,Li2021c,Qian2021,Christensen2021,Tan2021,Park2021,Wu2021,Scammell2022,Zhou2022,Tazai2022,DiSante2022,Nguyen2022}.

We determine the phase diagram of the system as a function of chemical potential and interaction strength. Employing physically realistic  parameters, we find three adjacent superconducting phases -- one first-order topological $p+ip$ intervalley, and two higher-order topological: $s_\tau$ intervalley, and $p+i\tau p$ intravalley -- with spin-orbit coupling entangling the valley and spin polarisation of the Cooper pairs. The $s_\tau$ state is similar to the $s_\pm$ state discussed in the context of iron-based superconductors, which consists of $s$-wave pairing but with a gap that has opposite signs at the hole and electron pockets \cite{Seo2008,Hanaguri2010,Hirschfeld2011,Bang2017,Zhang2019b,Hu2021}; here, the valley structure imposes that $s$-wave state changes sign under exchange of the valleys.

 The $p+i\tau p$ and $s_\tau$ pairing instabilities satisfy a simple criterion for second-order topology derived from symmetry-based indicators \cite{Shiozaki2019, Geier2020, Ono2020}: by counting the inversion eigenvalues of the valence and conduction bands in the normal state, we prove that if a superconducting instability with odd inversion parity opens a full excitation gap in a hole-doped Kane-Mele honeycomb system, then the resulting superconducting state must have second-order topology, hosting Kramers pairs of Majorana corner modes. This conclusion holds for weak spin-orbit coupling much smaller than the bandwidth, and for the onset of superconductivity where the superconducting order parameter is the smallest energy scale. The second-order topological phase persists as long as increasing the superconducting order parameter does not close the bulk excitation gap.

 %%%%%%%%%%%%%%%%%
\begin{figure}[t!]
{\includegraphics[width=80mm]{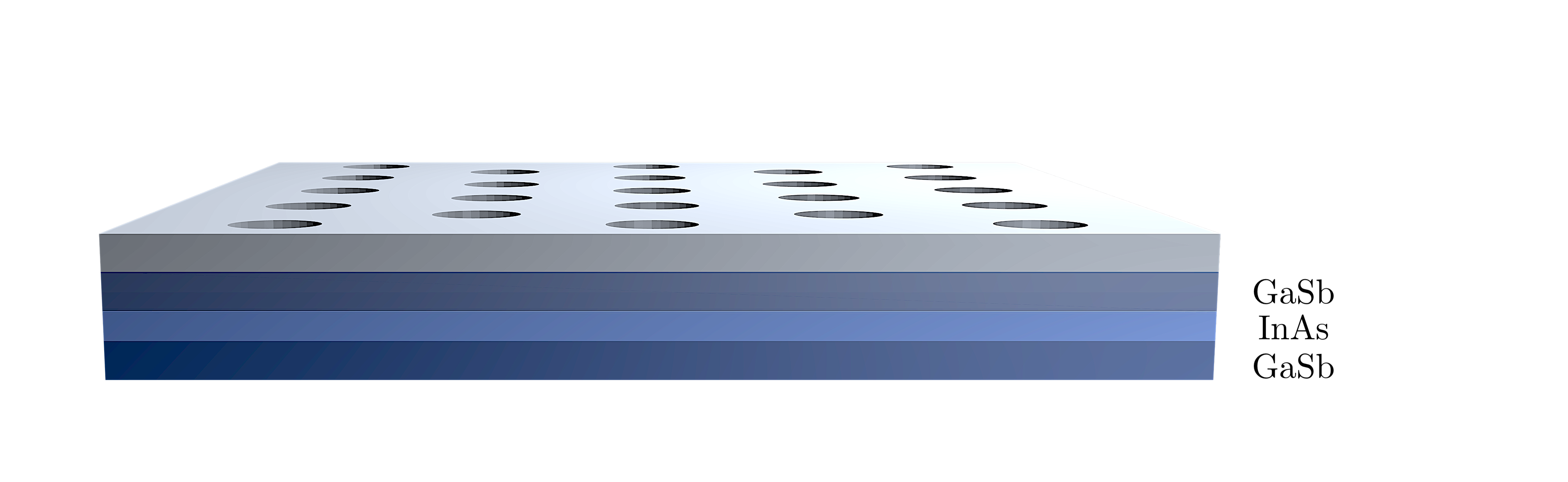}}
{\includegraphics[height=40mm]{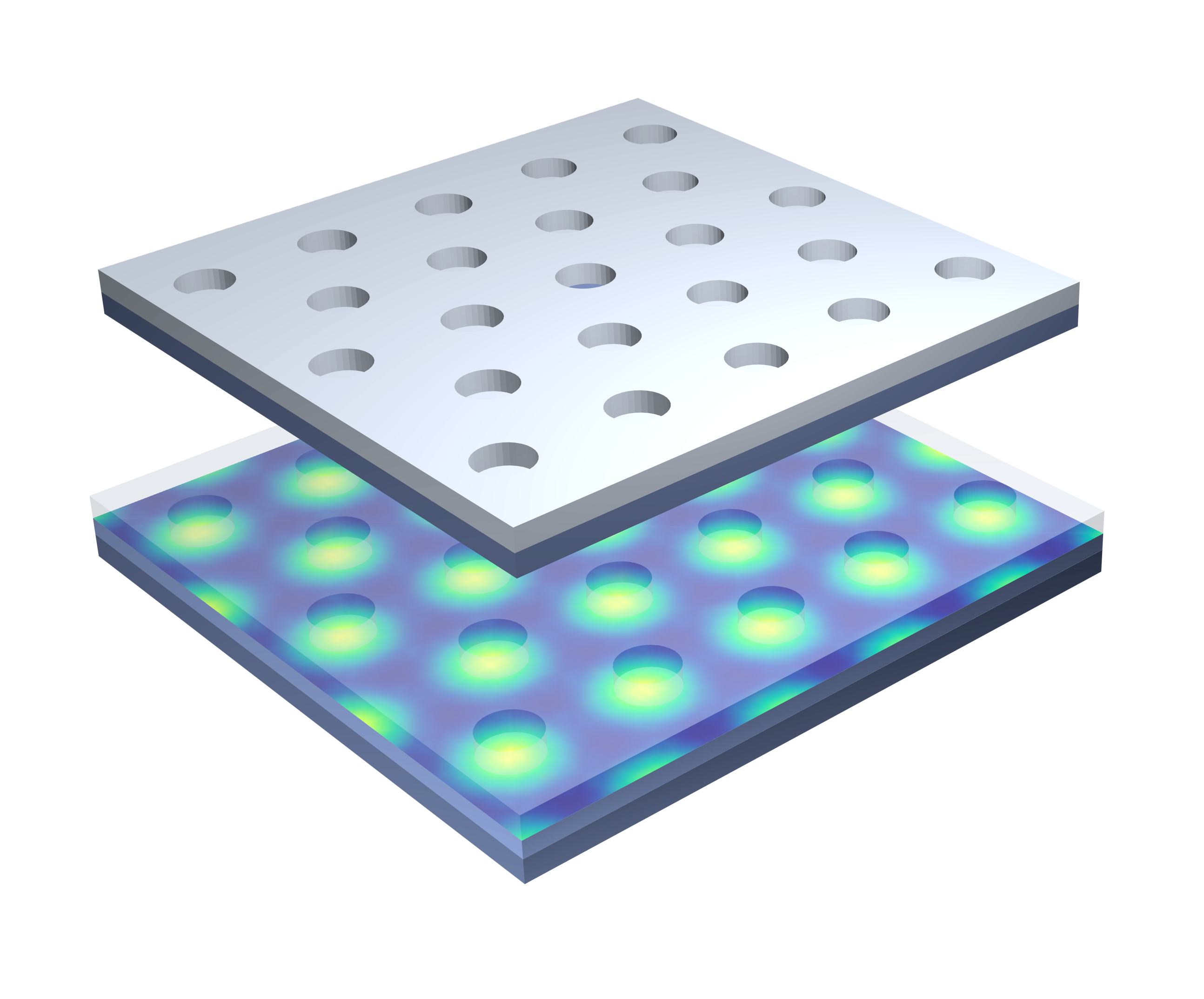}}
{\includegraphics[height=37mm]{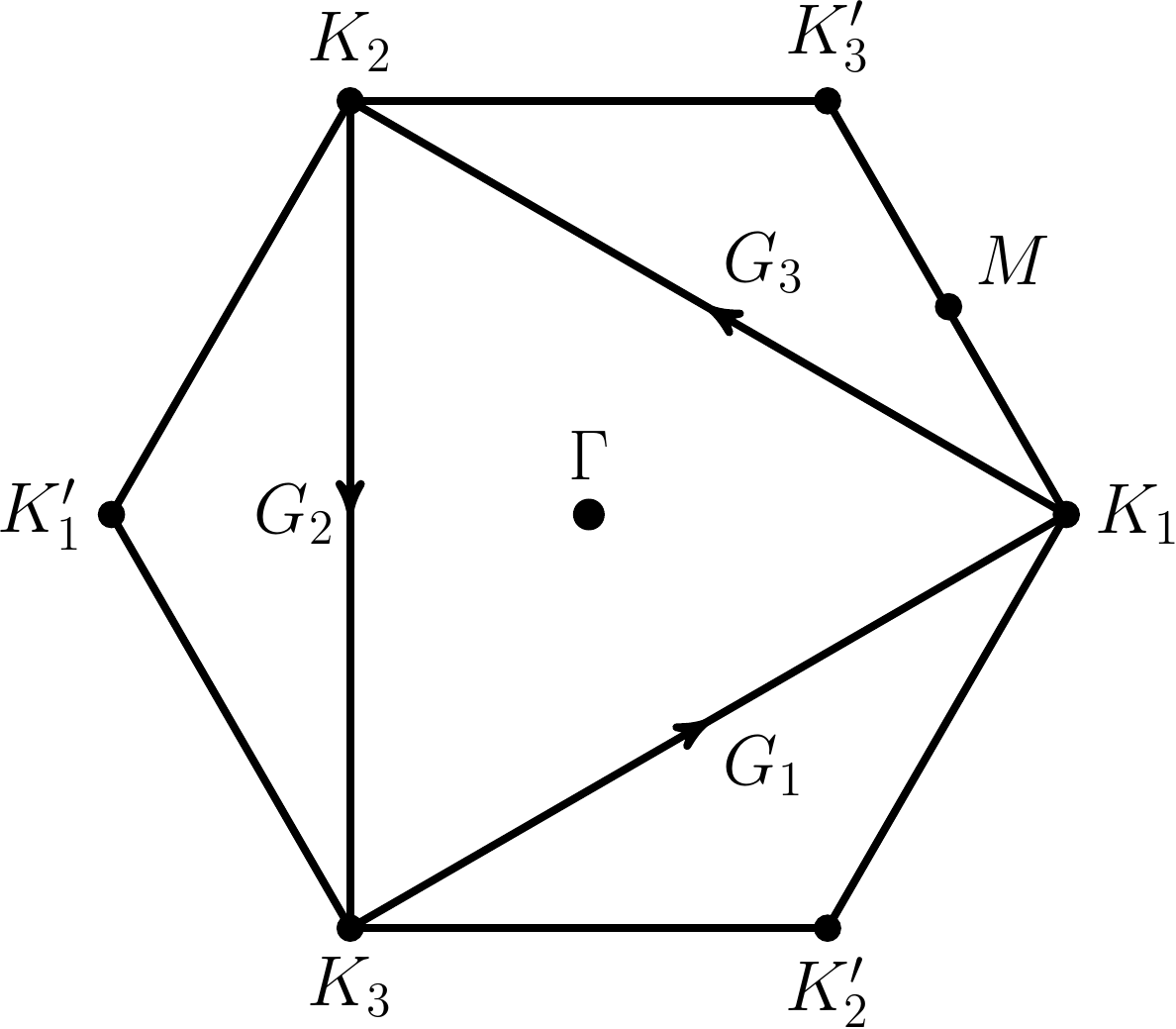}}
 \caption{ Schematic view of the honeycomb superlattice patterned on the heterostructure 2DHG -- a patterned dielectric or gate is placed on a quantum well e.g. GaSb-InAs-GaSb. Superlattice Brillouin zone: the reciprocal lattice vectors $\bm G_i$  connect zone corners corresponding to $\bm K_j$, and connect corners corresponding to $\bm K_j'$, the parity reflections of $\bm K_j$. 
 }
\label{fig:Schematics}
\end{figure}
%%%%%%%%%%%%%%%%% 

In Section \ref{hamiltonian}, we will outline how the effective Dirac theory arises from the superlattice imposed on the 2D hole gas. In Section \ref{bare}, we will discuss the form of the symmetry-allowed interactions for the effective Dirac system. Particularly important are the pseudospin-dependent Hubbard interactions; we present numerical results for these parameters. In Section \ref{screening}, we will analyse the screening properties of this system -- screening plays a crucial role for superconducting pairing mechanism, as discussed in earlier work for electrons, without spin-orbit coupling \cite{Li2020}. It was shown that the pseudospin--dependent Hubbard interactions are antiscreened (enhanced) by many-body effects; we analyse this phenomenon in the presence of spin-orbit coupling. In Section \ref{gapeqsect}, we present the solution to the BCS equations using the screened form of the interactions, and present a phase diagram. In Section \ref{topology}, we discuss the phenomenology of the possible superconducting phases, and present numerical results describing the edge physics as well as symmetry indicators which confirm the higher topology of the $p+i\tau p$ and $s_\tau$ states.

\section{Single particle effective Hamiltonian}\label{hamiltonian}

In this section, we will present the effective Dirac theory that arises for a particular honeycomb superlattice system -- $p$-type artificial graphene -- though aspects of the model apply generally. We briefly outline the schematics of artificial graphene, and in doing so establish the key parameters which may be tuned in experiment.

\subsection{Spin-orbit coupled honeycomb superlattice}
We consider a $p$-type quantum well, having in mind for e.g. a  GaSb-InAs-GaSb heterojunction (see Fig. \ref{fig:Schematics}). The hole gas experiences a potential well, arising from the band-bending along the growth direction of the heterojunction, confining the holes along the $z$-direction leaving a two dimensional hole gas (2DHG) unconfined in the $xy$ plane. The hole states are formed from $p_{\frac{3}{2}}$ orbitals and can be described by the Luttinger Hamiltonian involving spin-$\frac{3}{2}$ operators $\bm{S}$ in the axial approximation, i.e. $U(1)$ symmetry in-plane \cite{Winkler2003}. Ignoring the cubic anisotropy of the zincblende lattice, which has a weak effect for the carrier densities we consider, the Hamiltonian is
\begin{gather}
\label{HL}
H_{2DHG}=
\frac{1}{2m_e}\left[(\gamma_1 + \frac{5}{2}\gamma_2)\bm{p}^2 - 2\gamma_2 (\bm{p}\cdot\bm{S})^2\right]
 +W_c(z)
\end{gather}
The $\gamma_i$ are the Luttinger parameters; in what follows we shall use parameters for InAs, presented in Table \ref{t:table}. In this work we model the confinement as a rectangular infinite well of width $d$,
\begin{gather}
\label{sqconfine}
W_c(z) =  
\begin{cases}
  0, & z\in(-d/2,d/2)\\    
 \infty, &\text{otherwise.}    
\end{cases}
\end{gather}
The Hamiltonian (\ref{HL}) satisfies time-reversal and inversion symmetry, so each 2D subband is twofold degenerate. We consider densities for which only the lowest pair of subbands is occupied, and introduce an effective spin-$\frac{1}{2}$ degree of freedom with Pauli matrices $s_\mu$.

\begin{table}[b!]
\label{table}
\caption{Physical parameters for InAs.} % title of Table
\centering % used for centering table
\begin{tabular}{|c c c|} % centered columns (4 columns)
\hline\hline %inserts double horizontal lines
Parameter  \ \  \  & \ \  \  Details \ \  \ & \ \  \  Value \ \  \   \\ [0.5ex] % inserts table
%heading
\hline % inserts single horizontal line
$\gamma_1$ & Luttinger parameter & 20.4\\
$\gamma_2$ & Luttinger parameter & 8.3\\
$\gamma_3$ & Luttinger parameter & 9.1\\
$m_H$ & Effective mass: $m_e(\gamma_1 + \gamma_2)^{-1}$ &  $0.0348m_e$ \\
$\epsilon_r$ & Dielectric constant & 14.6\\
\hline %inserts single line
\end{tabular}
\label{t:table} % is used to refer this table in the text
\end{table}

Next, we consider the influence of  a periodic electrostatic potential, with honeycomb symmetry, on the 2DHG, i.e. the superlattice. Experimentally, this may be implemented by etching the pattern onto a metal plate or dielectric on top of the 2DHG. A minimal model of the superlattice is given by \cite{Tkachenko2015},
\begin{align}
\label{potential}
W(\bm r)&=2W_0 \sum_i \cos(\bm G_i\cdot \bm r),
\end{align}
where $\bm G_1=\bm K_2-\bm K_1, \ \bm G_2=\bm K_3-\bm K_2,\ \bm G_3=\bm K_1-\bm K_3; \bm K_1=\frac{4\pi}{3L}(1,0),\ \bm K_2=\frac{4\pi}{3L}\frac{1}{2}(-1,\sqrt{3}),\ \bm K_3=\frac{4\pi}{3L}\frac{1}{2}(-1,-\sqrt{3})$, with (super)-lattice constant $L$, and magnitude of the electrostatic potential $W_0$. The separation along the $z$-axis of the superlattice top gate from the 2DHG is $z_0$. Although $z_0$ plays a role \cite{Tkachenko2015}, we will fix its value and not consider it further. Moreover, we employ a minimal three $K$-point grid for numerical diagonalisation of $H_{2DHG}+W(\bm r)$, which is used to estimate the couplings entering the effective Dirac Hamiltonian \eqref{Hdirac}. In this scheme, $W_0$ scales out, and so will not explicitly appear as a free parameter in our analysis. The shortcomings of this approximation are discussed further in Section \ref{PhaseDiagram} in relation to the phase diagram.

%%%%%%%%%%%%%%%%%%%%%
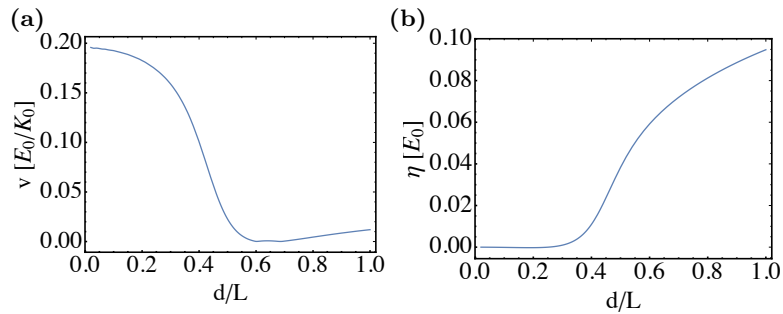
\begin{figure}[t!]
\includegraphics[width=49mm]{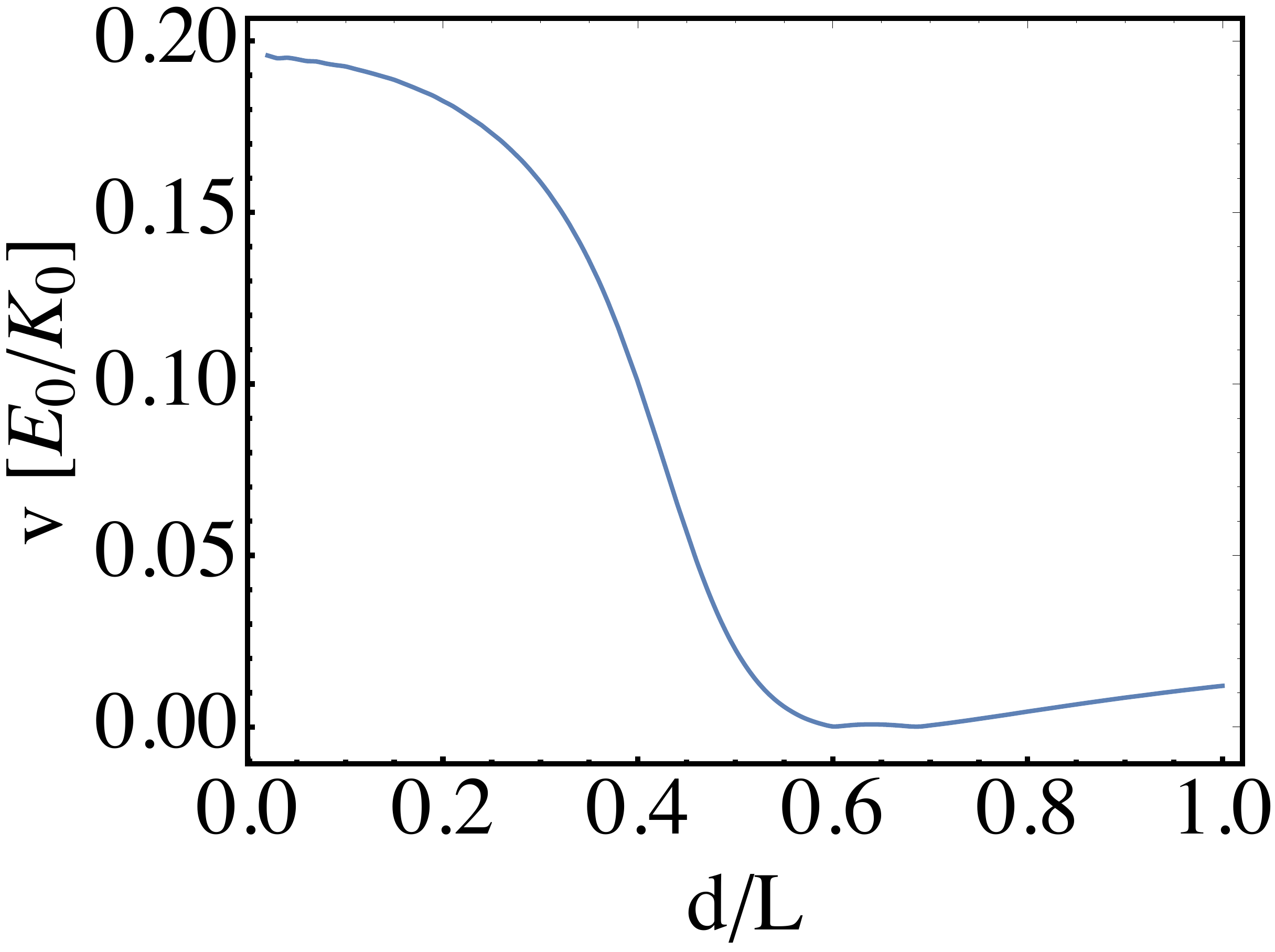}\hspace{0.1cm}
\raisebox{-0.1cm}{\includegraphics[width=50mm]{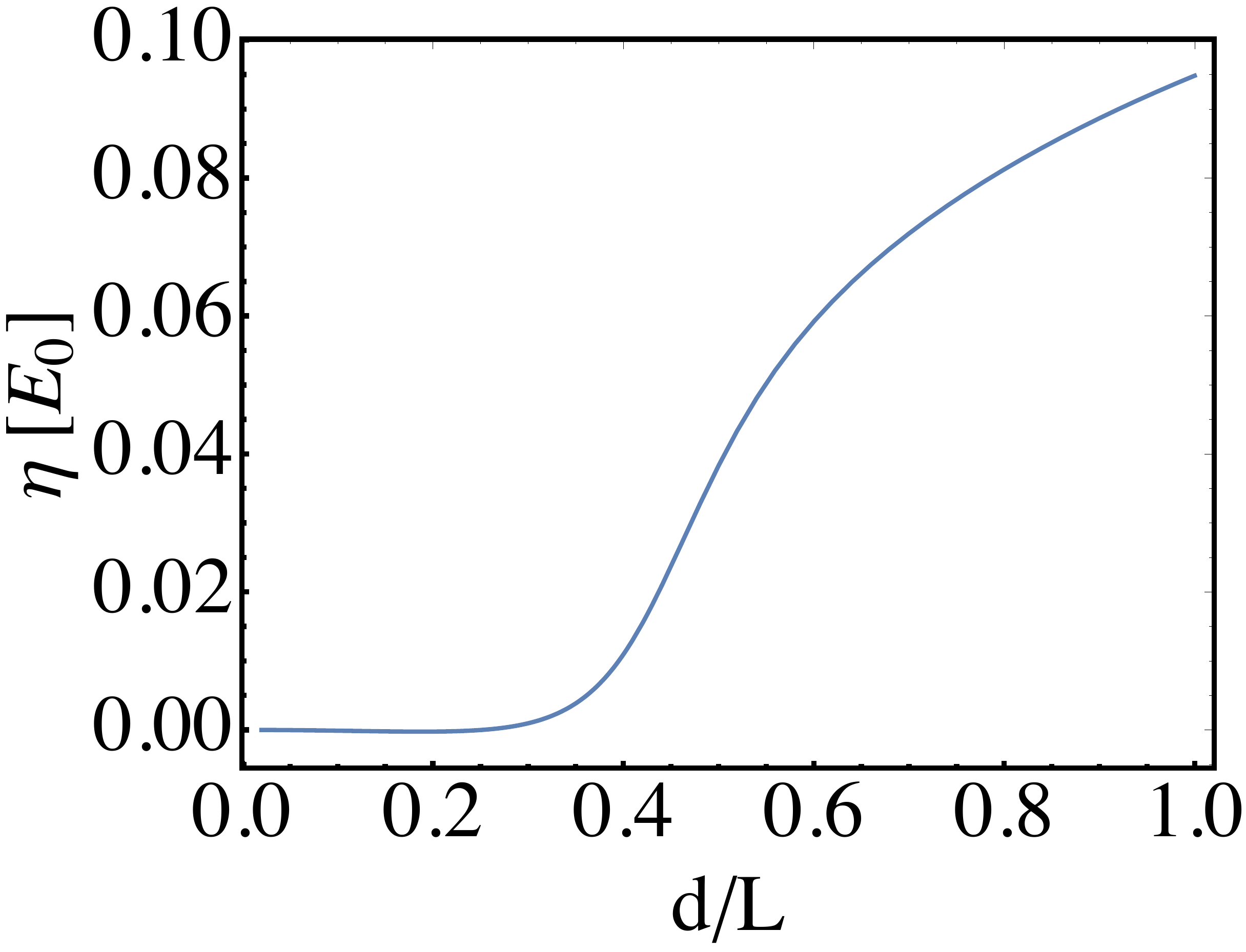}}
\caption{{Parameters of the effective Dirac Hamiltonian \eqref{Hdirac}. (a) Dirac velocity $v$, in units of $E_0/K_0$.  (b) spin-orbit gap $\eta$, in units of $E_0$.} \label{f:H0}}
\begin{picture}(0,0) 
\put(-145,175){\textbf{(a)}} 
\put(-3,175){\textbf{(b)}} 
\end{picture}
\end{figure}
%%%%%%%%%%%%%%%%%%%%%

\subsection{Effective Dirac Hamiltonian}
Since the periodic potential $W(\bm r)$ has the same symmetries as the atomic potential in graphene, the bandstructure of the hole gas with superlattice, i.e. $H_{2DHG} + W(\bm r)$, features Dirac cones at the high symmetry points $\bm K_i$. Performing this diagonalisation explicitly (see Appendix \ref{hamiltonian2}), and expanding the resulting Hamiltonian about the Dirac points, we arrive at the effective Dirac Hamiltonian with  Kane-Mele mass term \cite{Kane2005}
\begin{align}
\label{Hdirac}
H_0&=\sum_{\bm p} \psi_{\bm p}^\dag \left(v(\bm\sigma \cdot \bm{p}) \tau_z - \mu +  \eta \sigma_z  s_z\right) \psi_{\bm p}.
\end{align}
the Pauli matrices $\sigma_i$, $\tau_i$ and $s_i$ act on sub-lattice, valley, and the effective spin-$1/2$, and a chemical potential $\mu$ describes doping beyond the Dirac points. For $\tau=1$, pseudospin up (down) corresponds to sublattice $A$ ($B$), while at the opposite valley $\tau=-1$, pseudospin up (down) corresponds to sublattice $B$ ($A$). One may perform a unitary transformation so that the pseudospin has the same definition at $\tau=-1$ as it does at $\tau=1$, but intermediate calculations are made more simple in the basis of \eqref{Hdirac}. At the end of Section \ref{screening}, we shall change to the alternative basis as it makes aspects of our final results clearer.

The symmetries of the system are $2\pi/3$ and $\pi$ rotations, and time reversal.  The resulting transformation properties of the operators $\sigma_i$, $\tau_i$ and $s_i$ are given in Table \ref{T:transforms}.

The time-reversal invariant mass term $\eta \sigma_z s_z$ arises from the spin-orbit interaction and is absent in $n$-type artificial lattices. This term gives rise to a $\mathbb{Z}_2$ topological insulating state. In the Appendix we show that, in an effective tight-binding description of the artificial lattice, this term arises due to a spin-dependent complex next nearest neighbor hopping which is equivalent to two copies of the Haldane model. While the effective Dirac theory is identical to that of the Kane-Mele model, the hopping phases in the real space description are different, due to the fact that the mass term arises from a spin-orbit interaction quadratic in momentum, rather than a linear Rashba spin-orbit interaction.

\begin{table}[t!]
{
\caption{Transformation properties of operators $\sigma_i$, $\tau_i$ and $s_i$ under the symmetries of the system:  $2\pi/3$ and $\pi$ rotations $C_{3z}$, $C_{2z}$, $C_{2x}$ (for completeness we include $C_{2y}$), and time reversal $\mathcal{T}$.}

\begin{tabular}{ | c | c c c c c |  }
\hline\hline 
 & \ \ \label{T:transforms} $C_{3z}$ & \ \ $C_{2z}$ &\ \  $C_{2x}$ &\ \  $C_{2y}$ &\ \   $\mathcal{T}$ \\ 
 \hline
\ \ $s_z$ \ \ & \ \  $s_z$ &\ \  $s_z$ &\ \ $-s_z$ &\ \ $-s_z$ &\ \  $-s_z$ \\ 
 \ \ $\sigma_z$ \ \ & \ \  $\sigma_z$ &\ \  $\sigma_z$ &\ \ $-\sigma_z$ &\ \ $-\sigma_z$ &\ \  $-\sigma_z$ \\ 
 \ \ $\tau_z$ \ \  & \ \ $\tau_z$ &\ \  $-\tau_z$ &\ \  $\tau_z$ &\ \ $-\tau_z$ &\ \  $-\tau_z$ \\ 
  \ \ $\tau_\pm$ \ \ & \ \ $\tau_\pm$ &\ \  $\tau_\mp$ &\ \   $\tau_\pm$ & \ \ $\tau_\mp$ & \ \ $\tau_\mp$ \\ 
   \ \  $\sigma_\pm$ \ \ & \ \ $e^{2i\theta_\pm}\sigma_\pm$  &\ \  $\sigma_\pm$ & \ \ $\sigma_\mp$ & \ \ $\sigma_\mp$ &\ \ $\sigma_\mp$ \\
\hline 
\end{tabular}}
\end{table}

Performing exact diagonalisation of the Luttinger Hamiltonian \eqref{HL}, with parameters for an InAs 2DHG, we numerically obtain the Dirac Hamiltonian (\ref{Hdirac}).  We plot the Dirac velocity $v$ and spin-orbit mass gap $\eta$ as a function of $d/L$ in Fig. \ref{f:H0}, in terms of the scale $E_0=K_0^2/(2m_H)$, with $K_0=|\bm K_i|$. There we see that the effective Dirac velocity $v$ and the spin-orbit mass gap $\eta$ depend strongly on the ratio $d/L$.  The Dirac velocity can be significantly reduced by increasing $d/L$, thereby enhancing interaction effects {(for a full treatment of the band structure see \cite{Scammell2019, Krix2022} and explicitly for the physics of this systems in the flatband limit see \cite{Krix2022}). We sketch here the reason that increasing $d/L$ decreases the velocity $v$ and increases SOC band gap $\eta$: in the 2DHG system, the strength of the spin-orbit is encoded in the admixture of the heavy-hole (lowest bands) with the light-hole (next lowest bands). The admixture becomes significant near $p\sim 1/d$, since this is the location of the anticrossing of these bands. Now, consider applying an external modulated potential on top of the 2DHG, and consider the effective theory at the Dirac cone, ie near momenta $K=4\pi/(3L)$. If the anticrossing is band-folded to the Dirac cone, i.e. $1/d \approx 4\pi/(3L)$, then there is `large' spin-orbit. On the other hand, if $1/d\gg 4\pi/(3L)$, then the approximately quadratic part of the underlying 2DHG heavy-hole band is band-folded to the effective Dirac point, in which case there is minimal admixture of the heavy-hole and light-hole states, and therefore there a `weaker' effective spin-spin orbit. }

Finally, we comment that since the ratio $d/L$ controls the effective Dirac velocity, it also controls how localised the orbitals are. We note that such a handle is not available in the analogous electron-based superlattice honeycomb systems \cite{Li2020,Du2021}, { and is highly desirable for generating correlated phases.}

\section{Coulomb Matrix Elements}
\label{bare}
In this section, we will discuss the form of the Coulomb interaction in the effective Dirac theory. By writing the Coulomb interaction in the basis of states near the $K$ and $K'$ points, we find that the Coulomb repulsion contains a short range Hubbard part, which depends on the pseudospin $\sigma$, valley $\tau$ and effective spin $s$,  extending earlier results on these models by including spin \cite{Li2021,Li2020}. The form of these Hubbard interactions are constrained by the symmetry transformations of Table \ref{T:transforms}. Here we numerically compute the values of the symmetry-allowed matrix elements.

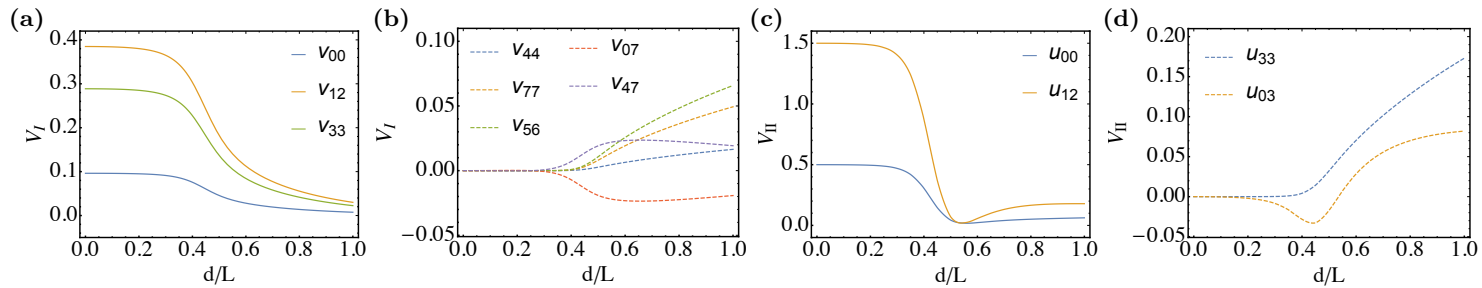
\begin{figure*}[t]
\includegraphics[width=45mm]{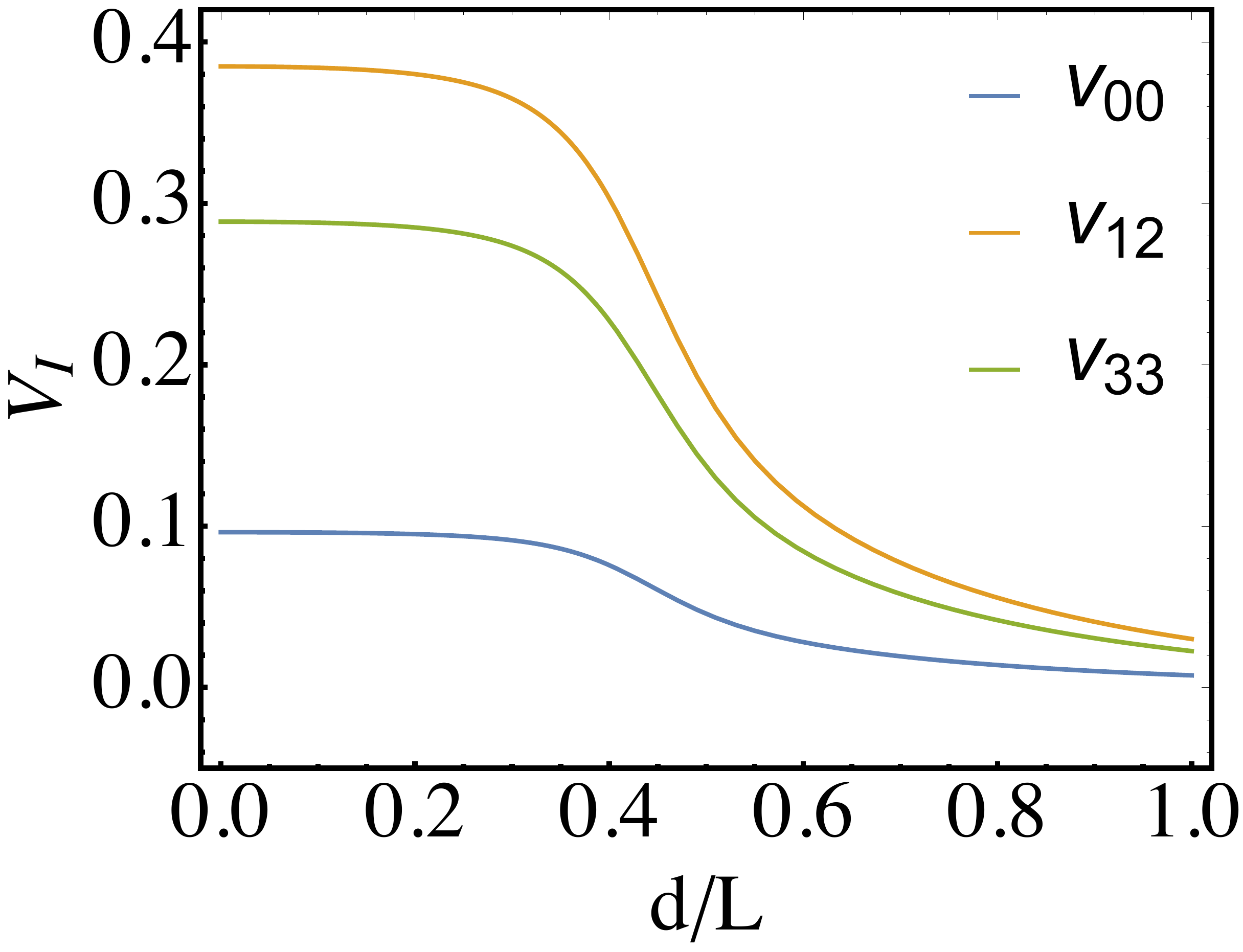}
\includegraphics[width=49mm]{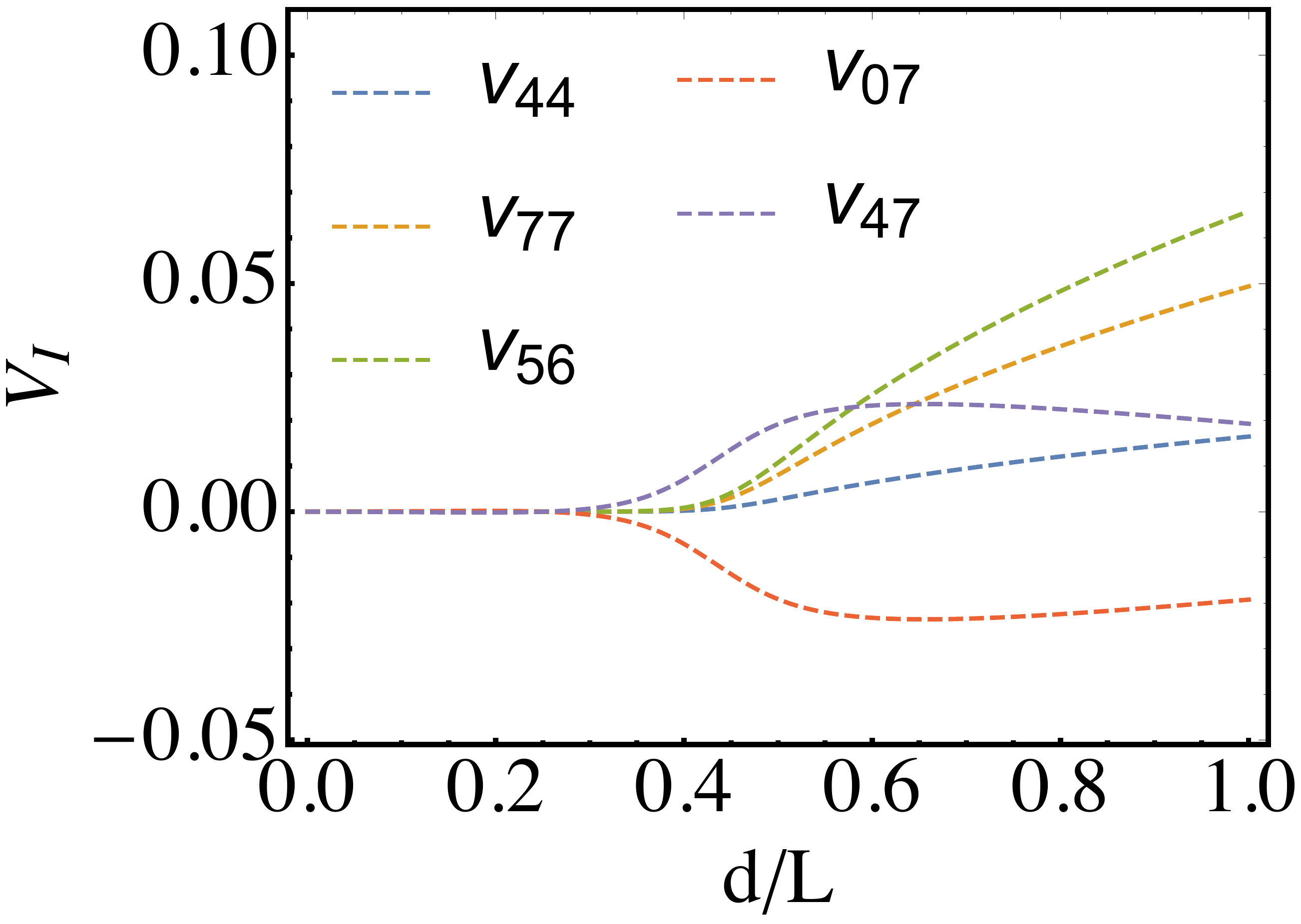}
\includegraphics[width=45mm]{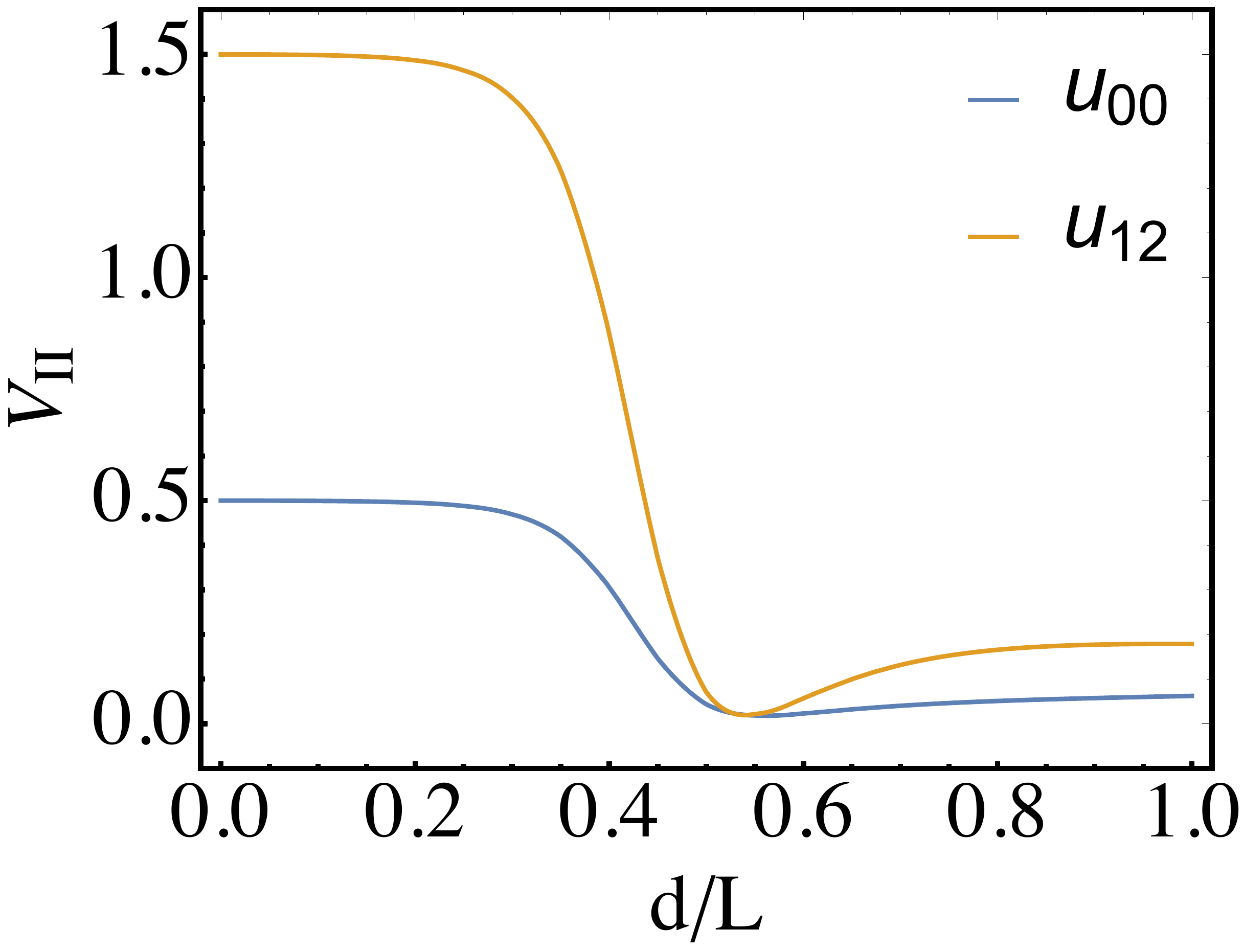}
\includegraphics[width=49mm]{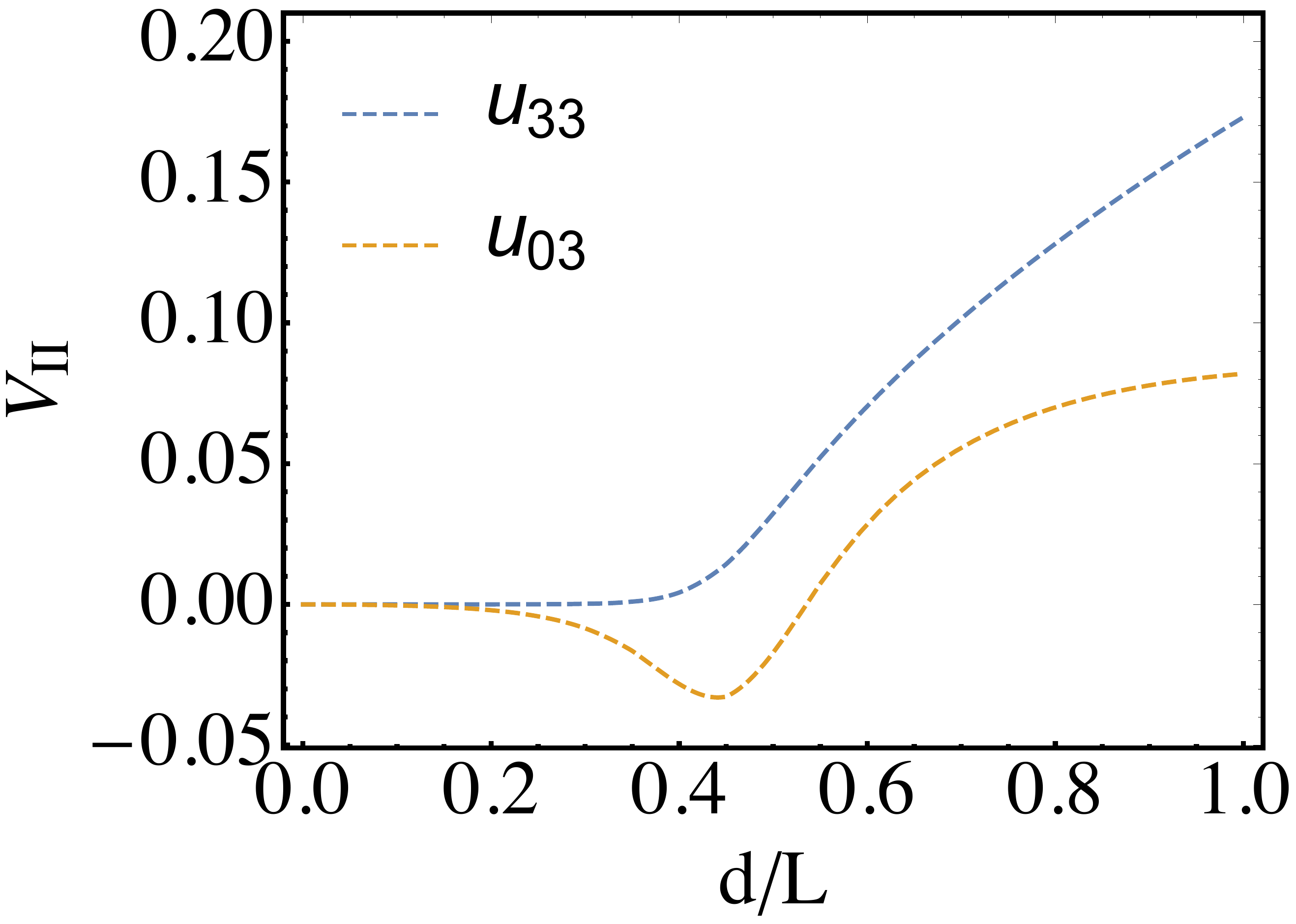}
\caption{{(a) Spin-independent matrix elements of $V_I$. Solid lines: blue, orange, green $= v_{00}, { v_{12}, v_{33}}$.  (b) Spin-dependent matrix elements of $V_I$. Dashed lines: blue, orange, green, red and purple $= v_{44}, v_{77} , v_{56} , v_{07} , v_{47}$. (c) Spin-independent matrix elements of $V_{II}$. Solid lines: blue, orange $=u_{00}, u_{12}$. (d) Spin-dependent matrix elements of $V_{II}$. Dashed lines: blue, orange $= u_{33}, u_{03}$.  In units of $2\pi e^2/(\epsilon_r K_0)$ } \label{f:bare}}
\begin{picture}(0,0) 
\put(-278,165){\textbf{(a)}} 
\put(-142,165){\textbf{(b)}} 
\put(-1,165){\textbf{(c)}} 
\put(128,165){ \textbf{(d)}} 
\end{picture}
\end{figure*}

Using the wavefunctions, $\ket{\bm{k},s,\sigma,\tau}$, obtained from diagonalisation of the InAs 2DHG subject to superlattice potential $W(\bm r)$ \eqref{potential}, i.e. $H_{2DHG} + W(\bm r)$, and expanding near the Dirac points, we explicitly compute the matrix elements of the Coulomb interaction,
\begin{widetext}
\begin{align}
\label{Vbare}
\hat{V}&={\bra{ \bm{k}_1,s_1,\sigma_1,\tau_1}\otimes\bra{\bm{k}_2,s_2,\sigma_2,\tau_2}\frac{e^2}{2\varepsilon_r |\bm r-\bm r'| }\ket{\bm{k}_3,s_3,\sigma_3,\tau_3}\otimes\ket{\bm{k}_4,s_4,\sigma_4,\tau_4}}  \equiv \frac{2\pi e^2}{\epsilon_r q} + \hat{V}_I + \hat{V}_{II},\\
\notag \hat{V}_I&=\left(v_{00} \sigma_0\otimes\sigma_0+ v_{44}   \tau_z s_z\otimes \tau_z s_z\right)+\left(v_{33} \tau_z\otimes\tau_z+ v_{77}  \ s_z\otimes s_z\right)\sigma_z\otimes\sigma_z +\left(v_{12} + v_{56} \tau_z s_z\otimes  \tau_z s_z\right)\left(\sigma_+\otimes \sigma_- + \sigma_-\otimes\sigma_+\right)\\
\notag  &+v_{07} (\sigma_0\otimes \sigma_z s_z+\sigma_z s_z \otimes \sigma_0)+v_{47} (\tau_z s_z \otimes \sigma_z s_z + \sigma_z s_z \otimes \tau_z s_z),\\
\notag\hat{V}_{II}&=\Big[u_{00} \sigma_0\otimes\sigma_0+ u_{33}s_z\sigma_z\otimes s_z\sigma_z + u_{12} \left(\sigma_+\otimes\sigma_- + \sigma_-\otimes\sigma_+\right)+u_{03} (\sigma_0\otimes \sigma_z s_z+\sigma_z s_z \otimes \sigma_0)\Big]\left(\tau_+\otimes\tau_- + \tau_-\otimes\tau_+\right),
\end{align}
\end{widetext}
Here $\bm q= \bm k_1-\bm k_3$, and subscripts $I$ and $II$ denote intravalley ($\tau$-diagonal) and intervalley ($\tau$-off-diagonal) interactions.  The vertices appearing in the bare interactions are $J_I^\mu\in\{\mathbbm{1}, \sigma_\pm, \tau_z \sigma_z,  \tau_z s_z, \tau_z s_z \sigma_\pm, \sigma_z s_z\}$,
 $J_{II}^\mu\in\{\mathbbm{1}, \sigma_\pm, \sigma_z s_z\}\otimes\tau_\pm$, which defines the adjoint basis. Using these vertices, the interactions are parametrized $\hat{V}^0_I= v_{\mu\nu} \ J_I^\mu \otimes J_I^\nu$ and $\hat{V}^0_{II}= u_{\mu\nu} \ J_{II}^\mu \otimes J_{II}^\nu$, which defines the notation in Eq. \eqref{Vbare}. In Figure \ref{f:bare} we plot the dependence of the coefficients $\{v_{\mu\nu}, u_{\mu\nu}\}$ on the spin-orbit parameter, $d/L$.

\section{Screening}
\label{screening}
In this section we discuss how the bare Coulomb interactions \eqref{Vbare} are modified by screening. A standard approach for analysing the feedback of many body effects on interactions is the Random Phase Approximation \cite{Li2021,Li2020,KotovN,Wang2006,Thakur2016,Pyatkovskiy2008}, which involves resumming the infinite series of bubble diagrams which contribute corrections to the bare Coulomb interaction.

 %%%%%%%%%%%%%%%%%%%%%
%%%%%%%%%%%%%%%%%%%%%

The resulting screened interactions $V^R_{\mu \nu}(p_0,\bm p)$ are given by
\begin{align}
\label{VRPA}
V^R_{\mu \nu}(p_0,\bm p) = V_{\mu \nu} + V_{\mu \alpha} \Pi^{\alpha \gamma} (p_0,\bm p) V^R_{\gamma \nu} (p_0,\bm p)
\end{align}
where $\Pi^{\alpha \gamma}$ is the particle-hole polarisation operator, given by 
\begin{gather}
i\Pi^{\alpha \gamma}(p_0,\bm p) =\text{Tr} \int{
J^\alpha G(q_0+p_0, \bm{q}+\bm{p}) J^\gamma G(q_0, \bm{q}) \frac{ dq_0 d^2\bm{q}}{(2\pi)^3}
} \ \ , \nonumber \\
G(q_0, \bm{q}) = \frac{1}{ q_0+\mu - v \tau_z \bm q  \cdot \bm \sigma-\eta \sigma^z s^z +i 0 \text{sgn}(q_0)}
\label{Pi}
\end{gather} 
where $G(q_0, \bm{q})$ is the single particle Green's function. In general, the vertices $J^\mu, J^\nu$ can be any matrix $\sigma^i \tau^j s^k$ which appears in the bare interactions of the form $ V_{\mu\nu} J^\mu \otimes J^\nu$.  In this paper we will restrict our attention to the case of static screening, so we neglect the frequency dependence of the polarisation operator and set $p_0=0$.

As shown in Section \ref{bare}, the vertices appearing in the bare interactions \eqref{Vbare} are $J_I^\mu\in\{\mathbbm{1}, \sigma_\pm, \tau_z \sigma_z,  \tau_z s_z, \tau_z s_z \sigma_\pm, \sigma_z s_z\}$,
 $J_{II}^\mu\in\{\mathbbm{1}, \sigma_\pm, \sigma_z s_z\}\otimes\tau_\pm$, which defines the adjoint basis. In this basis, the tensor form of the static polarisation operator becomes, 
\begin{align}
\notag\hat{\Pi}_I(p_0=0, \bm p)&=\Pi^{\mu\nu}(0, \bm p)J_I^\mu J_I^\nu, \\
\hat{\Pi}_{II}(p_0=0, \bm p)&=\Pi^{\mu\nu}(0, \bm p)J_{II}^\mu J_{II}^\nu. \
\end{align}
The quantities $\Pi^{\mu\nu}(0, \bm p)$ are evaluated in the Appendix \ref{a:pol}. We find that only { four} independent polarisation operators emerge. To gain some insight into their physical meaning, we shall discuss their behavior in the long wavelength limit, $q\rightarrow0$. First, we find a term $\Pi_0 \rightarrow -\mu N/(2\pi)$, this term corresponds to the usual density-density (Thomas-Fermi) screening, i.e. the vertices coupling to a negative polarisation operator are weakened by screening. Second,  $\Pi_z \rightarrow \mu N/(2\pi)$, which corresponds to a pseudospin dipole-dipole {\it anti}screening, first discussed in \cite{Li2020} -- the positive sign here causes an enhancement of the couplings $v_{33}(\tau_z\sigma_z\otimes\tau_z\sigma_z)$ and $v_{77}(s_z\sigma_z\otimes s_z\sigma_z)$, i.e. those proportional to $\sigma_z\otimes\sigma_z$, which as we shall later see promotes an intravalley $p+i\tau p$ higher topological superconductivity. Similarly, an antiscreening occurs for intervalley terms $u_{00} \sigma_0\tau_\pm\otimes \sigma_0\tau_\mp, u_{33} s_z\tau_\pm\otimes s_z\tau_\mp$, which acts to favour the higher topological intervalley $ s_\tau$ state. We shall elaborate on this phenomenon in the following subsection. Third, we have $\Pi_\eta =\rightarrow \eta N/(2\pi)$, a direct result of the spin-orbit coupling. Last, we have $\Pi_\pm \propto e^{i\theta_{\bm p}}$, this Hall-like response, with momentum dependence, promotes interaction matrix elements that were otherwise not present in the bare interaction structure (\ref{Vbare}).

Inverting the matrix equation (\ref{VRPA}) is now straightforward and gives,
\begin{align}
\notag 
\hat{V}^R_I&=v^{R}_{\mu\nu} J_I^\mu J_I^\nu,\ \  \hat{V}^R_{II}=u^{R}_{\mu\nu} J_{II}^\mu J_{II}^\nu, 
\end{align}
with superscript $R$ to denote the RPA renormalised values. Despite there being a closed form analytic expression, we do not provide the full expressions for matrix elements $v_{\mu\nu}, u_{\mu\nu}$ since they are lengthy and unenlightening. The expression (\ref{Vrpa}) defines the RPA-renormalised interaction structure which we use to search for superconducting and magnetic instabilities.

Up until this point we have worked in a particular basis for the single particle Hamiltonian (\ref{Hdirac}), which allowed for  straightforward evaluation of the polarisation operators. However, from this point on we work in a more physical basis, which will make our later discussion of the superconducting gap structure  more transparent. Performing a unitary transformation, with $P=\frac{1}{2}(\tau_0+\tau_z) + \frac{1}{2}(\tau_0-\tau_z)\sigma_x$, we obtain
\begin{align}
\label{HdiracDASH}
\widetilde H_0=P H_0 P^\dag&=v p_x\sigma_x\tau_z +v p_y\sigma_y {- \mu \sigma_0} + \eta s_z \sigma_z\tau_z.
\end{align}
The interactions transform as 
\begin{align}
\label{Vrpa}
 \hat{\widetilde V}^R_I &=v^R_{\mu\nu} \left(P J_I^\mu P^\dag\right) \left(P J_I^\nu P^\dag\right), \\ \notag \hat{\widetilde V}^R_{II} &=u^R_{\mu\nu} \left(P J_{II}^\mu P^\dag\right) \left(P J_{II}^\nu P^\dag\right),\\
\notag P J_I^\mu P^\dag &\in P\{\mathbbm{1}, \tau_z s_z,  \tau_z \sigma_z,  \sigma_z s_z, \sigma_\pm, \tau_z s_z \sigma_\pm\}  P^\dag\\
\notag & = \{\mathbbm{1},  \tau_z s_z,  \sigma_z,  \tau_z \sigma_z s_z, \sigma^{\tau}_\pm, \tau_z s_z \sigma^{\tau}_\pm\} ,\\
\notag P J_{II}^\mu P^\dag& \in P\{\mathbbm{1}, \sigma_z s_z, \sigma_+, \sigma_-\}\otimes\tau_\pm  P^\dag \\
\notag & =  \{\sigma_x, \pm i\sigma_y s_z, (\sigma_0 \pm \sigma_z)/2, (\sigma_0 \mp \sigma_z)/2 \}\otimes\tau_\pm.
\end{align} 
Here $\sigma_\pm^\tau \equiv  \sigma_x \pm  i\tau_z \sigma_y$, and in $P J_{II}^\mu P^\dag$, the $\pm$ indices in pseudospin and valley are connected.

{ `Pseudospin pairing' refers to the effective attraction mediated by the emergent quantum numbers $\sigma$ and $\tau$, as discussed in \cite{Li2020} and reviewed briefly in the next section. In the basis of \eqref{HdiracDASH}, the pseudospin is $\sigma_z$, but we refer to pseudospin pairing more loosely as including pairing from the $\tau_\pm$ interactions as these also act on emergent two component wavefunctions.}

\section{Superconducting instabilities}
\label{gapeqsect}
In this section we analyse superconductivity resulting from the renormalised pseudospin dependent couplings. At a finite doping away from the Dirac point, the states at the Fermi surface are not pseudospin eigenstates, but band eigenstates. The interactions (\ref{Vrpa}) are therefore be rewritten in the basis of band indices, and furthermore since we are only interested in Fermi surface instabilities, we project onto the upper band (i.e. only include states at the Fermi surface). The BCS gap equation is then used to calculate $T_c$ for pairing between these states.

\subsection{Interactions in the Cooper channel}
To find the superconducting instability, we are interested only in states near the Fermi surface, which participate in pairing. Hence, we keep only states in the upper band of (\ref{HdiracDASH}),
the eigenstates of which are given by
\begin{align}
|\bm k,\tau,{s}\rangle = \frac{1}{\sqrt{2}}e^{i\bm k \cdot \bm r} (w^a_{\tau,s}(k)|a\rangle + w^b_{\tau,s}(k)e^{i \tau \theta_{\bm{k}}} |b\rangle ) \ \ ,
\label{Diracwf}
 \end{align}
 where $|a \rangle,|b\rangle$ are the $\sigma^z$ eigenstates, which are localised on the $A$ and $B$ sites respectively, and the wavefunction components,  $w^a_{\tau,s}(k)= v k/\sqrt{2\epsilon_k(\epsilon_k - s\tau \eta)}$, $w^b_{\tau,s}(k)= (\epsilon_k-s\tau\eta) w^a_{\tau,s}(k)/(\tau v k)$. { Note that the `upper band' in question are the positive energy states of the Dirac theory. These originate from the lowest doubly-degenerate bands in the original Luttinger Hamiltonian Eq. \eqref{HL} (see the solid line in Appendix \ref{hamiltonian2} Fig. \ref{fig:2dhg}a).}

 To obtain the interactions between Cooper pairs, we perform the following process: ($i$) project the RPA interaction tensor  (\ref{Vrpa}) onto the upper band using (\ref{Diracwf}), ($ii$) impose the scattering conditions of the Cooper channel $\bm k_1=-\bm k_3$, $\bm k_2=-\bm k_4$, i.e. $\theta_{k_3}=\pi + \theta_{k_1}$, $\theta_{k_4}=\pi + \theta_{k_2}$, ($iii$) restrict all momenta to lie on the Fermi surface $|\bm k_i|=k_F$. The interactions then only have angular dependence, and we decompose the resulting Cooper interaction into partial waves with different angular momentum. The result is the coupling between Cooper pairs with a given angular momentum.
 
We arrive at the couplings in angular momentum channels $\ell=0,\pm 1$ (the $|\ell|>1$ channels are negligible or zero),
\begin{align}
\label{Vcooperl0}
\hat{\cal V}_{\ell=0}&= \tilde{g}_{0}  +  \tilde{g}_{1} \tau_z \otimes \tau_z  +\tilde{g}_{2} s_z \otimes s_z + \tilde{g}_{3} s_z \tau_z \otimes  s_z \tau_z\\
\notag & + \left(\tilde{j}_{0} + \tilde{j}_{1} s_z \otimes s_z \right)  (\tau_x \otimes \tau_x+\tau_y \otimes \tau_y)\\
\label{Vcooperl1}
\hat{\cal V}_{\ell=\pm 1}&= g_{0}  +  g_{1} \tau_z \otimes  \tau_z + g_{2} s_z \otimes  s_z +g_{3} s_z \tau_z \otimes  s_z \tau_z \\
\notag &+  \ell \left(g_{4}  +  g_{5} s_z s_z\right) (\tau_0 \tau_z + \tau_z \tau_0) \\
\notag  & + \left(j_{0} + j_{1} s_z \otimes  s_z + \ell j_{2} (s_0 \otimes  s_z+ s_z \otimes s_0)\right)  (\tau_x \otimes  \tau_x+\tau_y \otimes  \tau_y)
\end{align}
The coefficients $g_i, j_i, \tilde{g}_i, \tilde{j}_i$ are functions of chemical potential $\mu$ due to the screening effects, as well as the well width to lattice spacing ratio $d/L$, which controls the strength of the spin-orbit dependent couplings. The (un)tilded couplings correspond to the ($\ell=\pm 1$) $\ell=0$ partial wave channels. They also depend on the microscopic parameters of the 2DHG; we have evaluated these quantities numerically for an InAs 2DHG. The matrix elements $g_i$ denote intravalley scattering processes, while $j_i$ represent intervalley scattering.

\subsection{Gap equation}
The mean field Hamiltonian, which accounts for all pairing possibilities, is 
\begin{align}
\label{H_MF}
&{\cal H}_{MF}=\sum_{\bm k, s, \tau}\varepsilon_{\bm k} \widetilde{\psi}^\dag_{\bm k s\tau}\widetilde{\psi}_{\bm k s\tau} +\frac{1}{2}\sum_{\bm k,s, \tau,s', \tau'} \widetilde{\psi}^\dag_{\bm k s\tau} \left(\Delta_{\bm k} \right)_{s \tau, s' \tau'} \widetilde{\psi}^\dag_{-\bm k s' \tau'} + \text{h.c.} \nonumber \\
 &+ \frac{1}{2}( \Delta_{\bm k}^\dagger )_{s_1 \tau_1, s_3 \tau_3} \left({\cal V}^{-1}\right)_{\bm k ,\bm p;s_1\tau_1 s_2\tau_2s_3\tau_3s_4\tau_4}\left(\Delta_{\bm p} \right)_{s_4 \tau_4, s_2 \tau_2} 
\end{align}
where $\widetilde{\psi}^\dag_{\bm k s\tau}$ is the hole creation operator for the upper band. We parametrize the gap in the standard form, collecting the spin, valley and angular momentum structure into a tensor  $d_{\ell}^{\mu\nu}$, 
\begin{align}
\Delta_{\bm k}&=  \sum_{\mu\nu, l}d_\ell^{\mu\nu} {s}_\mu \tau_\nu e^{-i \ell \theta_{\bm k}}  \left( \tau_y s_y\right)
%\notag \Delta_{\bm k}\Delta_{\bm k}^\dag&=|d^{\mu\nu}|^2,\\
\end{align} 
The spin and valley structure follows from the usual singlet-triplet decomposition (dropping the angular momentum index $\ell$) \cite{Sigrist2005}, 
\begin{align}
\notag d^{\mu\nu}&=d_s^{\mu}\otimes d_\tau^{\nu},\\
\notag d_s^x&=\frac{1}{2}\left(\ket{\uparrow\uparrow}-\ket{\downarrow\downarrow}\right) \ \ &&d_s^y=\frac{1}{2i}\left(\ket{\uparrow\uparrow}+\ket{\downarrow\downarrow}\right)\\
 \notag d_s^z&=-\frac{1}{2}\left(\ket{\uparrow\downarrow}+\ket{\downarrow\uparrow}\right) \ \  &&d_s^0=\frac{1}{2}\left(\ket{\uparrow\downarrow}-\ket{\downarrow\uparrow}\right),\\
\notag d_\tau^x&=\frac{1}{2}\left(\ket{++}-\ket{--}\right) \ \ &&d_\tau^y=\frac{1}{2i}\left(\ket{++}+\ket{--}\right)\\
\notag  d_\tau^z&=-\frac{1}{2}\left(\ket{+-}+\ket{-+}\right) \ \ &&d_\tau^0=\frac{1}{2}\left(\ket{+-}-\ket{-+}\right).
\end{align}
where subscript $s$ indicates spin and $\tau$ indicates valley. The BCS gap equation is given by
\begin{align}
\notag d_\ell^{\mu\nu}&=-G^\ell_{\mu\nu;\delta\gamma} d_\ell^{\delta\gamma}\int_0^{\varepsilon_c}\frac{N d\varepsilon}{2\pi v^2} \frac{ \varepsilon }{2E(d_\ell^{\delta\gamma})}\tanh\left(\frac{E(d_\ell^{\delta\gamma})}{2T}\right),\\
E(d_\ell^{\delta\gamma})&= \sqrt{(\varepsilon-\mu)^2+|d_\ell^{\delta\gamma}|^2}
\end{align}
where the matrix  $G_{\mu\nu;\delta\gamma}$ is given by
\begin{align}
\label{Gtensor}
G^\ell_{\mu\nu;\delta\gamma}&\equiv \frac{1}{4}(\hat{s}_\mu\hat{s}_y)^\dag_{ac}(\hat\tau_\nu\hat\tau_y)^\dag_{a'c'}  (\hat{\cal V}_{\ell})_{abcd;a'b'c'd'} (\hat{s}_\delta\hat{s}_y)_{bd}(\hat\tau_\gamma\hat\tau_y)_{b'd'}   
\end{align}
To determine the dominant instability $d^{\mu\nu}$, we find the gap function with highest $T_c$ via the eigenvalue problem (with eigenvalue $\lambda_{\mu\nu}^\ell$)
\begin{align}
    G^\ell_{\mu\nu;\delta\gamma}d^{\delta\gamma}_\ell = \lambda^\ell_{\mu\nu} d^{\mu\nu}_\ell.
\end{align}
Substitution of the eigenvectors $d^{\mu\nu}$ into the gap equation then results in
\begin{gather}
  \notag  1 = -\nu_0 \lambda^\ell_{\mu\nu} L(T_c,\mu,\epsilon_c), \\
  \label{BCSgapeq}
     L(T_c,\mu,\epsilon_c)=\int_0^{\varepsilon_c} \frac{ d\varepsilon \varepsilon/\mu }{2|\varepsilon-\mu|}\tanh\left(\frac{|\varepsilon-\mu|}{2T_c}\right), \ \ \nu_0=\frac{N\mu}{2\pi v^2}, 
\end{gather}
where $\nu_0$ is the density of states at the Fermi level, and $\varepsilon_c$ is an ultraviolet cut-off. The logarithmic behavior of $L(T_c,\mu,\epsilon_c)$ gives rise to the exponential dependence of $T_c \sim \varepsilon_c e^{-1/(\nu_0 \lambda^\ell_{\mu\nu})}$ on the density of states $\nu_0$ and the eigenvalue $\lambda_{\mu\nu}$, which must be negative for the the gap equation to have a solution, corresponding to an attractive interaction. 

Using the explicit form of the interactions (\ref{Vcooperl0}) and (\ref{Vcooperl1}), we find the three dominant gap structures, with the following (negative) eigenvalues of ${G}$ (\ref{Gtensor}),
\begin{subequations}
\begin{align}
\label{gapeigs1}
\ d_{\ell=\pm1}^{x\mp}: \ \lambda^{\mp 1}_{x\pm}&= g_0 + g_1 + g_2 + g_3 - g_4 - g_5,\\
\label{gapeigs2}
\ d_{\ell=\pm1}^{zz}: \ \lambda^{\pm 1}_{zz}&= g_0 - g_1 -g_2 + g_3 +j_0+j_1,\\
\label{gapeigs3}
d_{\ell=0}^{z0}: \ \lambda^{0}_{z0}&= \tilde{g}_0 - \tilde{g}_1 -\tilde{g}_2 + \tilde{g}_2 - \tilde{j}_0-\tilde{j}_1.
\end{align}
\end{subequations}
These gap structures will be described in detail in Section \ref{PhaseDiagram}. 

We pause to discuss the mechanism of attraction explicitly in reference to these eigenvalues \eqref{gapeigs1}-\eqref{gapeigs3}. We focus on terms that do not contain $s_z$ in \eqref{Vcooperl0} and \eqref{Vcooperl1}, since these drive the transition, while $s_z$ dependent terms act to fix the spin orientation of the corresponding spin-triplet states.

For $\ell=0$, we identify the driving term for superconductivity as $\tilde{j}_0 (\tau^x\otimes \tau^x+\tau^y\otimes \tau^y)$, whereas for $\ell=\pm 1$, the driving term for superconductivity is $g_0 (\tau_0\otimes\tau_0)$. The coupling $\tilde{j}_0$ is positive, and antiscreening increases its magnitude as the chemical potential increases. Hence choosing a valley singlet structure generates a negative eigenvalue $-\tilde{j}_0$, analogous to how antiferromagnetism promotes spin singlet pairing. Antiscreening in $g_0$ manifests as a sign change -- for large enough chemical potential, $g_0$ is overscreened and becomes negative, as has been previously discussed in Refs \cite{Li2021,Li2020}.

\subsection{Explicit solution and phase diagram}\label{PhaseDiagram}

\begin{figure*}[t]
\includegraphics[width=64mm]{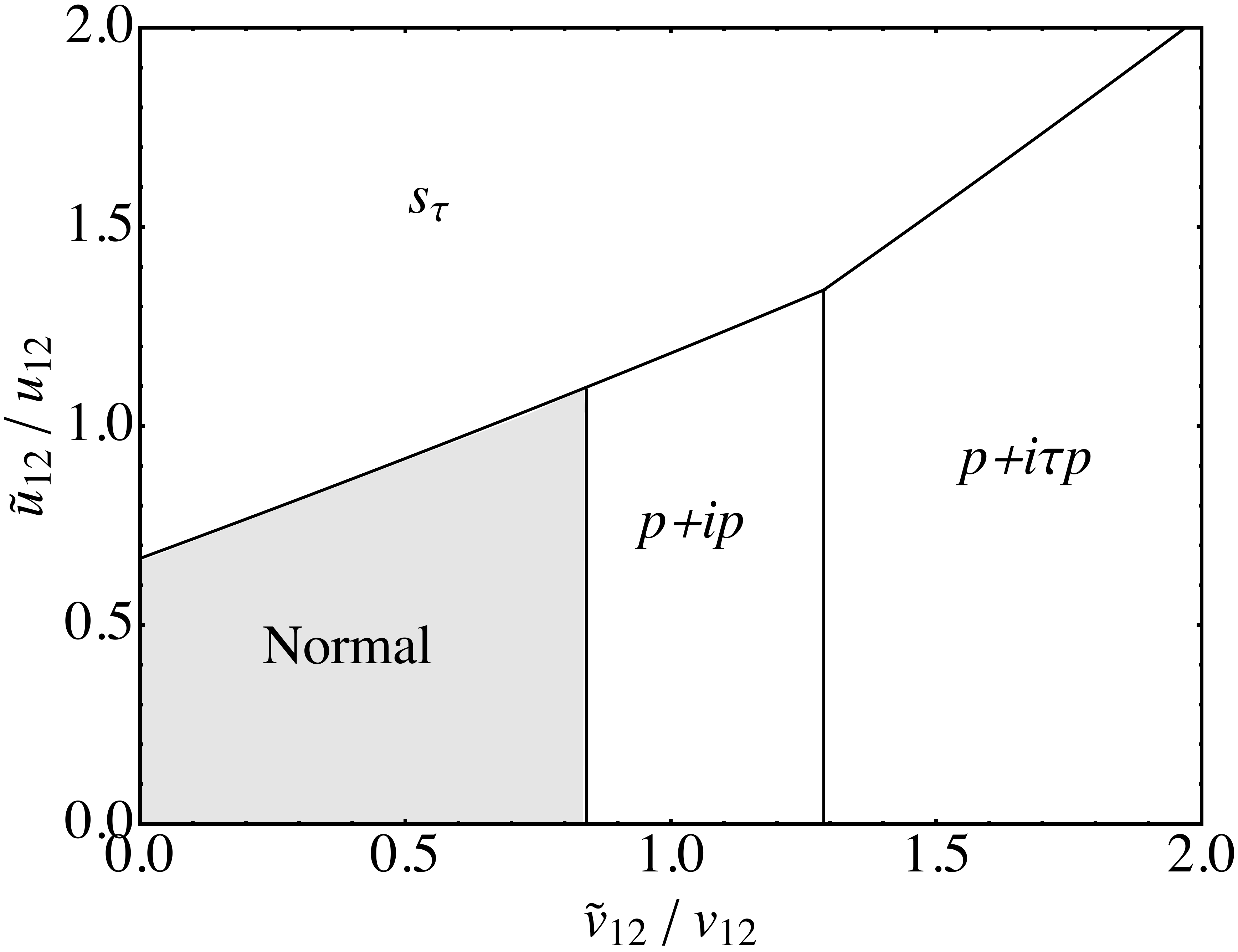}
\includegraphics[width=64mm]{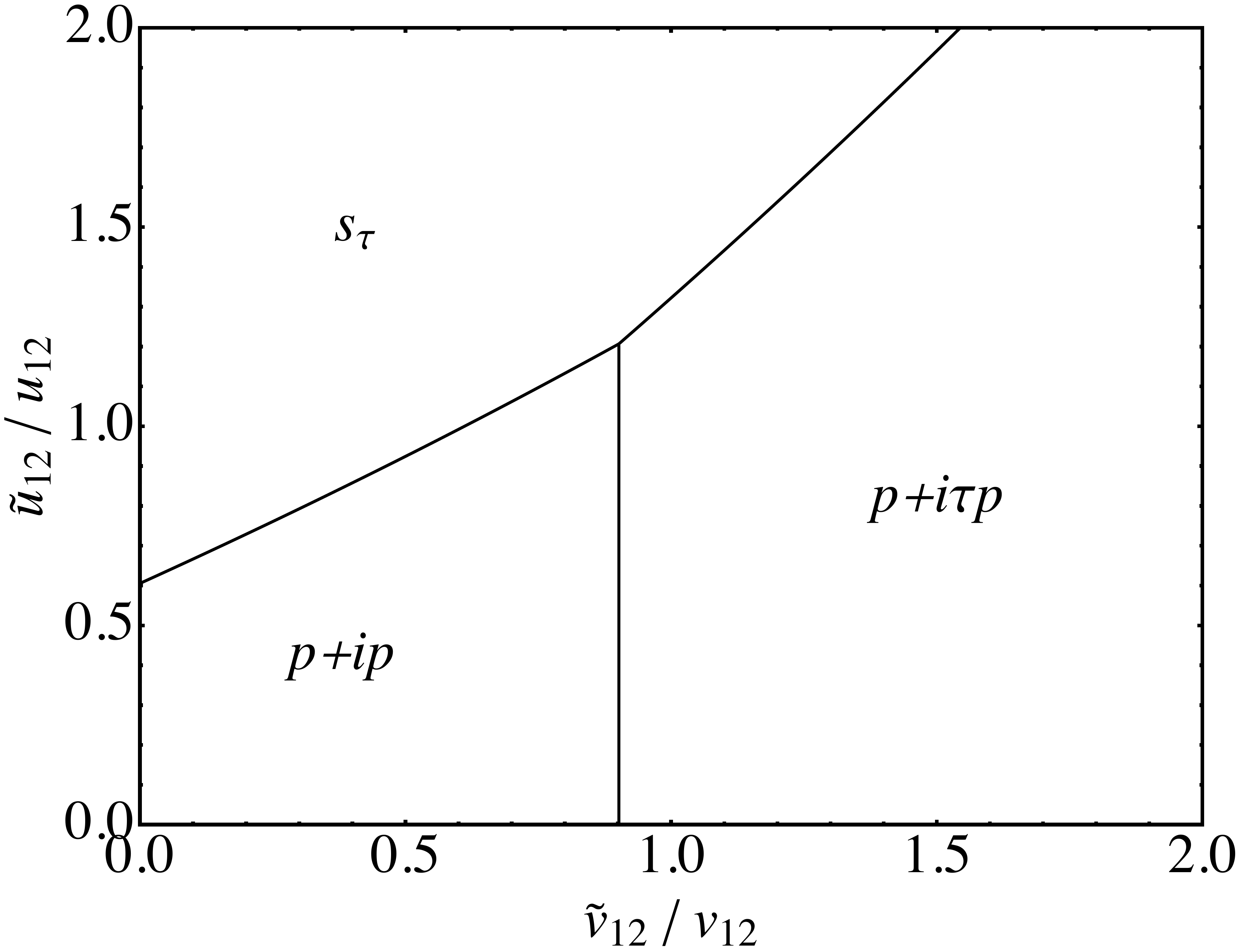}
\includegraphics[width=64mm]{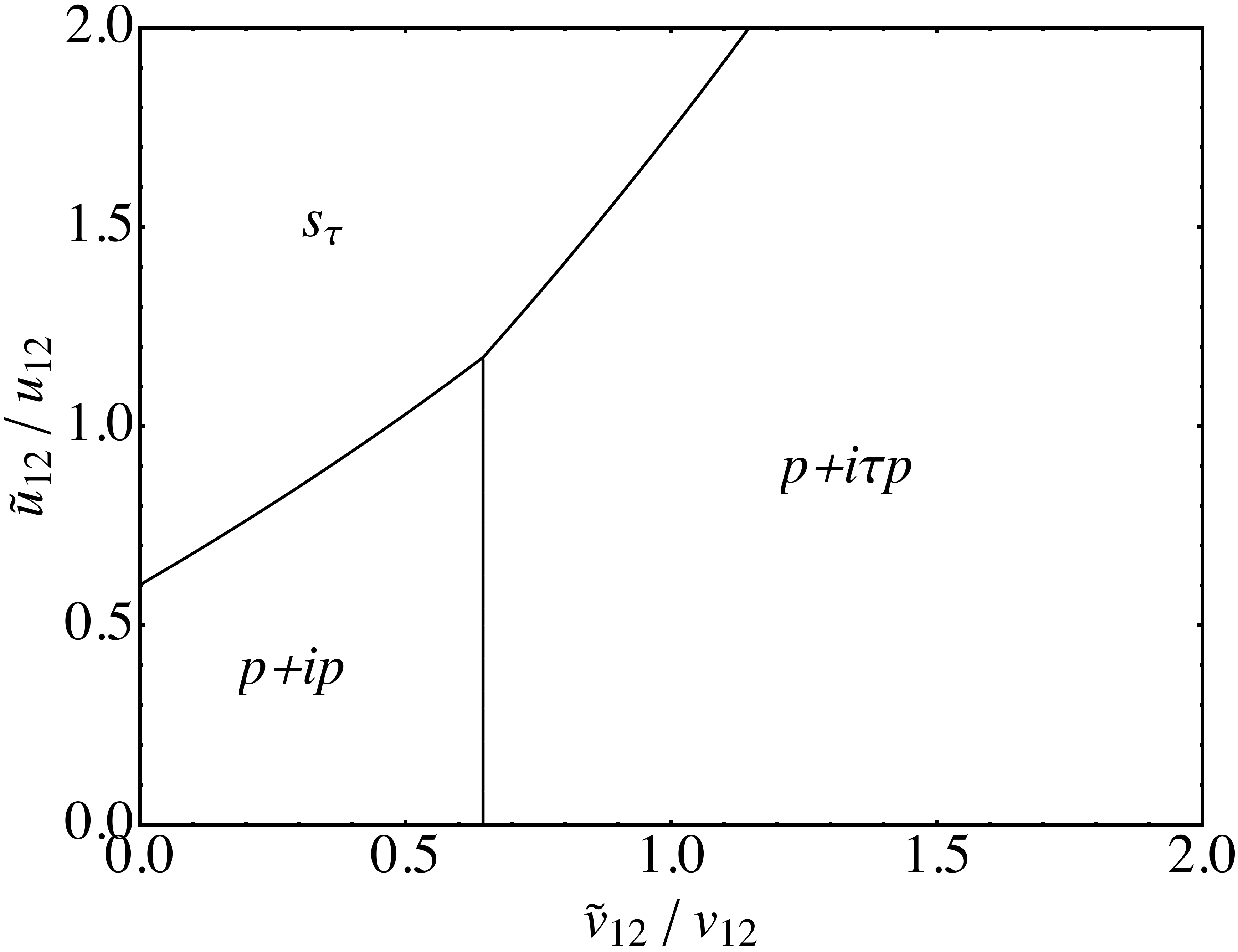}
\includegraphics[width=64mm]{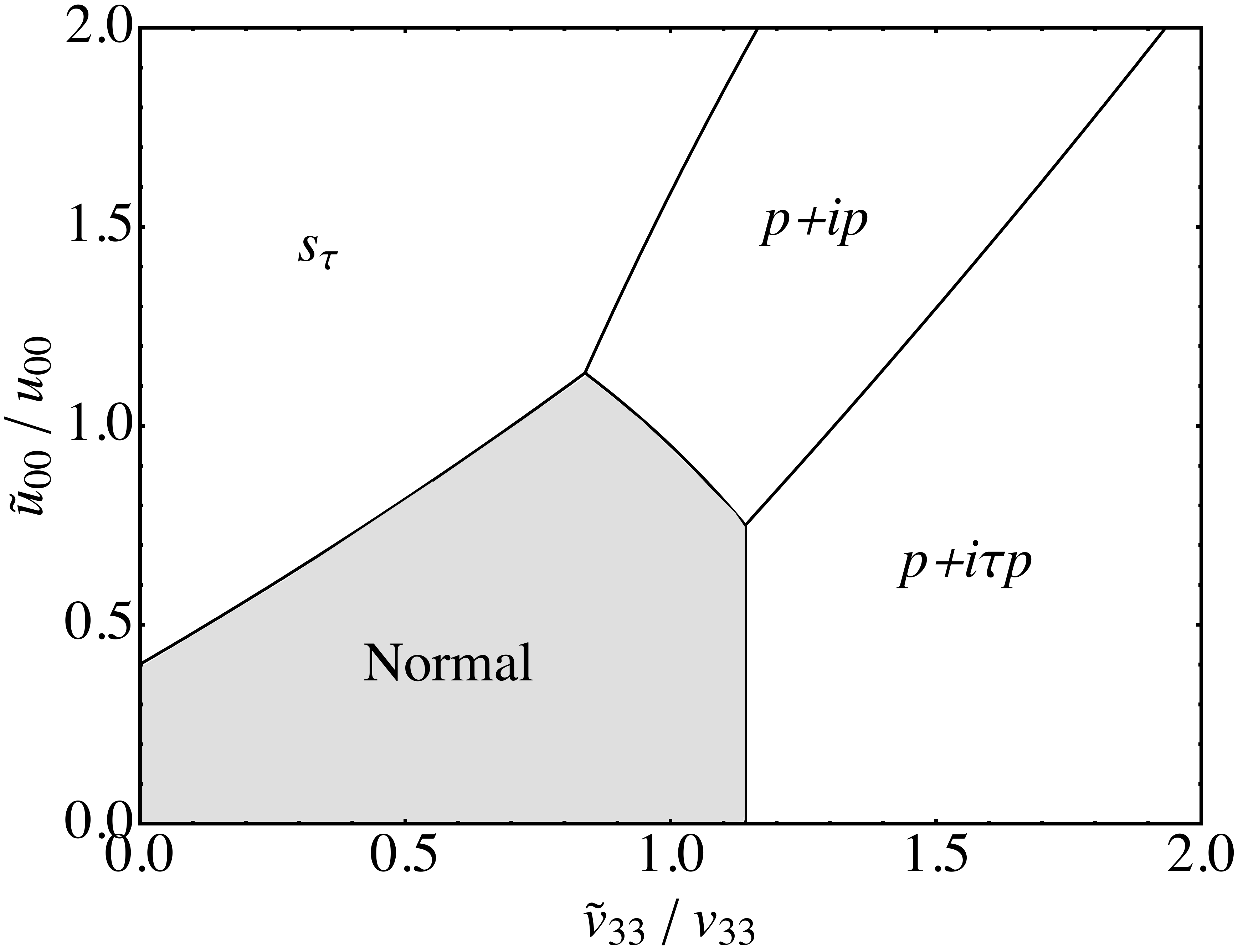}
\includegraphics[width=64mm]{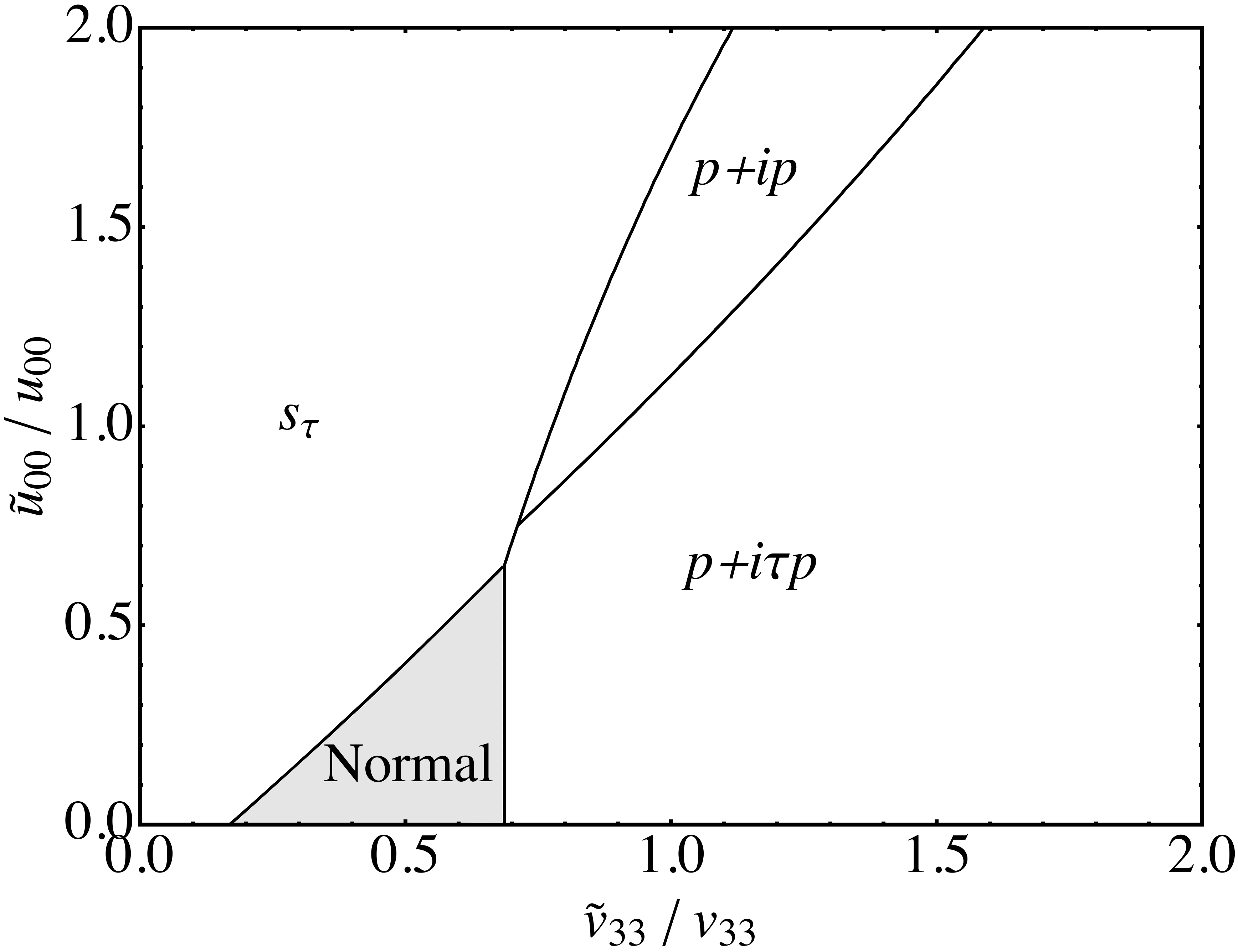}
\includegraphics[width=64mm]{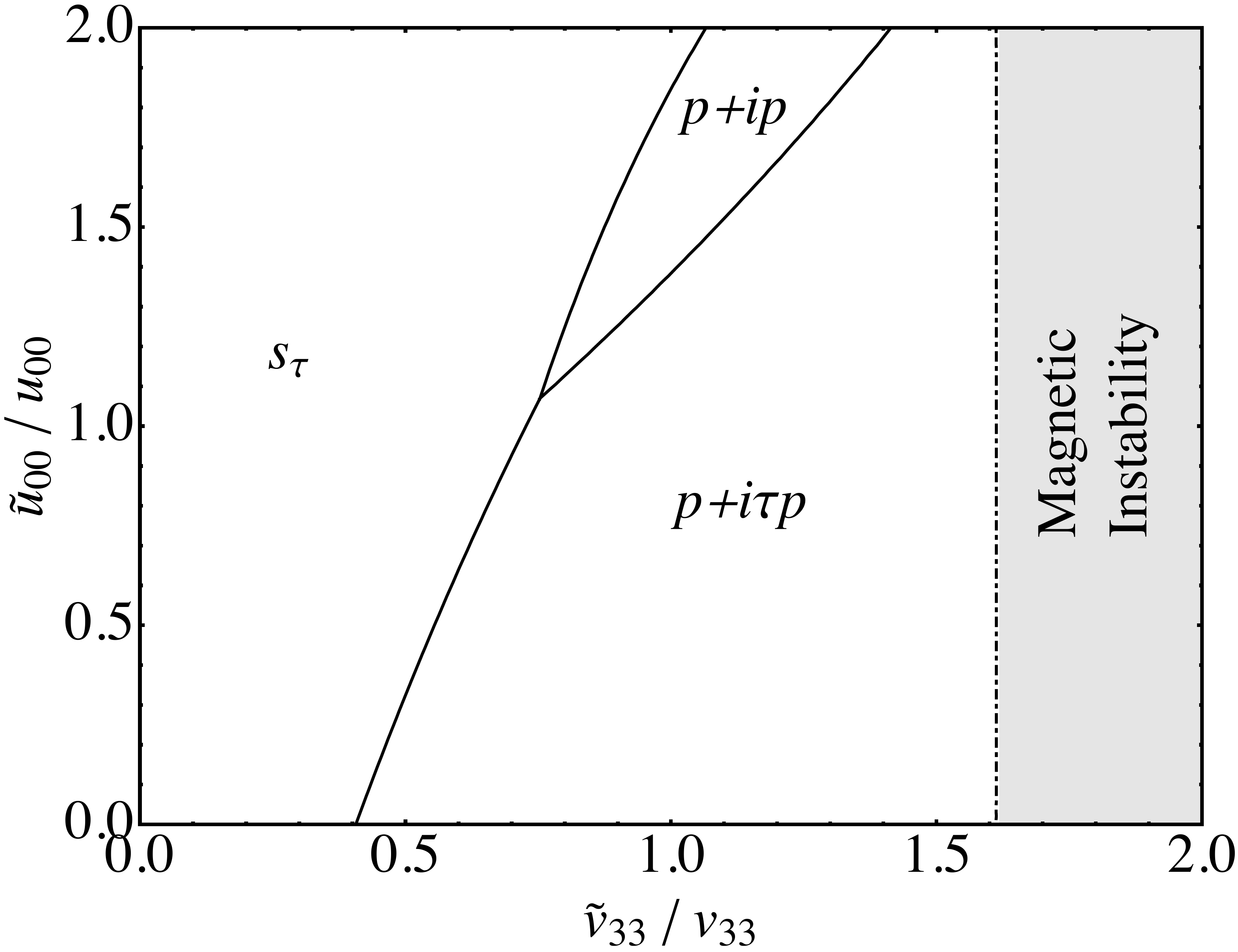}
\caption{{Phase diagram. The three superconducting phases $d^{\ell=0}_{z0}, d^{\ell=\pm}_{zz}, d_{\ell=\pm}^{\alpha\pm}$, which correspond to $s$-wave intervalley ($s_\tau$), $p+ip$-wave intervalley ($p+ip$), and $p+i\tau p$ intravalley ($p+i\tau p$), as well as the magnetic phase.  (a) Fixing the bare values $\{ v_{00} , v_{33} , v_{44}, v_{77} , v_{56} ,v_{07} ,v_{47} , u_{00}, u_{33}, u_{03}\}$ to those computed and shown in Figure \ref{f:bare}, while allowing for a variable $\tilde{v}_{12}$ and $\tilde{u}_{12}$, which are substituted into the interaction structure in place of $v_{12}$ and $u_{12}$. Here we show the variable values as ratio of the calculated values. (ai), (aii), (aiii) Show the same parameters but with increasing chemical potential $\mu/\mu_0=0.75, 1, 1.25$, respectively, with $\mu_0\equiv 0.025 v K_0$, with $d/L=0.375$, $L=30$nm. (b)  Same as (a), but allowing for a variable $\tilde{v}_{33}$ and $\tilde{u}_{00}$. (bi), (bii), (biii) Show the same parameters but with increasing chemical potential $\mu/\mu_0=0.75, 1, 1.25$. } \label{f:phase}}
\begin{picture}(0,0) 
\put(-280,410){\textbf{(ai)}} 
\put(-96,410){\textbf{(aii)}} 
\put(83,410){\textbf{(aiii)}}
\put(-280,268){\textbf{(bi)}} 
\put(-96,268){\textbf{(bii)}} 
\put(83,268){\textbf{(biii)}}
\end{picture}
\end{figure*}
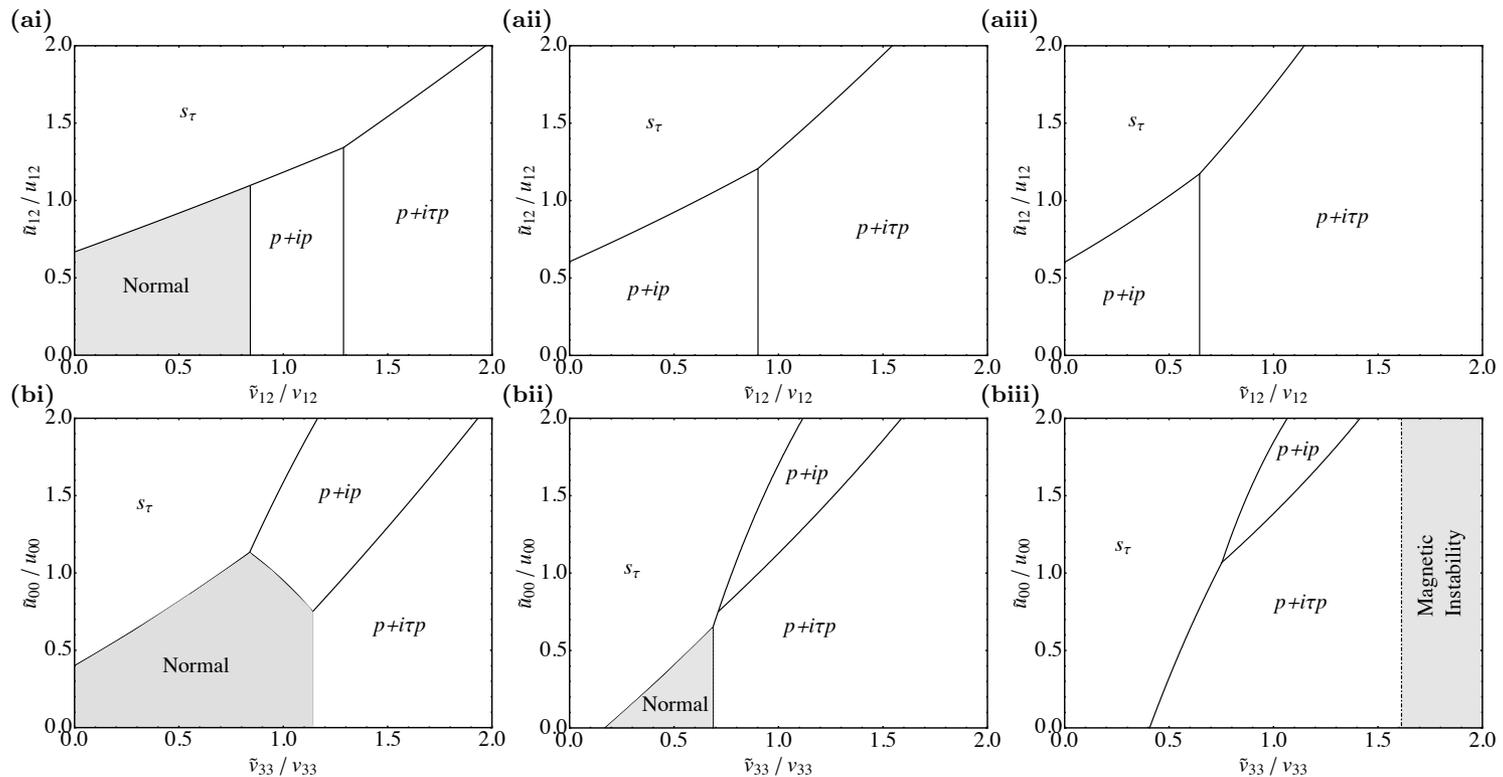

In this section we construct the phase diagram consisting of the three leading superconducting instabilities of \eqref{gapeigs1}-\eqref{gapeigs3}, as well as for competing charge and magnetic order, which will be described in Section \ref{competing_insta}. 

We specify the phase diagram as follows: ($a$) we choose to fix the ratio $d/L=0.375$, which as we have stated earlier quantifies the strength of spin-orbit coupling; { our motivation for this choice is that typical quantum wells are of width $d\approx 10$ nm. We allow for a physically achievable superlattice $L \approx 30$ nm -- several current superlattice devices have $L\approx 50$ nm \cite{AG, AG2,Du2021,FVergel2021,Gardenier2020,Chen2021ag,Freeney2022}. While, larger values of $d/L$ are desirable for resulting in flatter bands --- i.e. smaller velocity Fig. \ref{f:H0}a --- and therefore relatively stronger interactions, for presentation we constrain ourselves to the physically reasonable $d/L=0.375$. } ($b$) We designate a {\it critical doping} $\mu_c=0.025 v K_0$ and plot phase diagrams for $\mu/\mu_c=\{0.75,1,1,25\}$. ($c$) { Having computed the bare interaction matrix elements, $u_{\mu\nu}, v_{\mu\nu}$ -- i.e. $v_{12}, v_{33}, u_{00}, u_{12}$ from \eqref{Vbare} -- from exact diagonalization, we replace them with continuous tuning parameters $\tilde{u}_{\mu\nu}/u_{\mu\nu}, \tilde{v}_{\mu\nu}/v_{\mu\nu}$ which vary about the computed bare values. In Fig. \ref{f:phase}, we plot two sets of diagrams spanned by $(\tilde{v}_{12}/v_{12},\tilde{u}_{12}/u_{12})$ and $(\tilde{v}_{33}/v_{33},\tilde{u}_{00}/u_{00})$. }

The motivation for choice ($c$) is that one expects quantitative changes to the values of bare interaction matrix elements \eqref{Vbare}, shown in Fig. \ref{f:bare}, for four reasons: ($i$) inaccuracies of the microscopic modeling, such as those due to neglecting higher harmonics in \eqref{potential}, i.e. additional cosine terms which respect the honeycomb symmetry, as discussed in \cite{Tkachenko2015}; ($ii$) corrections to the infinite square well potential \eqref{sqconfine}; ($iii$) corrections of order $W_0/E_0$, which are not captured in the three $K$-point approach;  ($iv$) since we only present results for a InAs heterostructure, the variation in the calculated bare values may be very approximately linked to teasing out the phase diagram for other choices of semiconductor heterostructures. Hence, instead of incorporating all such corrections numerically, we will allow the bare interaction parameters to vary about the values presented in Fig. \ref{f:bare}. In this way we absorb uncertainty due to microscopic details of the superlattice potential into the numerical values of the bare interaction parameters $u_{\mu\nu},v_{\mu\nu}$ \eqref{Vbare}. The dominant bare interactions are found to be $v_{12}, v_{33}, u_{00}, u_{12}$, as shown in Figure \ref{f:bare}, and for the purposes of presentation, we choose to vary these four parameters, i.e. $\tilde{v}_{12}, \tilde{v}_{33},\tilde{u}_{00}, \tilde{u}_{12}$.

As anticipated in \eqref{gapeigs1}, \eqref{gapeigs2}, \eqref{gapeigs3}, three distinct gap structures appear in the phase diagram, which we describe here: 
\begin{itemize}
    \item 
  {\bf  Intravalley} {$\bm{p+ i \tau p }$} spin-triplet, valley-triplet,  
\begin{gather}
\Delta_{\bm{k}} = e^{i{\tau}_z(\phi-\theta_{\bm{k}})} (d^x_s{s}_x + d^y_s{s}_y) \tau_y \left( {\tau}_y s_y \right)
\end{gather}

The spin triplet vector is pinned in-plane, and the valley polarisation is coupled to the orbital angular momentum, i.e. $\ell=\pm1$ at valley $\tau=\mp1$.  This implies a chiral $p$-wave gap, with opposite chiralities in each valley, a state which respects time reversal symmetry. This phase exhibits a $U(1)\times U(1)$ symmetry breaking due to the presence of a relative phase $\phi$ between opposite valleys and a spin direction $\bm{d}_s = (d^x_s,d^y_s,0)$. The superconducting state is analogous to that of Ref. \cite{Li2020} but with $\bm{d}$ pinned in-plane. This state exhibits higher-order topology, as will be demonstrated in Section \ref{topology}.

\item
{{\bf  Intervalley} {$\bm{p+ i p }$}} spin-triplet, valley-triplet,
\begin{align}
 \Delta_{\bm k} &=  e^{\pm i \theta_{\bm{k}}} d_s^z {s}_z {\tau}_z  \left( {\tau}_y s_y \right)
\end{align}
 Here the chiral angular momentum states $\ell=\pm 1$ are degenerate. An analysis of the Landau-Ginzburg free energy is required to understand if these degenerate states compete or coexist. A simple computation gives the Landau-Ginzburg free energy for the two order parameters $e^{\pm i\theta_{\bm{k}}} {s}_z {\tau}_z\equiv \phi_\pm$, 
 \begin{align}
\ \ \ \ \  {\cal F}[\phi_\pm]&=-s (\phi_+^2 + \phi_-^2) +\alpha (\phi_+^4 + 4\phi_+^2\phi_-^2+\phi_-^4)
 \end{align}
 The quartic term breaks the $SO(2)$ rotational symmetry in the isospin space $(\phi_+,\phi_-)$, and the order parameters $\phi_\pm$ act like an Ising degree of freedom; the system must spontaneously choose a chirality ($\ell=\pm 1$), and therefore spontaneously break time reversal symmetry. This phase possesses a nontrivial first-order topological invariant which manifests as chiral modes propagating along the edge, as we discuss in Section \ref{topology}.

\item
{{\bf  Intervalley} {$\bm{s}_\tau$}} spin-triplet, valley-singlet,  
\begin{align}
\Delta_{\bm{k}} =  d^z_s {s}_z \tau_0 \left( {\tau}_y s_y \right)
\end{align}
The spin triplet vector is pinned out-of-plane along $\bm{z}$. As shown in \cite{SamajdarSceurer2020PRB}, owing to the valley singlet structure, this spin triplet phase satisfies an ``Anderson theorem'', which provides protection against non-magnetic disorder, provided the disorder does not induces intervalley scattering. Quite unexpectedly, this phase hosts a second-order topological invariant,  to be described in Section \ref{topology}.

\end{itemize}

As can be seen in Figure \ref{f:phase}, for each superconducting state there is a critical $\mu_c$ such that for $\mu>\mu_c$ the system become superconducting, which as discussed earlier reflects the fact that as the chemical potential is increased, screening becomes more efficient, causing the pseudospin and/or valley dependent interactions to become attractive. 

Note that the phase boundaries between normal and superconducting states are second-order, while the phase boundaries between distinct superconducting states are first-order, which in principle leaves open the possibility of coexistence between these superconducting phases. However, a straightforward Landau-Ginsburg analysis shows that all coexistence is energetically penalised.

The rightmost portion of Figure \ref{f:phase} contains a region labelled as a ``magnetic instability''. In this region, we find that magnetic insulating states can compete with superconductivity, as we discuss in Appendix \ref{competing_insta}. In short, the antiscreened couplings contribute only to ferromagnetic and spin-density wave order, which are nearly degenerate and eventuate through a Stoner instability for strong coupling.

{ Finally, we state that in the presented phase diagram, the highest critical temperatures reached are on the order of $T_c\approx 0.2 \mu\approx 2$ K, which is estimated using the gap equation (\ref{BCSgapeq}), and explicitly taking $L=30$ nm. From this expression we see that going to larger doping $\mu$ is desirable to achieve larger critical temperatures, however, large values of $\mu$ enter the strong coupling regime, in which magnetic or charge instabilities are likely to compete.}

\section{Topological properties of the superconducting phases}
\label{topology}

In this section we will prove that all three superconducting phases are topological, and discuss their properties. For intervalley pairing, we have ${\Delta}_{\bm{k}} \propto {\tau}_y$ and ${\Delta}_{\bm{k}} \propto e^{\pm i \theta_{\bm{k}}} {\tau}_x$ for the $s_\tau$ and $p+ip$ phases respectively, while ${\Delta}_{\bm{k}} \propto e^{i\tau_z(\phi - \theta_{\bm{k}})}$ for the intravalley $p+i\tau p$ phase. Since ${\tau}_x, \hat{\tau}_z$ are even under inversion  ($\bm{r}\rightarrow -\bm{r}$) while ${\tau}_y$ is odd, we find that the gap is odd under inversion for both intervalley phases, while the intravalley $p+i\tau p$ phase is even for $\phi = n\pi$ and odd for $\phi = (n+\frac{1}{2})\pi$, with $n\in\mathbb{Z}$. A fundamental requirement for a non trivial topology hosting Majorana edge or corner modes is that the gap change sign under inversion\footnote{
A close examination of the classification presented in Refs. \cite{Geier2018, Trifunovic2019} reveals that when the system respects time-reversal symmetry, i.e. in Cartan class DIII, and the gap is even under inversion, the topological classification with inversion symmetry is trivial. This implies that a first-order topological phase hosting a helical Majorana edge mode, as well as a second-order topological phase hosting Kramers pairs of Majorana corner state, is prohibited. When more symmetries are included it is still possible that further topological phases appear, however they must have distinct boundary signatures from the ones mentioned.}. 
This is fulfilled for both the intervalley phases, as well as for the intravalley $p+i\tau p$ phase in the special case $\phi = (n+\frac{1}{2})\pi$. 

The time-reversal symmetry breaking intervalley $p+ip$ phase exhibits first-order topology; taking into account the $U(1)$ spin-rotation symmetry, we find that this system is in Cartan class A, which permits a Chern number in two dimensions \cite{Altland1997, Schnyder2008}. We find that this phase exhibits a pair of chiral Dirac modes propagating along the boundary, establishing it as a first-order topological superconductor. 

The intervalley $s_\tau$ phase is time-reversal symmetric, and accounting for the $U(1)$ spin-rotation symmetry, is in class AIII, which always implies trivial first-order topology in two dimensions.
The time-reversal symmetric intravalley $p+i\tau p$ satisfies a $\mathbb{Z}_2$ symmetry expressed by a combination of spin rotation and gauge transformation, such that the system is described by a BdG Hamiltonian in class D. For intervalley $s_\tau$ and intravalley $p+i\tau p$, a second-order topological phase protected by the crystalline symmetries is possible. We will establish the second-order topology for intravalley $p+i\tau p$ and intervalley $s_\tau$ pairing using symmetry-based indicators. Finally, we will present exact diagonalisation results for the Bogoliubov-de Gennes Hamiltonian for all three superconducting phases. These numerical results provide clear evidence for the suggested topology by demonstrating the corresponding anomalous edge and corner states.

{ The origins of protected corner modes in the higher-order topological phases $p+i\tau p$ and $s_\tau$ may be understood intuitively as follows. We find that, in both these phases, edge modes exist for certain parameters, which are gapped for certain edge geometries. Since these modes can be gapped, they are not protected by a topological bulk-boundary correspondence and can be continuously pushed into the bulk continuum; in cases where they do exist, we may introduce a 1D theory for the boundary modes. Since the superconducting gap is odd under inversion $\Delta \rightarrow -\Delta$, the gap is forced to vanish at inversion symmetric points, ie the corners of the sample. Hence, there are domain walls, or `kinks', in the superconducting gap function at the corners of the sample, which give rise to anomalous zero modes. These corner modes survive even when parameters are tuned so that the 1D modes are pushed into the bulk continuum. Hence, despite the non-existence of 1D edge modes generically, each phase is adiabatically connected to a model possessing gapped 1D modes, which must possess anomalous zero energy corner modes.}

It is first necessary to express the mean field Hamiltonian \eqref{H_MF} as a lattice model involving creation operators $c_{\bm{R},s}^\dag$ for Wannier orbitals localised at the sites $\bm{R}$ of the artificial honeycomb lattice, $\mathcal{H}_{\text{MF}} = \mathcal{H}_{\text{nor.}} + \mathcal{H}_\Delta$. The normal state Hamiltonian $\mathcal{H}_{\text{nor.}}$ is equivalent to two copies of the Haldane model, consisting of a sum of spin-independent nearest neightbour hoppings and next nearest neighbour spin-dependent hopping terms,
\begin{gather}
\mathcal{H}_{\text{nor}.} = -\sum_{\langle \bm{R},\bm{R}'\rangle;s}{t c^\dag_{\bm{R},s} c_{\bm{R}',s}} - \sum_{\langle\langle \bm{R},\bm{R}'\rangle\rangle;s}{t' e^{\frac{2\pi i}{3}\sigma s} c^\dag_{\bm{R},s} c_{\bm{R}',s}} 
\label{eq:topo_H_norm}
\end{gather}
where the parameters of the Dirac model \eqref{HdiracDASH} are related to the hopping parameters via $v = \sqrt{3}at/2$ and $t' = 9\eta/2$.

The pairing term $H_\Delta$ is given by
\begin{gather}
H_\Delta = 
\sum_{\bm{R},\bm{R}'}{\Delta(\bm{R},\bm{R}')c^\dag_{\bm{R},\uparrow} c^\dag_{\bm{R}',\downarrow}}
\label{eq:topo_H_Delta}
\end{gather}
for the intervalley $p+ip$ and $s_\tau$ phases with pairing between opposite spins, and
\begin{gather}
H_\Delta = \sum_{\bm{R},\bm{R}'}{
\Delta(\bm{R},\bm{R}')\frac{1}{2}\left[e^{i\phi_s} c^\dag_{\bm{R},\uparrow} c^\dag_{\bm{R}',\uparrow} + e^{-i\phi_s} c^\dag_{\bm{R},\downarrow}c^\dag_{\bm{R}',\downarrow}\right]} 
\label{eq:topo_H_Delta2}
\end{gather}
 for the intravalley $p+i\tau p$ phase with equal spin pairing, where $(d_x,d_y,d_z) = (\sin\phi_s,\cos\phi_s,0)$. The form of the pairing function $\Delta(\bm{R},\bm{R}')$ may be derived by projecting the momentum-space expression for $H_\Delta$ in (\ref{H_MF}) onto the Wannier orbitals, and are derived in the Appendix. For the intervalley paired phases, $\Delta(\bm{R},\bm{R}')$ possesses the discrete translational symmetry of the lattice and changes sign under inversion, $\Delta(-\bm{R},-\bm{R}') = -\Delta(\bm{R},\bm{R}')$, while for the intravalley $p+i\tau p$ phase, the discrete translation symmetry of the lattice is spontaneously broken and $\Delta(\bm{R},\bm{R}')$ exhibits spatial modulations, oscillating as a function of $\bm{R}+\bm{R}'$ and, except at special values $\phi = n\pi/2$, also spontaneously breaks inversion symmetry.

\subsection{Symmetry-based indicators for $p+i\tau p$ and $s_\tau$ phases}

In this subsection, we prove that the superconducting states
with $p+i\tau p$ or $s_{\tau}$ pairing symmetry realise a second-order topological phase with Majorana Kramers pairs pinned to the corners by the crystalline point-group symmetries. We will first examine the symmetries of the system to determine under which conditions we may expect a second-order topological phase. Next, we apply the theory of symmetry-based indicators \cite{Shiozaki2019, Geier2020, Ono2020} to derive a simple, sufficient criterion for a transition into a second-order topological superconducting state when an infinitesimal pairing which is odd under inversion symmetry creates a full gap in the BdG spectrum. Finally, we show that this criterion is fulfilled for the $p+i\tau p$ and $s_\tau$ pairing instabilities in our honeycomb lattice model.

The symmetry-group of our hexagonal lattice is given by the direct product of translations in the $x,y$ plane and the crystalline point group $D_{6h}\simeq C_{6v}\otimes C_{i}$, where $C_{i}$ is generated by spatial inversion $\mathcal{I}:x,y,z\to-x,-y,-z$
and $C_{6v}$ is the point group of the hexagonal lattice in the $x,y$ plane. Furthermore, the normal-state Hamiltonian satisfies time-reversal symmetry $\mathcal{T}$ and $U(1)$ spin rotation symmetry $\mathcal{S}$ around the $s_{z}$ axis. A symmetric unit cell can be chosen to coincide with the hexagons in the hexagonal lattice, where the lattice sites are located on the threefold rotation symmetric corners of the hexagonal unit cell. Each site is occupied by one Kramers pair of fermionic orbitals, which, without loss of generality for the following discussion, can be chosen to be $s$-orbitals\footnote{The $s$-orbitals are even under inversion. Choosing different orbitals may change the representation of inversion symmetry that is carried through the calculation, but does not affect the conclusions.}. In the following, we argue that inversion symmetry is sufficient
to protect the second-order topological phase and prove its appearance from the symmetry-based indicator. Therefore, it is sufficient to consider the representations of time-reversal symmetry and inversion symmetry. In Bloch basis, these representations in the normal state can be written as
\begin{align}
u(\mathcal{T}) & =is_{y}\sigma_{0}\nonumber \\
u(\mathcal{I};\vec{k}) & =s_{0}\sigma_{x}e^{i(\vec{a}_{2}-\vec{a}_{1})\vec{k}}\label{eq:topo_reps}
\end{align}
with $s_{i}$, $\sigma_{i}$ the Pauli matrices in spin and sublattice space, respectively, and the Bravais lattice vectors $\vec{a}_{1}=\sqrt{3}a\hat{x}$, $\vec{a}_{2}=(\sqrt{3}a\hat{x}+3a\hat{y})/2$, where $a$ is the interatomic distance. Here we chose the center of the hexagons as the center of inversion.

The $p+i\tau p$ and $s_\tau$ superconducting orders preserve time-reversal symmetry. Out of the large symmetry-group containing the point group $D_{6h}$ and $\mathbb{Z}_2$ or $U(1)$ spin rotation symmetry, respectively, it is sufficient to preserve only a single crystalline symmetry element such as inversion, perpendicular twofold rotation, or mirror symmetry in order to protect a second-order topological phase \cite{Geier2018, Trifunovic2019}. Here, we focus on inversion symmetry, as it also allows us to write down a symmetry-based indicator as a topological invariant. By restricting the topological classification to inversion and time-reversal symmetry and neglecting the remaining symmetries, we resolve the topological phases in Cartan class DIII with inversion symmetry\footnote{Notice that previously, we took the $U(1)$ spin-rotation symmetry or $\mathbb{Z}_2$ combined spin-gauge symmetry into account to conclude that each of the spin-blocks is in Cartan class AIII or D, respectively. Here, we only utilize a minimal set of symmetries that is necessary to protect the second-order topological phase whose existence we want to prove, which does not require additional $U(1)$ or $\mathbb{Z}_2$ symmetry. Thus we may utilize the results for the less restrictive class DIII.}. 
The remaining symmetry elements apart from time-reversal and inversion may enrich these topological phases, either prohibiting or giving rise to further topological phases. For example, the $U(1)$ spin rotation symmetry prohibits the first-order topological
superconductor in Cartan class DIII with helical Majorana edge states. The mirror and sixfold rotation symmetry enrich the second-order topological phase protected by inversion, as the mirror symmetry pins the corner states to mirror-symmetric corners and at the same time requires a gapless anomalous edge state on mirror symmetric edges \cite{Langbehn2017,Geier2018}, while the sixfold rotation symmetry requires that on a sixfold symmetric sample, gapless states should exist on all six corners. 

The topological classification depends on whether the superconducting order parameter is even or odd under inversion; this parity determines the representation of inversion symmetry and its commutation relations with the particle-hole antisymmetry of the BdG Hamiltonian \cite{Geier2020, Ono2020}. In case the superconducting order parameter is even under inversion, the topological classification is trivial \cite{Geier2018,Trifunovic2019}. In case it is odd under inversion, the classification of topological phases with anomalous boundary states is $\mathbb{Z}_{4}$, where odd elements ``1'', ``3'' indicate a first-order topological superconductor hosting a helical Majorana edge mode, and the even element ``2'' is a second-order topological superconductor hosting Kramers pairs of Majorana corner states on an inversion symmetric sample \cite{Geier2018,Trifunovic2019}. 

The $p+i\tau p$-wave order parameter in Eq. \eqref{eq:topo_H_Delta} is spatially modulated \cite{Li2021,Li2020} such that it is even (odd) under inversion for $\phi=0$ ($\pi/2$). For other values of $\phi$, the system does not respect inversion symmetry. Following the arguments above, this implies that we may find a second-order topological phase hosting Kramers pairs of Majorana corner states only for $\phi=\pi/2$. However, the corner states may persist for a range of $\phi$ around $\phi=\pi/2$ until the surface gap closes \cite{Langbehn2017,Geier2018,Li2021}. The $s_\tau$-wave order parameter Eq. \eqref{eq:topo_H_Delta2} is odd under inversion, thus allowing a second-order topological phase.

Symmetry-based indicators are sufficient criteria for topological crystalline phases expressed in terms of symmetry-eigenvalues at a few high-symmetry momenta only. A particular strength of this formalism is that in the weak-pairing limit of an infinitessimal pairing strength $\Delta\to0$, the symmetry-based indicator can be expressed in terms of symmetry-data of the normal-state Hamiltonian only. The symmetry-based indicator takes the symmetry of the superconducting order parameter into account, as different symmetry-based indicators are defined depending on the irreducible representation of the order parameter. This allows one to formulate sufficient criteria for the topology of a superconducting phase depending on the pairing symmetry and band structure data of the normal state.

The symmetry-based indicator for the second-order topological phase
with inversion symmetry $\mathcal{I}$ and pairing symmetry $u(\mathcal{I})\Delta(-k_{x},-k_{y})u^{T}(\mathcal{I})=-\Delta(k_{x},k_{y})$ has been calculated as \cite{Geier2020}
\begin{equation}
z_{2}=\mathfrak{N}_{+}^{\Gamma}-\mathfrak{N}_{+}^{M}\mod4\label{eq:SI_inversion_odd-1}
\end{equation}
where $\mathfrak{N}_{+}^{\textbf{k}_{s}}$ is the number of Kramers pairs of eigenstates of the BdG Hamiltonian with negative energy and even inversion eigenvalue $+1$ at the inversion symmetric momenta $\textbf{k}_{s}=\Gamma,M$. Here, we used that sixfold rotation symmetry relates the three $M$ points in the hexagonal Brillouin zone, such that $\mathfrak{N}_{+}^{M}=\mathfrak{N}_{+}^{M_{1}}=\mathfrak{N}_{+}^{M_{2}}=\mathfrak{N}_{+}^{M_{3}}$. For the symmetry-based indicator, $z_{2}=1,3$ corresponds to a first-order topological superconductor with a helical Majorana edge state, and $z_{2}=2$ corresponds to the second-order topological superconductor. In the weak pairing limit of an infinitesimal order parameter $\Delta\to0$,
we can express the symmetry-based indicator in terms of the symmetry-data of the normal-state Hamiltonian only:
\begin{equation}
z_{2}^{\text{WP}}=n_{+}^{\Gamma}|_{\text{occ}}+n_{-}^{\Gamma}|_{\text{unocc}}-n_{+}^{M}|_{\text{occ}}-n_{-}^{M}|_{\text{unocc}}\mod4\label{eq:SI_inversion_odd_WP-1}
\end{equation}
where $n_{\pm}^{\textbf{k}_{s}}|_{\text{occ}}$ ($n_{\pm}^{\textbf{k}_{s}}|_{\text{unocc}}$),
are the occupied (unoccupied) Kramers pairs of bands with inversion
parity $\pm1$ at the high-symmetry momentum $\textbf{k}_{s}=\Gamma,M$. It is notable that this formula does not depend on the properties of the low-energy theory at the $K$, $K^{\prime}$ points.

\emph{$s_\tau$ pairing.} First, we evaluate the weak-pairing limit of the symmetry-based indicator for $s_\tau$-wave pairing. At the points $\Gamma,M$, the energy of the bands is of the order of the nearest neighbour hopping $t$, which is our largest energy scale, $t\gg t^{\prime},\mu,\Delta$. This allows one to neglect spin-orbit coupling when computing the inversion parities of the occupied and unoccupied bands. Without spin-orbit coupling, the Bloch Hamiltonian for the nearest neighbour hopping can be written as
\begin{align}
h_{0}(\vec{k})=ts_{0}\left(\begin{array}{cc}
0 & 1+e^{-i\vec{a}_{1}\vec{k}}+e^{-i\vec{a}_{2}\vec{k}}\\
1+e^{i\vec{a}_{1}\vec{k}}+e^{i\vec{a}_{2}\vec{k}} & 0
\end{array}\right)_{\sigma}
\end{align}
where we wrote the $2\times2$ matrix in sublattice space $\sigma$
explicitly. Together with the representation of inversion symmetry,
Eq. \ref{eq:topo_reps}, we find by simultaneously diagonalising $h_0(\vec{k})$ and $u(\mathcal{I};\vec{k})$ for the number of Kramers pairs resolved by their inversion parity $n_{+}^{\Gamma}|_{\text{occ}}=0,\ n_{-}^{\Gamma}|_{\text{unocc}}=0$,
$n_{+}^{M}|_{\text{occ}}=1,\ n_{-}^{M}|_{\text{unocc}}=1$, such that $z_{2}^{\text{WP}}=2$. Taking into account that the $s_\tau$ pairing opens a full excitation gap, the onset of this pairing instability is a second-order topological superconducting phase. 

\emph{$p+i\tau p$ pairing.} Due to the spatial modulation of
the $p+i\tau p$ superconducting order parameter, the
Dirac cones at the $K$ and $K^{\prime}$ points get folded onto the
$\Gamma$ point. For finite hole doping, the chemical potential lies
inside the valence band. Taking the band folding into account, we
find $n_{+}^{\Gamma}|_{\text{occ}}=0,\ n_{-}^{\Gamma}|_{\text{unocc}}=2$,
$n_{+}^{M}|_{\text{occ}}=2,\ n_{-}^{M}|_{\text{unocc}}=2$ such that
$z_{2}^{\text{WP}}=2$. As the $p+i\tau p$ pairing instability opens a full gap in the spectrum that is odd under inversion for $\phi=\pi/2$, it leads to a second-order topological phase for $\phi=\pi/2$.

\subsection{Exact diagonalisation results}
\label{sec:topo_ED}

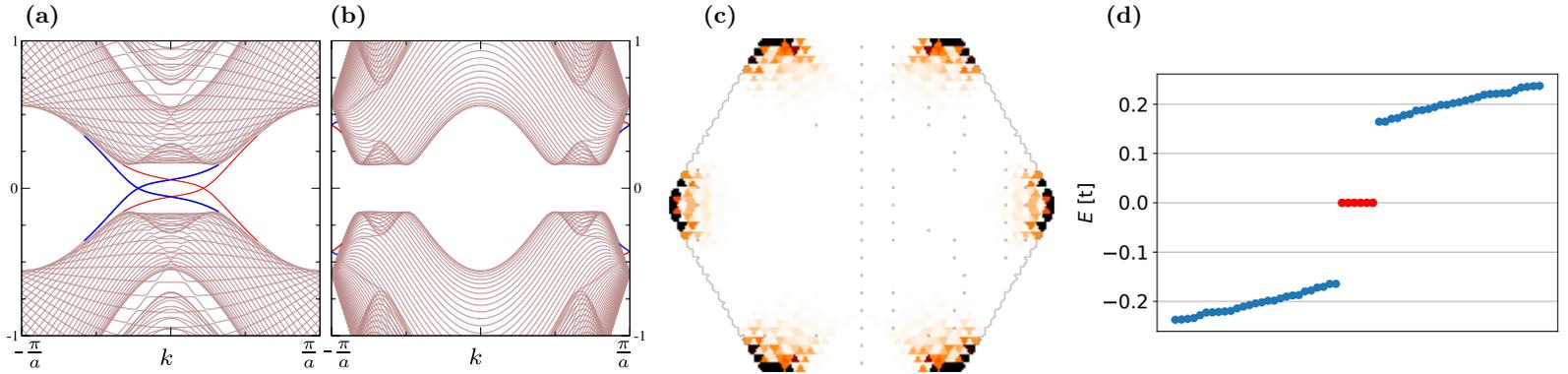
\begin{figure*}[t]
{{\includegraphics[width=0.4\textwidth]{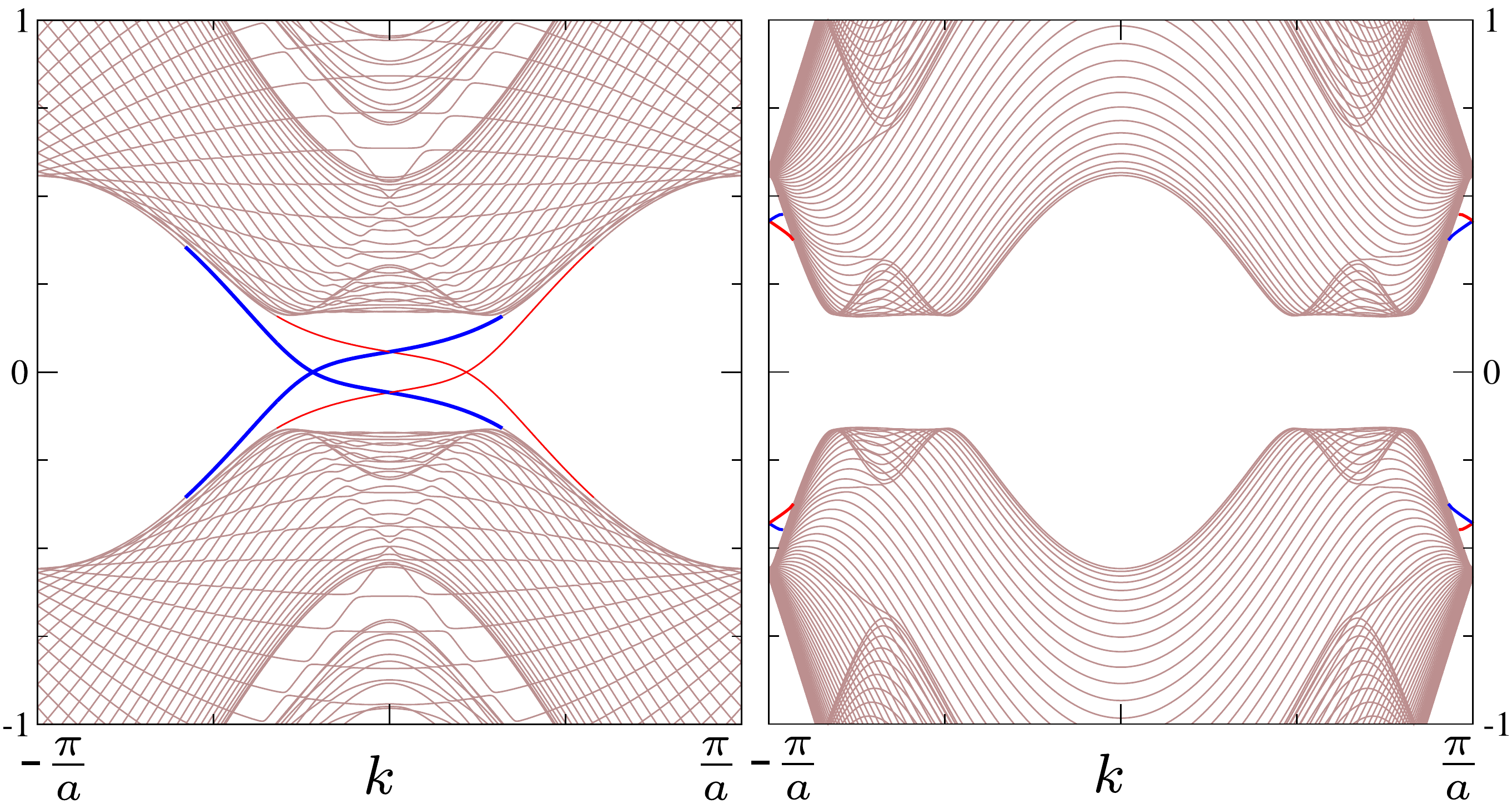}} 
\hspace{0.2cm}\includegraphics[width=0.24\textwidth]{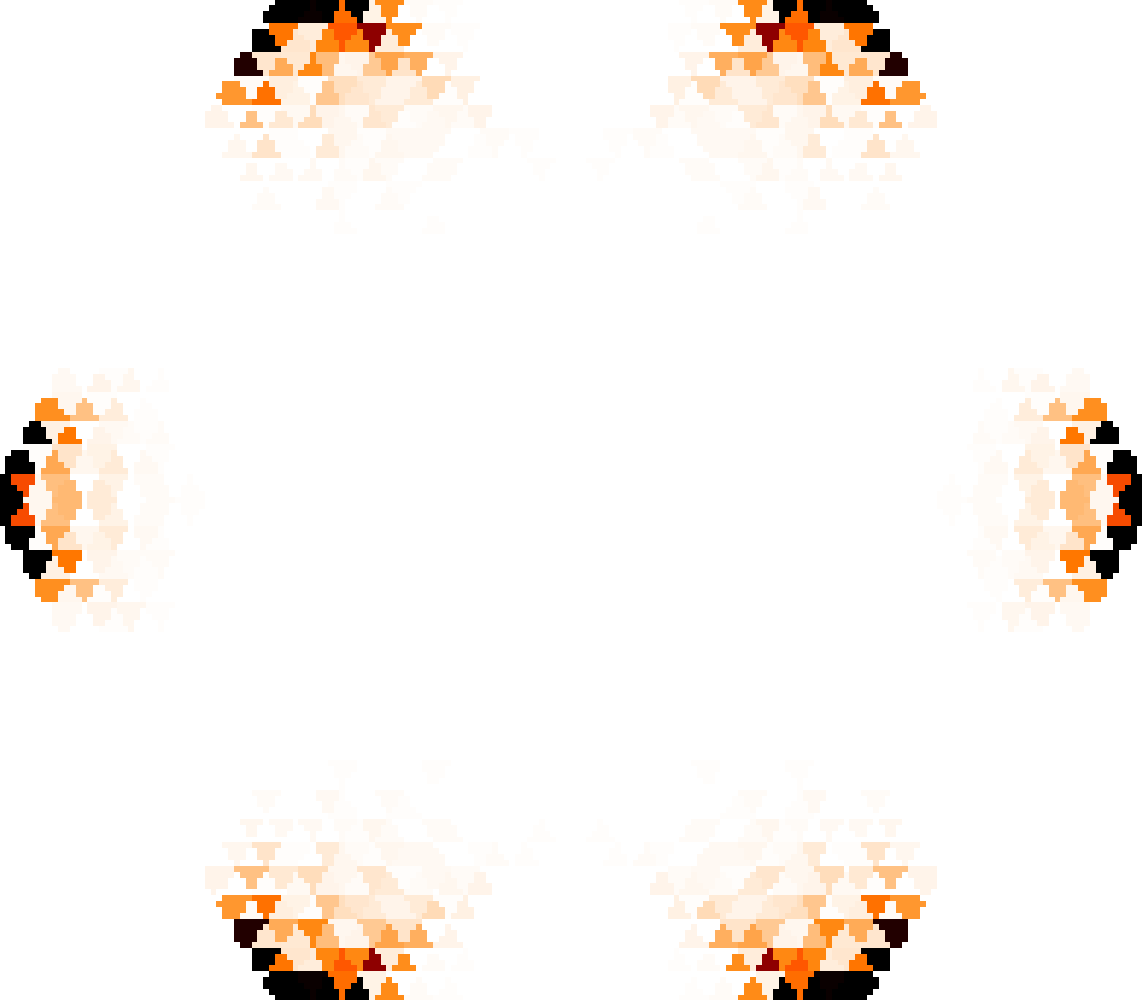} } { \raisebox{0.53cm}{\includegraphics[width=0.30\textwidth]{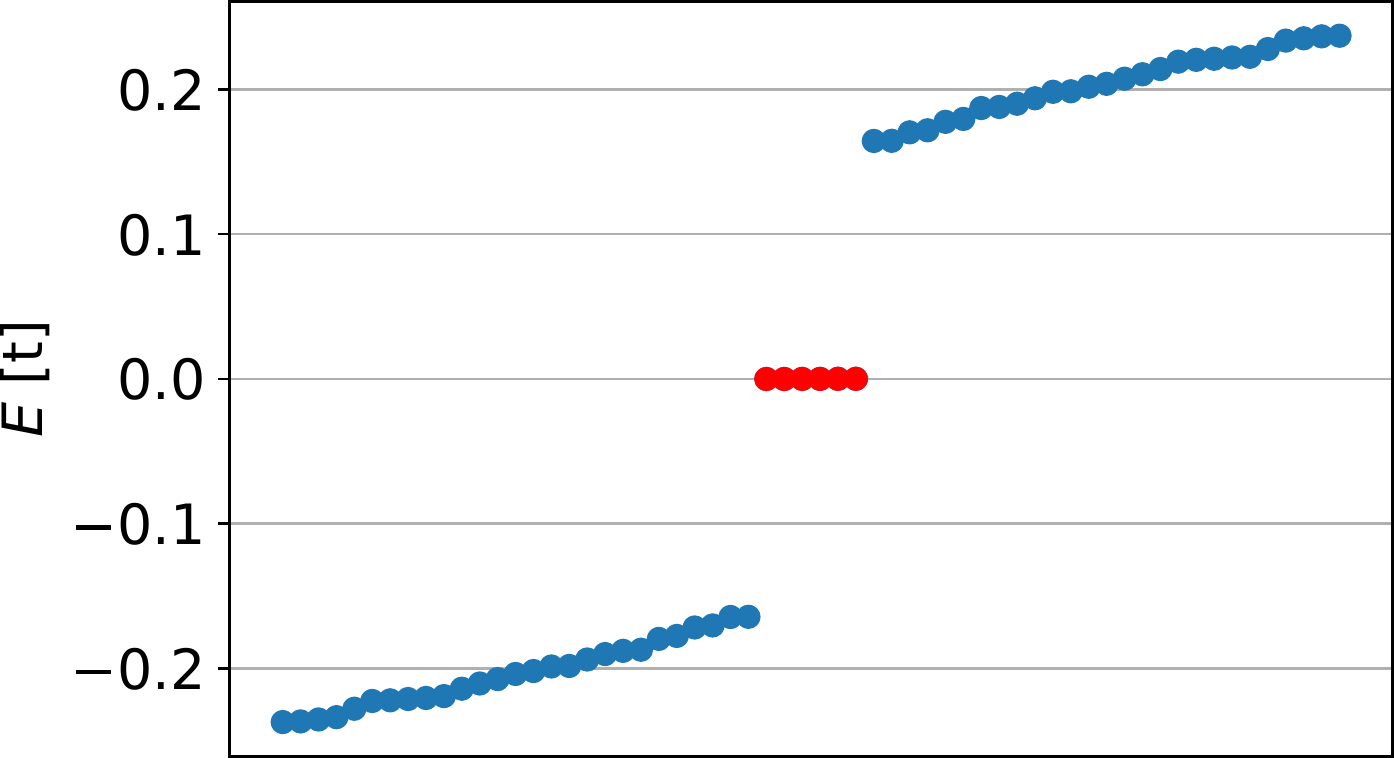}}}
\caption{Exact diagonalisation results for the second-order topological $s_\tau$ spin triplet phase. (a), (b) The 1D dispersion of infinite superconducting ribbons with (a) armchair and (b) zigzag terminations. Edge modes propagating along opposite edges are shown in different colors. (c) Wavefunction profile of the six zero energy eigenstates on a flake geometry. These are the subgap states marked in red in the corresponding spectrum displaying 60 eigenstates around zero in (d). Here we use the parameters $\eta = 0.2t$, $\mu = 0.4t$, $\Delta' = 0.033t$ corresponding to a bulk superconducting gap $\Delta \approx 0.16t$. }
\label{fig:intervalley_s}
\begin{picture}(0,0) 
\put(-288,230){\textbf{(a)}} 
\put(-172,230){\textbf{(b)}} 
\put(-30,230){\textbf{(c)}}
\put(123,230){\textbf{(d)}}
\end{picture}
\end{figure*}

\begin{figure*}
\hspace{-0.05cm}\includegraphics[width=0.4\textwidth]{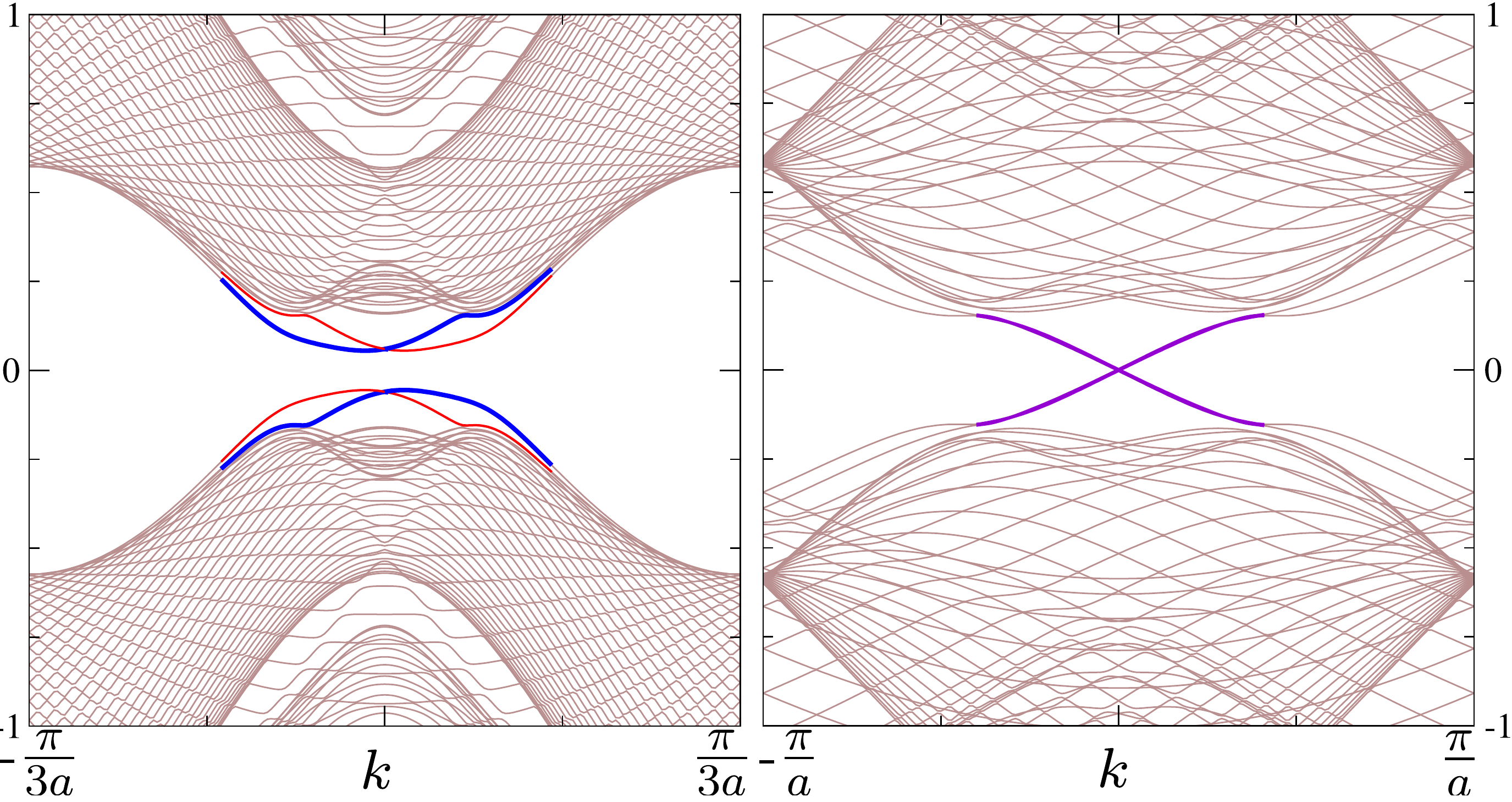}
\hspace{0.48cm}\includegraphics[height=0.24\textwidth]{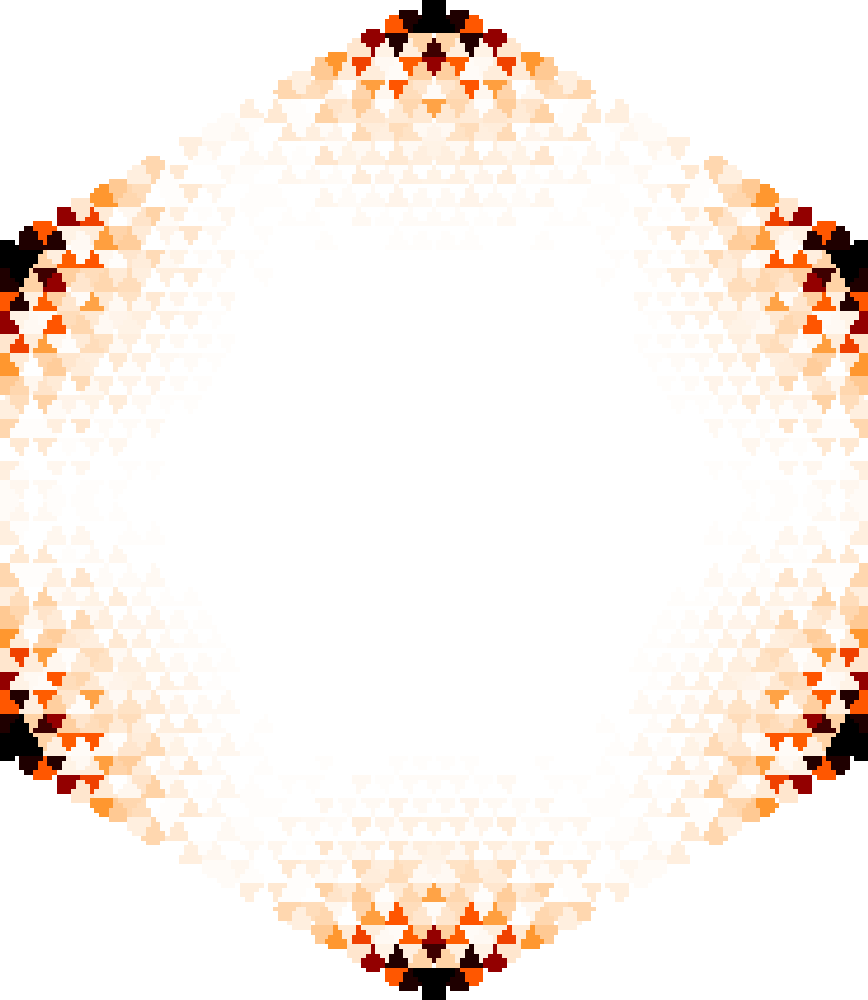}  \hspace{0.28cm}
\raisebox{0.44cm}{\includegraphics[width=0.30\textwidth]{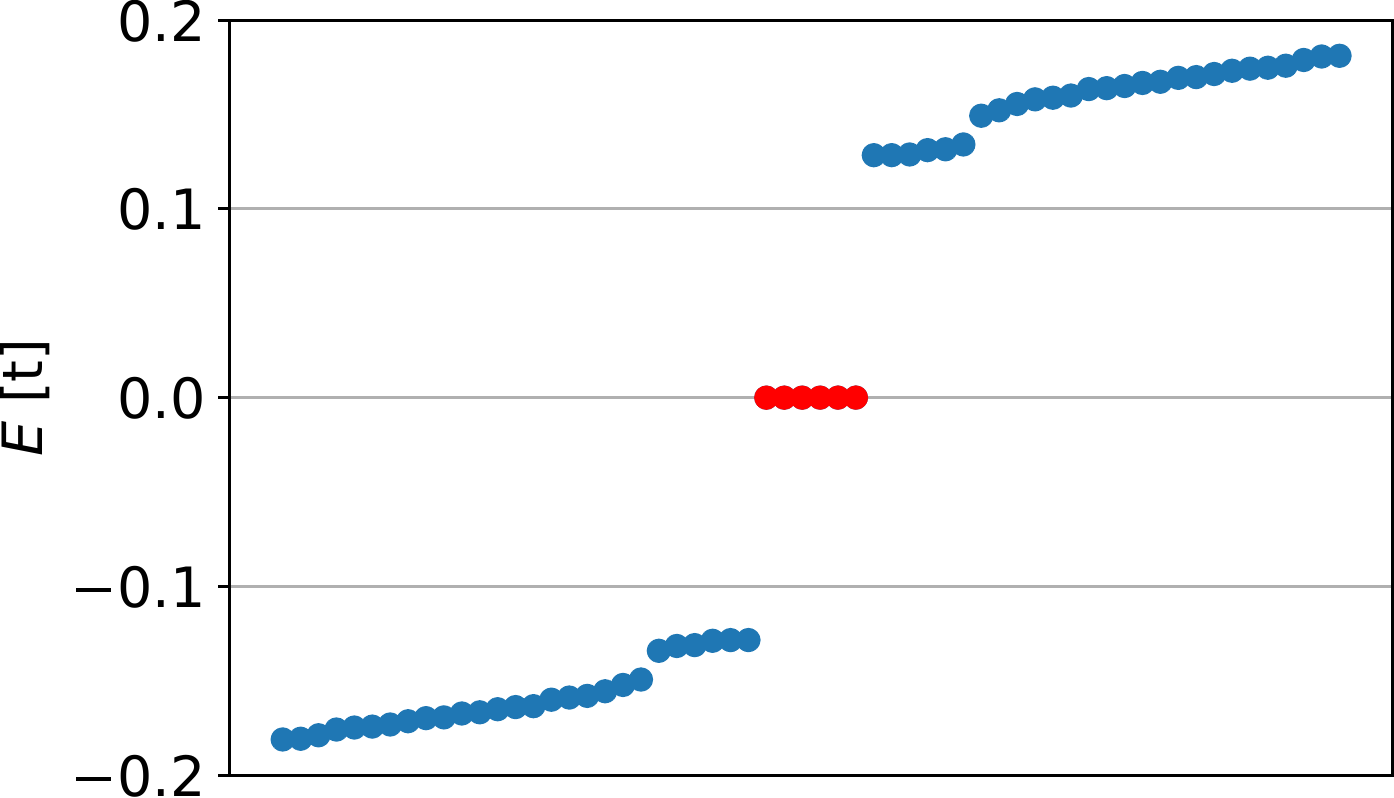}}
\caption{Exact diagonalisation results for the second-order topological $p+i\tau p$ intravalley spin triplet phase. (a), (b), The 1D dispersion of infinite superconducing ribbons with (a) armchair and (b) zigzag terminations. Edge modes propagating along opposite edges are shown in different colors. (c) Wavefunction profile of the six lowest absolute energy eigenstates on a flake geometry. These are the subgap states marked in red in the corresponding spectrum displaying 60 eigenenergies around zero in (d). Here we use the  parameters $\eta = 0.2t$, $\mu = 0.4t$, $\Delta' = 0.13t$, and $\phi = \pi/2$ corresponding to a bulk superconducting gap $\Delta \approx 0.16t$.}
\label{fig:intravalley_p}
\begin{picture}(0,0) 
\put(-288,230){\textbf{(a)}} 
\put(-172,230){\textbf{(b)}} 
\put(-30,230){\textbf{(c)}}
\put(123,230){\textbf{(d)}}
\end{picture}
\end{figure*}

We now present exact diagonalisation results, for which we have employed a simplified lattice model which accounts only for pairing between the closest sites for which the gap is nonvanishing. For the intervalley $p+ip$ spin triplet phase 
\begin{gather}
    \mathcal{H}_\Delta = \sum_{\langle \bm{R},\bm{R}'\rangle}{\Delta(\bm{R},\bm{R}') c^\dag_{\bm{R},\uparrow} c^\dag_{\bm{R}',\downarrow}} \nonumber \\ 
    \Delta(\bm{R},\bm{R}') = \begin{cases}
      \Delta' e^{i(\theta - \frac{\pi}{2})} \ \ \ \ \bm R' \in A \\
      \Delta' e^{i(\theta + \frac{\pi}{2})} \ \ \ \ \bm R' \in B 
    \end{cases}
\end{gather}
 where $\theta$ is the hopping direction. 

For the intervalley $s_\tau$ spin triplet phase, we find that pairing vanishes exactly between nearest neighbours, thus we consider only pairing between next nearest neighbours, 
\begin{gather}
    \mathcal{H}_\Delta =\sum_{\langle \langle \bm{R},\bm{R}'\rangle\rangle}{\Delta(\bm{R},\bm{R}') c^\dag_{\bm{R},\uparrow} c^\dag_{\bm{R}',\downarrow}} \nonumber \\ 
    \Delta(\bm{R},\bm{R}') = \begin{cases}
      +\Delta' \ \ \ \ \theta = 0, \pm \frac{2\pi}{3} \\
      -\Delta' \ \ \ \ \theta = \pi, \pm \frac{\pi}{3}
    \end{cases}
\end{gather} 
For the intravalley $p+i\tau p$ spin triplet phase, we consider pairing between nearest neighbors, 
\begin{gather}
    \mathcal{H}_\Delta = \frac{1}{2}\sum_{\langle \bm{R},\bm{R}'\rangle}{\Delta(\bm{R},\bm{R}')(e^{i\phi_s} c^\dag_{\bm{R},\uparrow}c^\dag_{\bm{R}',\uparrow} + e^{-i\phi_s}c^\dag_{\bm{R},\downarrow}c^\dag_{\bm{R}',\downarrow})} \nonumber \\
\Delta(\bm{R},\bm{R}') =\begin{cases}
+\Delta' & \bm{R}'\in A \\
-\Delta' & \bm{R}'\in B
\end{cases}
\end{gather}
For the intervalley $s_\tau$ and $p+ip$ spin triplet phases, we may write the Bogoliubov-de Gennes Hamiltonian in matrix form as 
\begin{align}
\label{eq:topo_BdG_stau_p+ip}
& \mathcal{H}_\text{nor.} + \mathcal{H}_\Delta = \\ \nonumber
& \sum_{\vec{R}, \vec{R}^\prime} \left(\begin{array}{cc}
c^\dag_{\vec{R},\uparrow} & c_{\vec{R},\downarrow}\end{array}\right)
\left(\begin{array}{cc}
H_{\uparrow \uparrow}(\vec{R}, \vec{R}^\prime) & \Delta_{\uparrow \downarrow} (\vec{R}, \vec{R}^\prime) \\
\Delta^\dag_{\uparrow \downarrow}(\vec{R}, \vec{R}^\prime) & - H^*_{\downarrow \downarrow}(\vec{R}, \vec{R}^\prime)
\end{array}\right)
\left(\begin{array}{c}
c_{\vec{R}^\prime,\uparrow} \\ c^\dag_{\vec{R}^\prime,\downarrow}
\end{array}\right)
\end{align}
where the normal-state Hamiltonian $\mathcal{H}_\text{nor.}$ is defined in Eq. \eqref{eq:topo_H_norm}.

Similarly, we may write the Bogoliubov-de Gennes Hamiltonian for the intravalley $p+i\tau p$ spin-triplet phase as
\begin{align}
\label{eq:topo_BdG_p+itaup}
& \mathcal{H}_\text{nor.} + \mathcal{H}_\Delta = \\ \nonumber
& \frac{1}{2} \sum_{\vec{R}, \vec{R}^\prime, s} \left(\begin{array}{cc}
c^\dag_{\vec{R},s} & c_{\vec{R},s}\end{array}\right)
\left(\begin{array}{cc}
H_{s, s}(\vec{R}, \vec{R}^\prime) & \Delta_{s,s} (\vec{R}, \vec{R}^\prime) \\
\Delta^\dag_{s,s}(\vec{R}, \vec{R}^\prime) & - H^*_{s,s}(\vec{R}, \vec{R}^\prime)
\end{array}\right)
\left(\begin{array}{c}
c_{\vec{R}^\prime,s} \\ c^\dag_{\vec{R}^\prime,s}
\end{array}\right)
\end{align}
%which satisfies a $\mathbb{Z}_2$ spin-gauge symmetry $\psi^\dagger_s \to (s_z)_{s,s} \psi^\dagger_s$.  This symmetry operation is a product of a spin-rotation $\psi^\dagger_s \to i(s_z)_{s,s} \psi^\dagger_s$ and a $U(1)$ electromagnetic gauge transformation $\psi^\dagger_s \to - i \psi^\dagger_s$.
Here the two blocks with opposite $s_z$ eigenvalue are related by time-reversal symmetry $\mathcal{T} = i s_y K$, while each block separately satisfies particle-hole symmetry.
The block-diagonal form allows us to perform the exact diagonalisation in only one of the two spin blocks, and infer the results in the other block by its relation required by time-reversal symmetry or particle-hole antisymmetry.

\begin{figure}
\includegraphics[width = 0.45\textwidth]{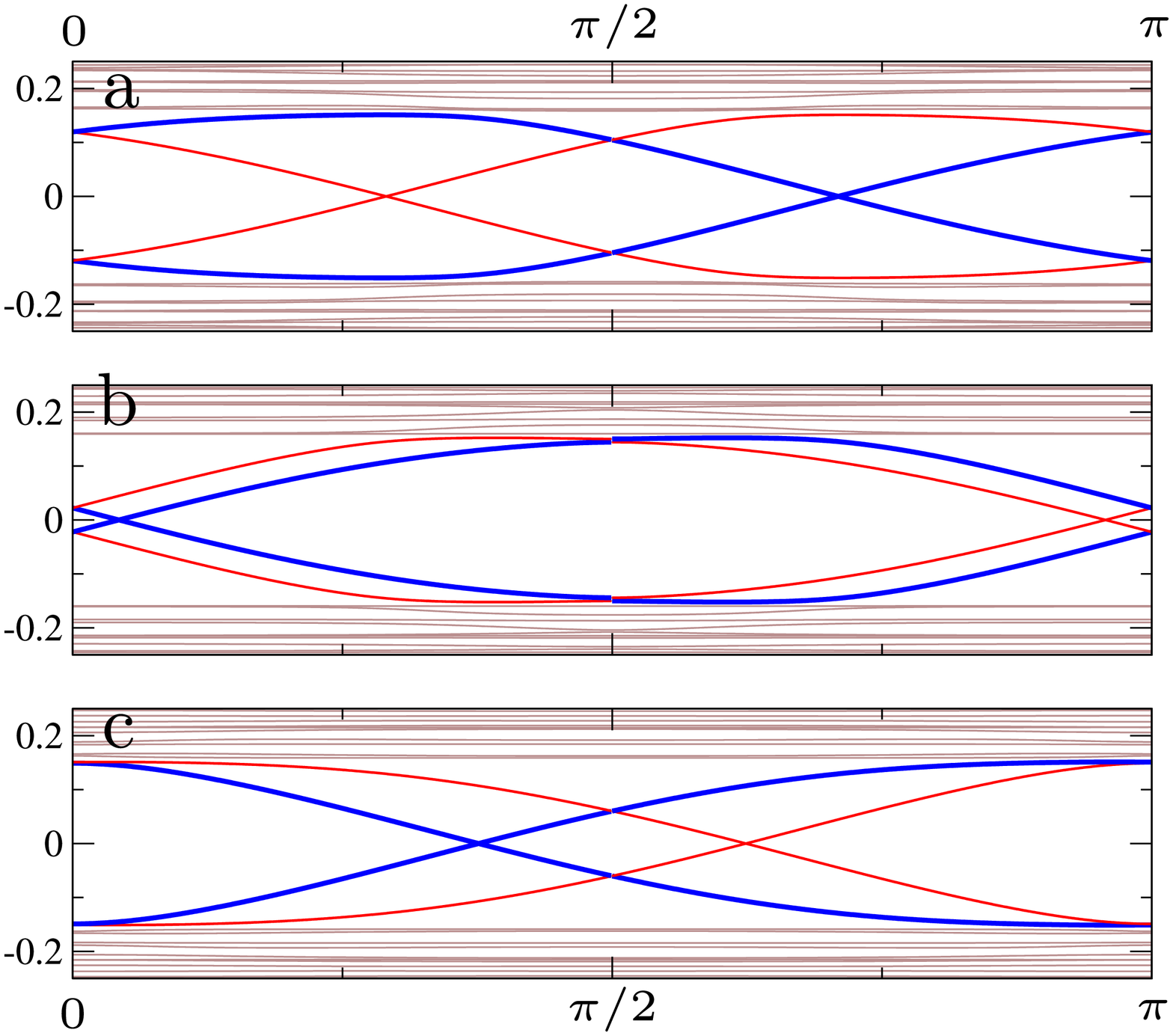}
\caption{Dependence of the spectrum (in units of $t$) on the pair density wave order parameter $\phi$ in the $p+i\tau p$ state, for ribbons of various width. The energy spectrum at $k=0$ of an infinite superconducting ribbon with armchair termination of width (a) 35, (b) 36, (c) 37 unit cells as a function of $\phi$, for parameters $\eta = 0.2t, \mu = 0.4t$ and $\Delta' = 0.13t$ corresponding to a bulk superconducting gap $\Delta \approx 0.16t$.}
\label{fig:phi}
\begin{picture}(0,0) 
\put(-140,360){\textbf{(a)}} 
\put(-140,280){\textbf{(b)}} 
\put(-140,203){\textbf{(c)}}
\end{picture}
\end{figure}

\begin{figure}
\includegraphics[width = 0.5\textwidth]{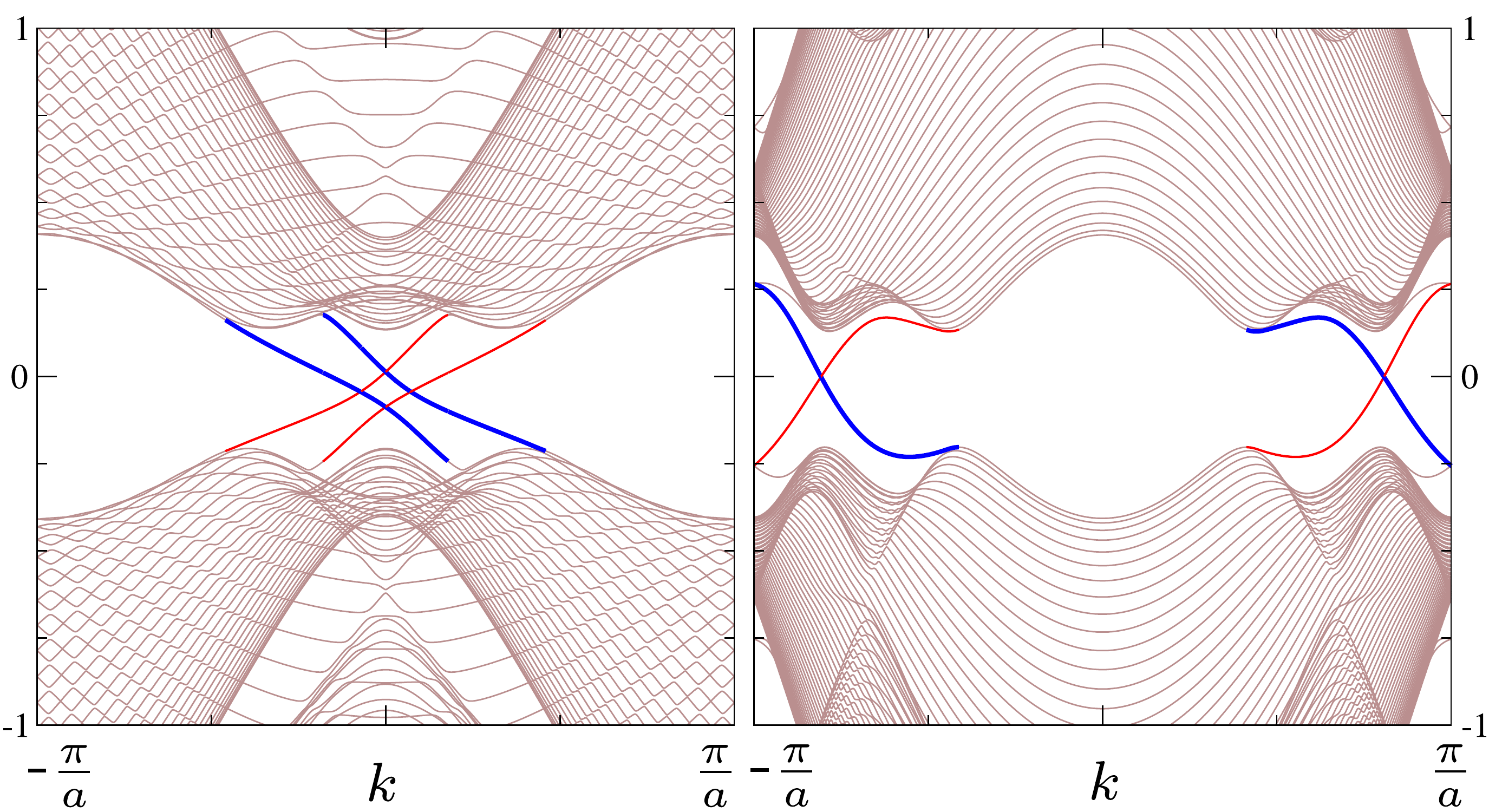}
\caption{The 1D dispersion of infinite superconducting ribbons with (a) armchair and (b) zigzag terminations in the $p+ip$ intervalley spin triplet phase, with parameters $\eta = 0.2t$, $\mu = 0.4t$, and $\Delta' = 0.2t$ corresponding to a bulk superconducting gap $\Delta \approx 0.18t$. Edge modes propagating along opposite edges are shown in different colors.}
\label{fig:intervalley_p}
\begin{picture}(0,0) 
\put(-140,276){\textbf{(a)}} 
\put(5,276){\textbf{(b)}} 
\end{picture}
\end{figure}
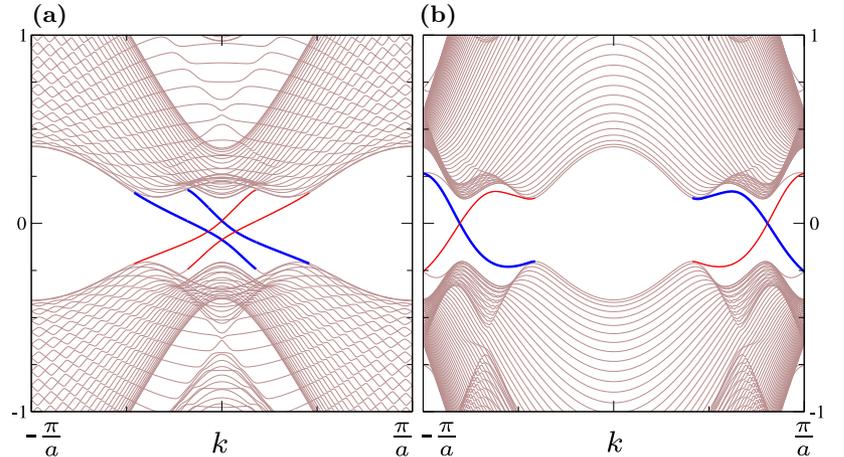

We plot the spectrum of infinite superconducting ribbons in the intervalley $s_\tau$, intravalley $p+i\tau p$ and intervalley $p+ip$ spin triplet phases in Figs. \ref{fig:intervalley_s}, \ref{fig:intravalley_p}, \ref{fig:intervalley_p}, as a function of momentum $k$ along the ribbon respectively. We observe anomalous edge features in all three cases. In most cases, the 1D dispersion of modes propagating along opposite edges is split, and the lines of different color and thickness indicate opposite edge modes. { We note that the Bogoliubov-de Gennes Hamiltonian \eqref{eq:topo_BdG_stau_p+ip} does not exhibit a redundancy associated with particle-hole doubling, thus the quasiparticle spectra are not particle-hole symmetric for the intervalley phases.}

To demonstrate the second-order topology of the intervalley $s_\tau$ and intravalley $p+i\tau p$ spin triplet phases, we show the wavefunction profile of the six lowest energy eigenstates forming the Majorana corner modes on a hexagonal flake geometry, and corresponding spectrum in Figs. \ref{fig:intervalley_s} and \ref{fig:intravalley_p}. 
The exact diagonalisation of the BdG Hamiltonians was performing within a spin-block for both intervalley $s_\tau$ and intravalley $p+i\tau p$ spin-triplet phases, c.f. Eqs. \eqref{eq:topo_BdG_stau_p+ip} and \eqref{eq:topo_BdG_p+itaup}, so the Majorana corner modes in both cases have a degenerate Kramers partner in the opposite spin block.

For the intervalley $s_\tau$ spin triplet phase, in which pairing occurs between opposite spins, we plot the spectrum of the non-redundant BdG Hamiltonian, so that each energy eigenvalue corresponds to a quasiparticle whose antiparticle is identical to its Kramers partner. Decomposing each zero energy mode into two Majorana modes, we find one Majorana Kramers pair at each corner which are protected by time-reversal symmetry. Two gapless counterpropagating modes are observed on each edge for the armchair geometry, but we find no edge states for the zigzag geometry, as shown in Fig. \ref{fig:intervalley_s}. On the flake geometry, Majorana corner states appear on corners between zigzag edges. The Majorana corner modes are a signature of the intrinsic second-order topology of the crystalline bulk superconductor, because these corner modes can not be removed without breaking the symmetries or closing the bulk gap \cite{Geier2018}. 

For the intravalley $p+i\tau p$ phase, in which pairing occurs for equal spins, we plot the spectrum for the BdG Hamiltonian within a single spin block, so that each energy eigenvalue corresponds to a quasiparticle with a Kramers partner in the opposite spin block. For a hexagonal flake with an armchair boundary, at $\phi=\pi/2$, we find one Majorana Kramers pair at each corner of the flake, protected by time-reversal symmetry. These results confirm the existence of a second-order topology which we concluded in the previous section via the symmetry-based indicators. We observe gapless counterpropagating modes along each edge for the zigzag geometry, however the edge behavior for an armchair ribbon is sensitive to the width of the ribbon as well as the value of the pair density wave order parameter $\phi$,  as shown in Fig. \ref{fig:intravalley_p}. Both the armchair ribbon and flake have a width of 35 unit cells. In this case, the plotted ribbon dispersion with armchair edges is gapped, and the flake exhibits Kramers pairs of Majorana corner states on corners between armchair edges at $\phi = \pi/2$. For $\phi = \pi/2$, the flake is inversion symmetric and the zero energy corner modes are a signature of the intrinsic second-order topology of the bulk superconductivity \cite{Li2021,Geier2018}.

The behavior of edge modes exhibits a threefold periodicity in the ribbon width. In Fig. \ref{fig:phi} we show the level spectrum at $k=0$ as a function of $\phi$ for an armchair ribbon of various widths, (a) 35, (b) 36, (c) 37 unit cells. In all cases, the modes propagating along the left and right edges have distinct 1D dispersions, except at values $\phi = n\pi/2$, with $n\in \mathbb{Z}$ for which the gap function $\Delta(\bm{R},\bm{R}')$ is reflection symmetric about the center of the ribbon. We find that the edge is always gapped at $\phi = \pi/2$, when $\Delta(\bm{R},\bm{R}')$ is odd under inversion. At $\phi=0$, the gap function is even under inversion and no higher-order topology is possible. The edge gap closes for a value of $\phi$ between $\phi = 0$ and $\phi = \pi/2$, corresponding to the transition at which the Majorana corner modes disappear. 

For the intervalley $p+ip$ spin triplet phase, two co-propagating modes are observed on each edge for both the armchair and zigzag geometries, as illustrated in Fig. \ref{fig:intervalley_p}. Keeping in mind that these modes have been obtained from the BdG Hamiltonian in the form of Eq. \eqref{eq:topo_BdG_stau_p+ip}, the chiral edge modes are Dirac fermions, i.e. they are not their own antiparticle.

\section{Discussion}
\label{disc}

In this paper we considered the phase diagram of an interacting artificial honeycomb superlattice, with Fermi pockets around the Dirac $K$, $K'$ points, subject to intrinsic spin-orbit coupling, i.e. a doped two-dimensional topological insulator.

We have shown that first and second-order topological superconductivity arises purely due to the Coulomb repulsion, an effect which is enhanced in the limit of localised atomic orbitals.

The mechanism has been elucidated for general lattice models with $C_{6v}$ point group symmetry, $SU(2)$ spin rotation and time-reversal symmetry in \cite{Li2021}, and is extended here in two ways: ($i$) the influence of intrinsic spin-orbit coupling is incorporated, breaking spin $SU(2)\to U(1)$ generating a $\mathbb{Z}_2$ topological bandgap, and ($ii$)  microscopic modeling for a specific, experimentally promising, material is presented.  However, while our field theory treatment is generic, we present results specifically for a model of an artificial honeycomb lattice based on a nanopatterned hole-doped semiconductor quantum well, having in mind the fact that in this situation there is a high degree of experimental control over the electron-electron interaction as well as the band structure.

    Our microscopic modeling shows that three distinct (first and second-order) topological superconducting phases emerge for realistic material parameters, and moreover, that these instabilities are the leading weak coupling instabilities of the Fermi surface -- with magnetic and charge ordering only setting in at larger interaction strengths.  The superconducting phases are:
    \begin{enumerate}
        \item $p+ip$ intervalley, which admits a first-order topological invariant and therefore hosts gapless chiral edge modes; we have shown from numerical calculations that this phase hosts two co-propagating chiral Dirac fermionic edge modes.
        \item $p+i\tau p$ intravalley, a spatially modulated pair density wave which hosts two Majorana edge modes of opposite chirality due to the opposite pairing in the two valleys. Hybridization of the edge modes may give rise the Kramers pairs of Majorana corner modes. We have confirmed the corresponding second-order topology of the bulk superconductor for $\phi=\pi/2$, where $\phi$ is the phase of the pair density wave, using an argument from symmetry-based indicators in addition to our exact diagonalisation results.
        \item $s_\tau$ intervalley, which is also second-order topological, but with a bulk order that has a different spatial structure to the $p+i\tau p$ state. It is interesting to note that despite the $s$-wave nature of the $s_\tau$ state, the phase exhibits nontrivial higher-order topology. This is due to the fact that while the gap is $s$-wave, it has different signs on the two Fermi surfaces at each valley, since the gap is proportional to $\tau_y$. 
    \end{enumerate}

    The boundary physics of the superconducting state could be probed in experiment through STM \cite{Gray2019}, or through measurements of the Josephson critical current \cite{Choi2020}. It has been proposed that higher-order topological superconductors host Majorana states at disinclinations and defects \cite{Teo2013,Benalcazar2014,Zhu2018,Geier2021,Roy2021}, a phenomenon currently unexplored experimentally, which could offer another signature of higher-order topology.

    Recent progress in $n$-type semiconductors patterned with a honeycomb superlattice \cite{AG,AG2,Du2021} has clearly demonstrated Dirac band structure features.  Our findings show that $p$-type semiconductor patterned with a honeycomb superlattice is an enticing avenue towards topological superconducting phases. The $p$-type semiconductor allows for strong intrinsic spin-orbit coupling, which is otherwise negligible in $n$-type. Stronger spin-orbit coupling reduces the effective Dirac velocity, flattening the bands and enhancing interaction effects compared to the $n$-type scenario. We find that having spin-orbit coupling as an additional handle, we are more readily able to realise the necessary conditions for the pairing mechanism discussed here.

 { It would be interesting to further pursue the possible coexistence between the magnetic instabilities and the superconducting phases explored here. A similar coexistence has been exhibited in twisted trilayer graphene subject to proximity induced spin-orbit coupling \cite{LinZFDE}, which was further argued to introduce other exotic transport signatures, such as the zero-field superconducting diode effect \cite{ScammellScheurer2022}. }

 Other superconducting superlattice systems in which spin-orbit coupling is present intrinsically include, e.g. twisted transition metal dichalogenides \cite{Wang2020}, and Ba$_6$Nb$_{11}$S$_{28}$ \cite{Devarakonda2020}; { or extrinsically, via proximity to a transition metal dichalogenide, include twisted multi-layer graphene systems \cite{Siriviboon, Arora2020}.} Non-superlattice materials featuring superconductivity and Dirac physics, localised orbitals and spin-orbit coupling include Pb$_{1/3}$TaS$_2$ \cite{Yang2021}, few-layer stanene \cite{Liao2018}, monolayer TMDs \cite{Barrera2018,Lu2018,Lu2015,Yang2018,Ye2012}, doped topological insulators \cite{Yonezawa2019,Kreiner2011,Sasaki2012,Liu2015,Sato2013,Novak2013,Fu2010,Fatemi2018,Sajadi2018}, and recently discovered vanadium-based kagome metals \cite{Ortiz2020,Zhu2021,Chen2021,Ortiz2021,Ni2021,Chenb2021,Liang2021,Zhao2021,Kang2021,Jiang2021,Li2021b,Ortiz2019,Zhao2021b,Li2021c,Qian2021,Christensen2021,Tan2021,Park2021,Wu2021,Scammell2022,Zhou2022,Tazai2022,DiSante2022,Nguyen2022}. Many of these systems exhibit superconductivity at relatively low carrier densities, and a phase diagram as a function of density similar to the one predicted here. It is our hope that the present study offers a new perspective on the results of these experiments, and suggests new directions to explore.

\section*{Acknowledgements}
H. D. Scammell acknowledges funding from ARC Centre of Excellence FLEET. MG acknowledges support by the European Research Council (ERC) under the European Union’s Horizon 2020 research and innovation program under grant agreement No.~856526, and from the Deutsche Forschungsgemeinschaft (DFG) Project Grant 277101999 within the CRC network TR 183 "Entangled States of Matter" (subproject A03 and C01), and from the Danish National Research Foundation, the Danish Council for Independent Research | Natural Sciences.

\widetext

\appendix

\begin{center}
\textbf{\large Appendices}
\end{center}
\section{Deriving the effective Dirac Hamiltonian}
\label{hamiltonian2}

\subsection{2D hole gas}

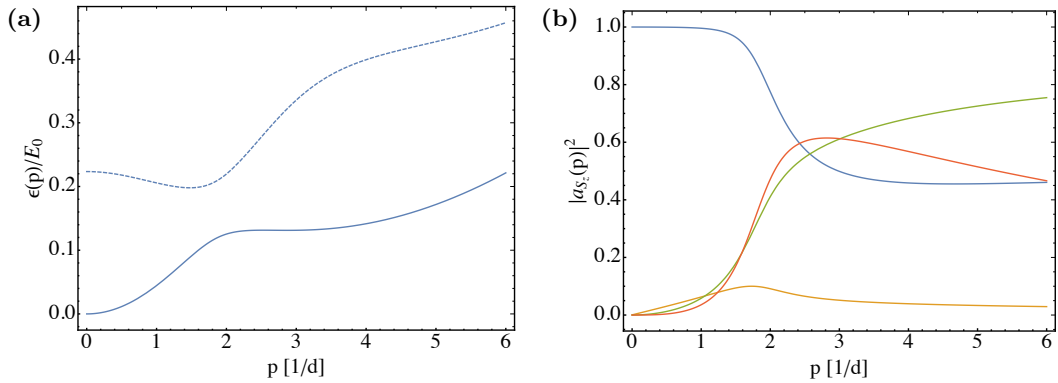
\begin{figure*}[t!]
\includegraphics[width=65mm]{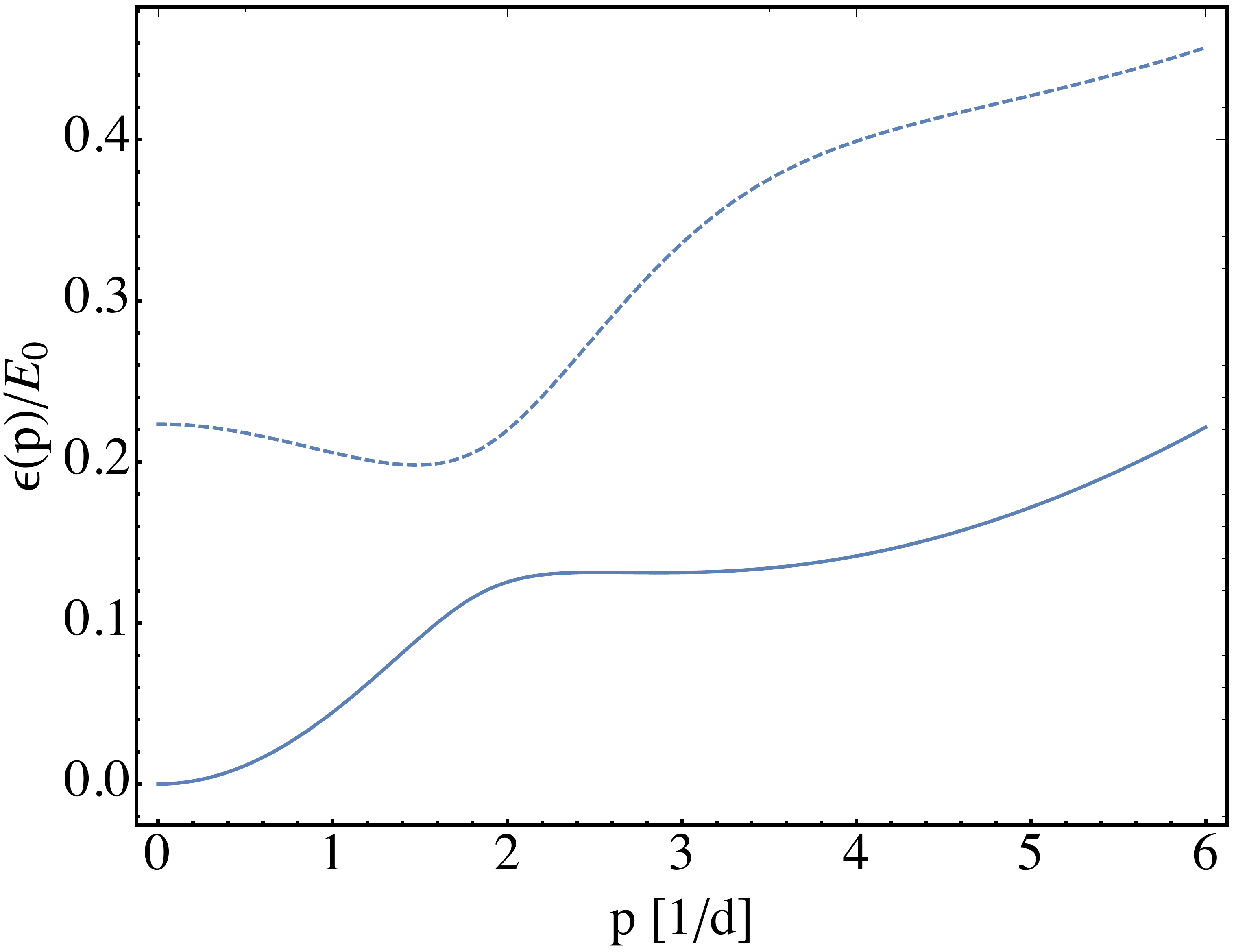}\hspace{0.5cm}
\includegraphics[width=65mm]{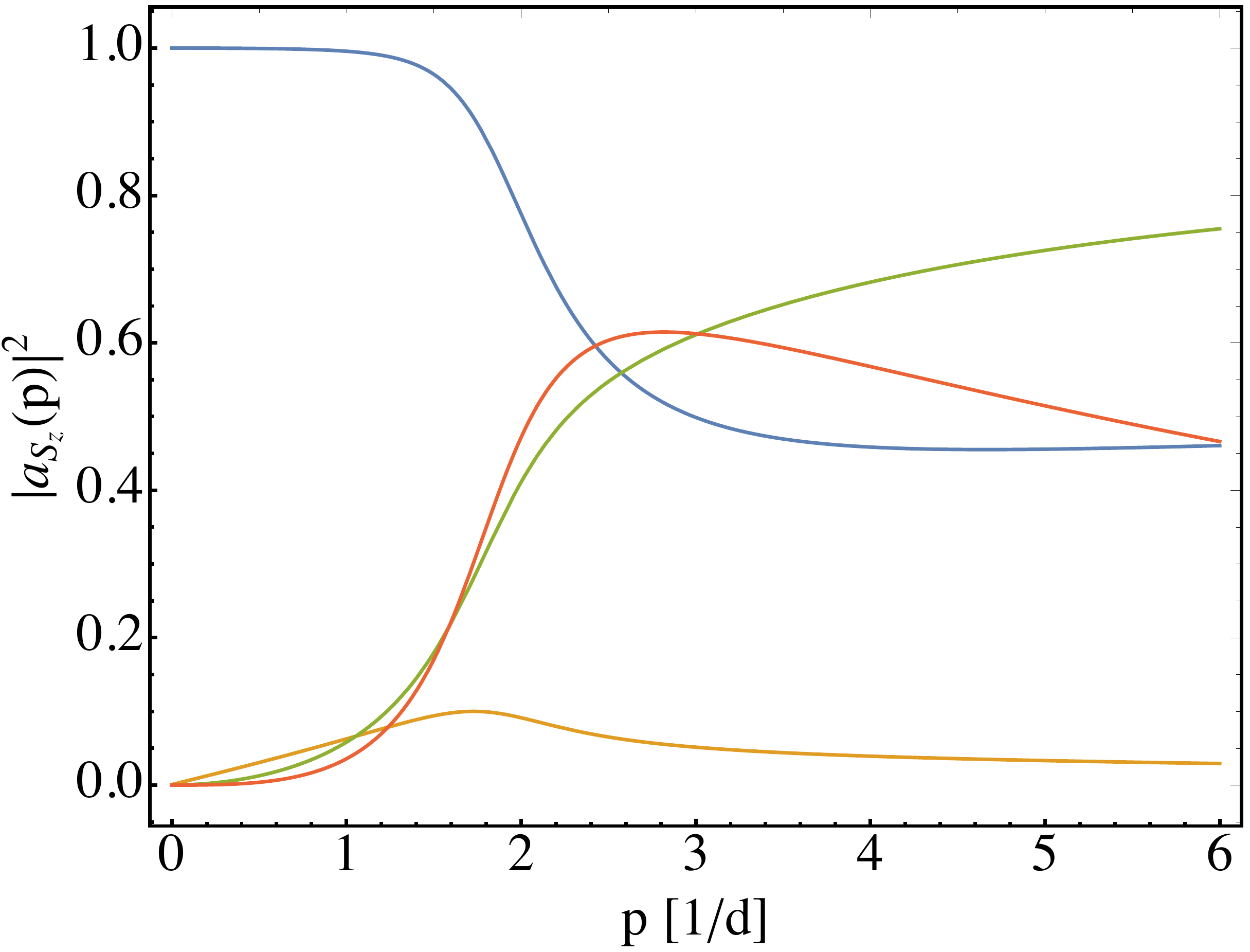}
 \caption{(a) 2DHG spectrum, $\epsilon(p)/E_0$ for InAs. Solid blue line corresponds to the doubly degenerate spectrum that enters the computations of the effective Dirac Hamiltonian \eqref{Hdirac}, the dashed line is the next highest subband, which is ignored in our approximations.  (b) Probability densities $|a_{S_z}(p)|^2$ of each physical spin component $S_z$, presented in \eqref{A:wavefunc2}.}
\label{fig:2dhg}
\begin{picture}(0,0) 
\put(-200,200){\textbf{(a)}} 
\put(0,200){\textbf{(b)}} 
\end{picture}
\end{figure*}

{ The two dimensional hole gas can be described by the Luttinger Hamiltonian in the axial approximation, i.e. $U(1)$ symmetry in-plane,
%\begin{widetext}
\begin{align}
\label{a:HL}
H_L&=H_0 + V\\
\notag H_0&=\left(\gamma_1+2\gamma_2\left(\frac{5}{2}- S_z^2\right)\right)\frac{p_z^2}{2m}+W(z) +\left(\gamma_1-\gamma_2\left(\frac{5}{4}- S_z^2\right)\right)\frac{\bm p^2}{2m},\\
\notag V&=-\frac{\gamma_2+\gamma_3}{8m}\left(p_+^2S_-^2 + p_-^2S_+^2\right) - \frac{\gamma_3}{4m}\{p_z,\{S_z,p_+S_- + p_-S_+\}\},
\end{align}
%\end{widetext}
where $S_x, S_y, S_z$ are angular momentum 3/2 operators, $S_{\pm}=S_x \pm S_y$ and we use bold font to express the in-plane momenta $\bm p =(p_x,p_y)$, and $p_{\pm}=p_x \pm p_y$. The axial approximation is useful for quasi-2D systems with frozen dynamics along one direction -- in the present case, the $z$-axis.

We perform exact diagonalization of $H_L$ in the basis of wavefunctions obtained from $H_0$. We use the lowest lying states in the quantum well to define effective spin $s =\uparrow,\downarrow$ states, which corresponds to the doubly degenerate band of Figure \ref{fig:2dhg}, and are related by the following transformation,
\begin{align}
\label{spinPsi}
\hat{\Psi}_{\uparrow}(\bm p)&=\bar{A}_\uparrow(\bm p)e^{i\bm p\cdot \bm r}=\begin{pmatrix}a_{3/2}(p)\hat{o}\\ ia_{1/2}(p)e^{i\phi} \hat{e}\\ a_{-1/2}(p)e^{2i\phi} \hat{o}\\  -ia_{-3/2}(p)e^{3i\phi}\hat{e}\end{pmatrix}e^{i\bm p\cdot \bm r}, \\ 
\hat{\Psi}_{\downarrow}(\bm p)&=\bar{A}_\downarrow(\bm p)e^{i\bm p\cdot \bm r}=\begin{pmatrix}i a_{-3/2}(p)e^{-3i\phi}\hat{e}\\ a_{-1/2}(p)e^{-2i\phi} \hat{o}\\ -ia_{1/2}(p)e^{-i\phi} \hat{e}\\  a_{3/2}(p)\hat{o}\end{pmatrix}e^{i\bm p\cdot \bm r}.
\end{align}
The complex phase is given by the in-plane momenta $e^{i\phi}=(p_x+ip_y)/p$, with $p=|\bm p|$ and the coefficients $a_{S_z}(p)$ are found numerically via exact diagonalisation of the Luttinger Hamiltonian \eqref{a:HL}, shown in Fig. \ref{fig:2dhg}(b). Finally, we have introduced two orthogonal vectors $\hat{e}, \hat{o}$, which account for the even and odd parity (inversion in $z$-axis) of the wave function/spin components, $a_{S_z}(p, z)$. 
}

\subsection{Superlattice potential}
For a superlattice placed on top of the 2DHG heterostructure, the superlattice potential has a $z$ dependence,
\begin{align}
W(\bm r,z)&=2W_0 \sum_i \cos(\bm G_i\cdot \bm r)e^{-(z+z_0)G_0},
\end{align}
where ${\bm G}_i$ are the reciprocal lattice vectors connecting corners of the hexagonal Brillouin zone Figure \ref{fig:Schematics}.  Here, $z=0$ is the center of the quantum well and $z_0$ is the distance from the superlattice to the center of the quantum well. This top-gate superlattice breaks inversion symmetry, but one may argue that parity breaking effects are exponentially suppressed and may be ignored in the regime where $z_0G\ll 1$. Alternatively, we can envisage placing a superlattice on both the top and bottom gates -- preserving parity. This is captured by, 
\begin{align}
W(\bm r,z)&=2W \sum_i \cos(\bm G_i\cdot \bm r)e^{-z_0G_0}\cosh(z G_0).
\end{align}
Henceforth we remove explicit parity breaking in the superlattice potential, working with the expression given in the main text Eq. \eqref{potential}.

\subsection{Effective Dirac Hamiltonian}
{ We describe the problem by the Hamiltonian operator,
\begin{align}
\label{Hamiltonian}
\hat{H}&=\hat{\cal E} + \hat{U}(\bm r),
\end{align}
whereby $\hat{\cal E}$ represents the kinetic energy operator.
To proceed we need to project the Hamiltonian operator onto an appropriate basis, which we generate directly from the wave functions of the 2D confined Luttinger Hamiltonian (\ref{a:HL}), i.e. $\ket{l, \sigma_l, \bm k}$, with an extra index $i$ added, which labels sites in the momentum grid $\bm k_i=\bm k + \bm g_i$, where the discrete momentum space grid $\bm g_i \in \{n_1 \bm G_1 + n_2 \bm G_2 + n_3 \bm G_3 : n_i\in Z\}$, is the space of degenerate momentum points. Therefore we write $\ket{l, \sigma_l, \bm k}\to\ket{l, \sigma_l, \bm k, i}$, and project the Hamiltonian operator onto this basis.

We generate the Hamiltonian by projecting onto the 2DHG wavefunctions, and enumerated by the lattice momentum $K_i$,
\begin{align}
\bar{A}_{s}(\bm p +\tau \bm K_j) &\equiv \bar{A}_{j, s,\tau}(\bm p),\\
\braket{\bm r |j, s,\tau,\bm p}&= \bar{A}_{j, s,\tau}(\bm p) e^{i\tau \bm K_j\cdot \bm r} e^{i\bm p\cdot \bm r},\\
\
\label{Hgrid}
{\cal H}_{is\tau;js'\tau'}(\bm p)&\equiv \bra{i,s,\tau,\bm p}H_\text{2DHG} + U(\bm r)\ket{j,s',\tau',\bm p}\\
\notag &={\cal E}_{is\tau;js'\tau'}(\bm p)+ U_{is\tau;js'\tau'}(\bm p).
\end{align}
We further decouple the potential into a spin (and valley) independent and dependent terms,
\begin{align}
U_{is\tau;js'\tau'}=U^0_{i;j}\delta_{s,s'}\delta_{\tau,\tau'} + \delta U_{is\tau;js'\tau'}
\end{align}
The lowest energy, doubly-degenerate subspace of $U^0_{ij}$ defines the pseudospin index $\sigma=\pm$; we project onto this subspace. Explicitly, we work with a weak potential, small $W$, such that we keep only the first three $K$-points, i.e. $K_i\in\{K_1, K_2, K_3\}$ The pseudospin eigenfunctions (i.e. of $U^0_{ij}$) are then obtained analytically,
\begin{align}
C_{\sigma,j}=\frac{1}{\sqrt{3}} e^{i2\pi(j-1)\sigma/3}.
\end{align}
Finally, projecting the Hamiltonian \eqref{Hgrid} into this psuedospin ($\sigma=\pm$) subspace, we arrive at the effective single-particle Hamiltonian near the Dirac point ($\bm p=\bm 0$) }
\begin{align}
\label{A:wavefunc1}
\braket{\bm r | s,\tau,\sigma,\bm p}&=\sum_{j=1}^3 \bar{A}_{j,s,\tau} C_{\sigma,j} e^{i\tau \bm K_j\cdot \bm r} e^{i\bm p\cdot \bm r},\\
\notag \left({\cal H}_0\right)_{s,\tau,\sigma;s',\tau',\sigma'}&=\bra{s,\tau, \sigma, \bm p}{\cal H} _{2DHG+W}\ket{s',\tau',\sigma', \bm p}\\
&=\left(v(\bm\sigma \cdot \bm{p} ) \tau_z - \mu +  \eta \sigma_z  s_z\right)_{s,\tau,\sigma;s',\tau',\sigma'}.
\end{align}
We have allowed for a chemical potential $\mu$ which can be tuned using e.g. gates.  Explicitly, we take
\begin{align}
\label{A:wavefunc2}
\bar{A}_{j, s=\uparrow, \tau}&=\begin{pmatrix}a_{3/2}(K_0)\hat{o}\\ ia_{1/2}(K_0)\tau e^{i2\pi(j-1)/3} \hat{e}\\ a_{-1/2}(K_0)e^{i4\pi(j-1)/3}\hat{o}\\  -ia_{-3/2}(K_0)\tau e^{i2\pi(j-1)}\hat{e}\end{pmatrix},\  \bar{A}_{j, s=\downarrow, \tau}=\begin{pmatrix}i a_{-3/2}(K_0)\tau e^{-2\pi(j-1)}\hat{e}\\ a_{-1/2}(K_0)e^{-i4\pi(j-1)/3} \hat{o}\\ -ia_{1/2}(K_0)\tau e^{-i2\pi(j-1)/3} \hat{e}\\  a_{3/2}(K_0)\hat{o}\end{pmatrix}.
\end{align}
For completeness we also specify the symmetry properties of the wavefunctions. The symmetries of a honeycomb system are $2\pi/3$ and $\pi$ rotations, reflections, and time reversal. At the high symmetry points $\bm p=\bm 0$, using the explicit form of the wavefunctions (\ref{A:wavefunc1}) and (\ref{A:wavefunc2}) the transformations are found to be
\begin{align}
%\notag D(R_{2\pi/3}) \ket{s,\tau,\sigma} &=\sum_{i=1}^3 \bar{A}_{R_{2\pi/3}^{-1}\bm K_i,s,\tau} C_{\sigma,i} e^{i\tau R_{2\pi/3}^{-1}\bm K_i\cdot \bm r} \\
%\notag &=\sum_{i=1}^3 \bar{A}_{\bm K_{i-1},s,\tau} C_{\sigma,i} e^{i\tau \bm K_{i-1}\cdot \bm r}\\
\notag D(C_{3z}) \ket{s,\tau,\sigma} &=-e^{2i\pi\sigma/3}\ket{s,\tau,\sigma} \\
\notag D(C_{2z}) \ket{s,\tau,\sigma} &=-i s\ket{s,-\tau,\sigma}   \\
\notag D(C_{2x}) \ket{s,\tau,\sigma} &=i \ket{-s,\tau,-\sigma}   \\
\notag D(C_{2y}) \ket{s,\tau,\sigma} &=-s\ket{-s,-\tau,-\sigma}   \\
D({\cal T}) \ket{s,\tau,\sigma} &= -s \ket{-s,-\tau,-\sigma} .
\end{align}
%We choose a natural basis for the pseudospin so that $C_3$ is represented by $e^{-2\pi i \sigma^z/3}$. Then $\sigma^z$ and $\tau^z$ are odd under time reversal. 
The resulting transformation properties are summarized in Table \ref{T:transforms}.

 %%%%%%%%%%%%%%%%%
%%%%%%%%%%%%%%%%% 

\newpage
\section{Polarization Operators}
\label{a:pol}

\subsection{Preliminaries}
The polarisation operators are given by (setting velocity $v=1$ for ease of notation)
\begin{align}
\label{Pi}
\notag i\Pi_{\mu\nu}(p_0, \bm p)&=\int \frac{dq_0 d^2q}{(2\pi)^{3}} \frac{\text{Tr}\left[J^\mu h^\alpha J^\nu h^\beta\right]}{\tilde{q}_0^2-{\bm q}^2-\eta^2}\frac{\tilde{q}_\beta (\tilde{q}+\tilde{p})_\alpha}{(\tilde{q}_0+p_0)^2-({\bm q+ \bm p})^2-\eta^2}
\end{align}
with modified momenta $\tilde{q}_\alpha \in\{q_0 e^{i\epsilon}+\mu, \bm q, \eta\}$,  $\tilde{p}_\alpha\in\{p_0e^{i\epsilon}+\mu, \bm q, 0\}$ with infinitesimal $\epsilon$, and vertices:
$h^\alpha \in\{\mathbbm{1}, \sigma_x, \tau_z \sigma_y, s_z \tau_z \sigma_z\}$, pertaining to the Hamiltonian, and $J_I^\mu\in\{\mathbbm{1}, \sigma_\pm, \tau_z \sigma_z,  \tau_z s_z, \tau_z s_z \sigma_\pm, \sigma_z s_z\}$,
 $J_{II}^\mu\in\{\mathbbm{1}, \sigma_\pm, \sigma_z s_z\}\otimes\tau_\pm$, pertaining to the interactions. 
 
By evaluating the frequency integral by residues, the expression reduces to
\begin{align}
\notag i\Pi_{\mu\nu}(p)&=\int \frac{d^2q}{(2\pi)^{3}} \frac{\text{Tr}\left[J^\mu h^\alpha J^\nu h^\beta\right]}{q_0^2-{\bm q}^2-\eta^2}\frac{q_\beta (q+p)_\alpha}{(q_0+p_0)^2-({\bm q+ \bm p})^2-\eta^2} \left[ 1 - (\Theta(\mu - \varepsilon_q) + \Theta(\mu - \varepsilon_{\bm q + \bm p}))\right]\\
&\equiv i\Pi^{(0)}_{\mu\nu}(p) +i\delta \Pi_{\mu\nu}(p)
\end{align}
where $\Pi^{(0)}_{\mu\nu}(p)$ is defined as the polarization operator at zero chemical potential $\mu=0$, i.e. the interband polarization operator \cite{Li2020}. The remaining contribution $\delta \Pi_{\mu\nu}(p)$ is referred to as the intraband polarization operator.

This function is divergent and requires regularization; we shall use dimensional regularization. Returning to the original expression Eq. \eqref{Pi} and setting $\mu=0$ gives us the intraband contribution; regulating this quantity gives a finite total result. Using the Schwinger parametrization,
\begin{align}
\frac{1}{AB} = \int_0^1\frac{dy}{(yA+(1-y)B)^2}
\end{align}
the function $i\Pi^{(0)}_{\mu\nu}(p)$ becomes
\begin{align}
\label{A:Pi0}
i\Pi^{(0)}_{\mu\nu}(p) \notag &=\int_0^1 dy  \int \frac{dq_0 d^2q}{(2\pi)^{3}} \frac{\text{Tr}\left[J^\mu h^\alpha J^\nu h^\beta\right]q_\beta (q+p)_\alpha}{\left[y(q_0^2-{\bm q}^2-\eta^2)+(1-y)((q_0+p_0)^2-({\bm q+ \bm p})^2-\eta^)\right]^2} \\
%\notag &=\int_0^1 dy  \int \frac{d^3l}{(2\pi)^3} \frac{\text{Tr}\left[J^\mu h^\alpha J^\nu h^\beta\right](l-yp)_\beta (l+(1-y)p)_\alpha}{\left[l_0^2-\bm l^2-\eta^2 + y(1-y)(p_0^2-\bm p^2)\right]^2}\\
&=\int_0^1dy\int \frac{ d^3l}{(2\pi)^3} \frac{\text{Tr}\left[J^\mu h^\alpha J^\nu h^\beta\right](l-yp)_\beta (l+(1-y)p)_\alpha}{\left[l^2-\Delta(p_0,\bm p,y)\right]^2}  
\end{align}
with $\Delta(p_0,\bm p,y)=\eta^2-y(1-y)(p_0^2-\bm p^2)$, where we Wick-rotated to Euclidean momentum $l_\mu$. The expression \eqref{A:Pi0} is evaluated analytically, at zero frequency $p_0=0$, and for all $\mu,\nu$; the results are printed in  Appendix \ref{A:calcs}. The second contribution $\delta \Pi_{\mu\nu}(p)$, which depends on chemical potential $\mu$, can be evaluated through the rearrangements,
\begin{align}
\notag \delta \Pi_{\mu\nu}(p)&=\int \frac{d^2q}{(2\pi)^{2}} \frac{\text{Tr}\left[J^\mu h^\alpha J^\nu h^\beta\right]q_\beta (q+p)_\alpha \Theta(\mu - \varepsilon_q)}{2\varepsilon_q\left[(\varepsilon_q+p_0)^2-({\bm q+ \bm p})^2-\eta^2\right]} + \frac{\text{Tr}\left[J^\mu h^\alpha J^\nu h^\beta\right]q_\beta (q+p)_\alpha \Theta(\mu - \varepsilon_{\bm q + \bm p})}{2\varepsilon_{\bm q + \bm p}\left[(\varepsilon_{\bm q + \bm p}-p_0)^2-q^2-\eta^2\right]}\\
%\notag &=\int \frac{d^2q}{(2\pi)^{2}} \frac{T^{\mu\alpha\nu\beta}q_\beta (q+p)_\alpha \Theta(\mu - \varepsilon_q)}{2\varepsilon_q\left[(\varepsilon_q+p_0)^2-({\bm q+ \bm p})^2-\eta^2\right]}\\
%\notag & + \frac{\text{Tr}\left[J^\mu h^\alpha J^\nu h^\beta\right](q-p)_\beta q_\alpha \Theta(\mu - \varepsilon_q)}{2\varepsilon_q\left[(\varepsilon_q-p_0)^2-|\bm q - \bm p|^2-\eta^2\right]}\\
%\notag &=\sum_{\sigma=\pm}\int \frac{d^2q}{(2\pi)^{2}}\frac{T^{\mu\alpha\nu\beta}q_\beta (q+\sigma p)_\alpha}{\mathbb{D}(\pm \bm p)}\Theta(\mu- \varepsilon_q)\\
&=-\sum_{\sigma=\pm}\sigma\int \frac{dq d\theta}{(2\pi)^{2}}\frac{\text{Tr}\left[J^\mu h^\alpha J^\nu h^\beta\right]q_\beta (q+\sigma p)_\alpha}{\cos\theta + a_\sigma}\frac{\Theta(\mu- \varepsilon_q)}{4\varepsilon_q p},
\end{align}
where,
\begin{align}
\notag a_\sigma&=\frac{q^2+p^2+\eta^2-(\varepsilon_q + \sigma p_0e^{i0})^2}{2qp \sigma}, \ a_\sigma^0\equiv a_\sigma(p_0=0) = \sigma \frac{p}{2q}
\end{align}
We specialise to the static limit $p_0=0$, which means we only need to keep the principle value of $1/(a_\sigma + \cos\theta)$. 

\subsection{Results}
\label{A:calcs}
We decomposed the polarization operator into interband and intraband contributions,
\begin{align}
\Pi_{\mu s \tau}(p_0,\bm p)&=\Pi^0_{\mu s \tau}(p_0,\bm p) + \delta\Pi_{\mu s \tau}(p_0,\bm p).
\end{align}
indices $\mu=0, x,y,z$, $s=0,s_z$, $\tau=0,\tau_z$. 
We denote the relatively few distinct non-zero polarization operators as
\begin{align}
\notag \Pi_\pm(\bm p) & = \Pi_{z00,x00}(0,\bm p) \pm i \Pi_{z00,y00}(0,\bm p)= \Pi_{zs\tau,xs\tau}(0,\bm p) \pm i \Pi_{zs\tau,ys\tau}(0,\bm p),\\
\notag \Pi_0(\bm p)&=\Pi_{0s\tau,0s\tau}(0,\bm p),\\
\notag \Pi_z&=\Pi_{zs\tau,zs\tau}(0,\bm p),\\
\Pi_\eta&= \Pi_{zs_z0;0s_z\tau}(\bm p)= \Pi_{z0\tau_z;000}(\bm p)
\end{align}
Here we factor out $N=8$ coming from the trace (spin $\times$ pseudospin $\times$ valley). Subscripts $x,y,z$ correspond to pseudospin, while $s,\tau$ correspond to spin $s_z$ and valley $\tau_z$. We then evaluate the trace in Eq. \eqref{A:Pi0}, which results in the following integrals which are straightforwardly evaluated:
\begin{align}
\notag i\Pi^{(0)}_{0s;0s}(\bm p)
\notag &=N \int_0^1 dy\int \frac{d^dl}{(2\pi)^{d}} \frac{l_0^2+l_x^2+l_y^2 - y(1-y)\bm p^2 +\eta ^2}{\left[l^2-\Delta(p_0,\bm p,y)\right]^2}\\
\notag &=N   \int_0^1 dy \left[\frac{1}{2}\frac{i \Gamma(1-d/2)}{(4\pi)^{d/2}\Gamma(2)} \Delta^{d/2-1} +  \left(\eta^2-y(1-y)\bm p^2\right) \frac{i \Gamma(2-d/2)}{(4\pi)^{d/2}\Gamma(2)} \Delta^{d/2-2}  \right]\\
\notag &=N  \frac{i}{8\pi}\int_0^1 dy \left[-(\eta^2 + y(1-y)\bm p^2)^{1/2} + \left(\eta^2-y(1-y)\bm p^2\right) (\eta^2 + y(1-y)\bm p^2)^{-1/2}  \right]\\
\notag &=N  \frac{i}{8\pi}\int_0^1 dy \left[-2y(1-y)\bm p^2(\eta^2 + y(1-y)\bm p^2)^{-1/2}  \right]\\
&=N  \frac{i}{8\pi}\left[-\eta + \frac{1}{2p}(4\eta^2-p^2)\arcsin\left[\frac{p}{\sqrt{4\eta^2+p^2}}\right]  \right]
\end{align}
and similarly,
\begin{align}
\notag i\Pi^{(0)}_{zs;zs}(\bm p)&=N  \frac{i}{8\pi}\left[2\eta + \frac{1}{p}(4\eta^2+p^2)\arcsin\left[\frac{p}{\sqrt{4\eta^2+p^2}}\right]   \right],\\
\notag i\Pi^{(0)}_{xs;xs}(\bm p)
\notag&=N  \frac{i}{8\pi} \frac{p_y^2}{p^2} \left[\eta - \frac{1}{2p}(4\eta^2-p^2)\arcsin\left[\frac{p}{\sqrt{4\eta^2+p^2}}\right] \right],\\
i\Pi^{(0)}_{ys;ys}(\bm p) 
\notag&=N  \frac{i}{8\pi} \frac{p_x^2}{p^2} \left[\eta - \frac{1}{2p}(4\eta^2-p^2)\arcsin\left[\frac{p}{\sqrt{4\eta^2+p^2}}\right] \right],\\
\notag  i\Pi^{(0)}_{xs;ys}(\bm p)&=i\Pi^{(0)}_{ys;xs}(\bm p)
= N  \frac{i}{8\pi} \frac{p_x p_y}{p^2} \left[-\eta + \frac{1}{2p}(4\eta^2-p^2)\arcsin\left[\frac{p}{\sqrt{4\eta^2+p^2}}\right] \right],\\
\notag i\Pi^{(0)}_{x00;0s_z\tau_z}(\bm p)&=-i\Pi^{(0)}_{0s_z\tau_z;x00}(\bm p)=i\Pi^{(0)}_{xs\tau;000}(\bm p)=-i\Pi^{(0)}_{000;xs\tau}(\bm p)\\
\notag &= -\frac{i N}{8\pi} \frac{i p_y}{p} \left[2\eta \arcsin\left[\frac{p}{\sqrt{4\eta^2+p^2}}\right]\right] ,\\
\notag i\Pi^{(0)}_{y00;0s_z\tau_z}(\bm p)&=-i\Pi^{(0)}_{0s_z\tau_z;y\bar{s}}(\bm p)=i\Pi^{(0)}_{ys\tau;000}(\bm p)=-i\Pi^{(0)}_{000;ys\tau}(\bm p)\\
&=\frac{i N}{8\pi} \frac{i p_x}{p} \left[2\eta \arcsin\left[\frac{p}{\sqrt{4\eta^2+p^2}}\right]\right]
\end{align}

We now consider the contribution explicitly dependent upon the chemical potential. 
Again with $p_0=0$, one finds
\begin{align}
\notag \delta\Pi_{0s;0s}(\bm p)&=\frac{N}{8\pi}\left[-2\mu+\eta-\frac{1}{2 p}(4\eta ^2 -p^2)\arcsin \left(\frac{p}{\sqrt{4 \eta
   ^2+p^2}}\right)\right]\\
\notag \delta\Pi_{x;x}(\bm p)&=\frac{N}{8\pi}\frac{p_y^2}{p^2}\left[-\eta+\frac{1}{2 p}(4\eta ^2 -p^2)\arcsin \left(\frac{p}{\sqrt{4 \eta
   ^2+p^2}}\right)\right]\\
\notag \delta\Pi_{y;y}(\bm p)&=\frac{N}{8\pi}\frac{p_x^2}{p^2}\left[-\eta+\frac{1}{2 p}(4\eta ^2 -p^2)\arcsin \left(\frac{p}{\sqrt{4 \eta
   ^2+p^2}}\right)\right]\\
\notag \delta\Pi_{z;z}(\bm p)&=\frac{N}{8\pi}\left[2\mu-2\eta-\frac{1}{p}(4\eta ^2 +p^2)\arcsin \left(\frac{p}{\sqrt{4 \eta
   ^2+p^2}}\right)\right]\\
\notag \delta\Pi_{x;y}(\bm p)&=\frac{N}{8\pi}\frac{p_x p_y}{p^2}\left[\eta - \frac{1}{2p}(4\eta^2-p^2)\arcsin \left(\frac{p}{\sqrt{4 \eta
   ^2+p^2}}\right)\right]\\
\notag \delta\Pi_{x;z}(\bm p)&=-\frac{N}{8\pi}ip_y\\
\notag \delta\Pi_{y;z}(\bm p)&=\frac{N}{8\pi} ip_x\\
\notag \delta\Pi_{x00;0\tau}(\bm p)&=\frac{N}{8\pi}\frac{i 2\eta  p_y }{p}\arcsin\left(\frac{\sqrt{p^2}}{\sqrt{4 \eta ^2+p^2}}\right)\\
\notag \delta\Pi_{xs\tau;000}(\bm p)&=\frac{N}{8\pi}\frac{i 2\eta  p_y }{p}\arcsin\left(\frac{\sqrt{p^2}}{\sqrt{4 \eta ^2+p^2}}\right)\\
\notag \delta\Pi_{y00;0s\tau}(\bm p)&=-\frac{N}{8\pi}\frac{i 2\eta  p_x }{p}\arcsin\left(\frac{\sqrt{p^2}}{\sqrt{4 \eta ^2+p^2}}\right)\\
\notag \delta\Pi_{ys\tau;000}(\bm p)&=-\frac{N}{8\pi}\frac{i 2\eta  p_x }{p}\arcsin\left(\frac{\sqrt{p^2}}{\sqrt{4 \eta ^2+p^2}}\right)\\
\delta \Pi_{zs_z0;0s\tau}(\bm p)&=\delta \Pi_{z0\tau_z;000}(\bm p) = \frac{N}{8\pi} 2\eta
\end{align}

Now (with  $s=0,s_z$), we relate to channel $I$ (here ordering of $\tau\bar\tau$ does not affect the result):
\begin{align}
\notag\Pi^{\tau\bar\tau}_{0;0}(\bm p)&= \frac{1}{2}\Pi_{z;z}(\bm p)\\
\notag\Pi^{\tau\bar\tau}_{x;x}(\bm p)&= \frac{1}{2}\Pi_{y;y}(\bm p)\\
\notag\Pi^{\tau\bar\tau}_{y;y}(\bm p)&= \frac{1}{2}\Pi_{x;x}(\bm p)\\
\notag\Pi^{\tau\bar\tau}_{z;z}(\bm p)&= \frac{1}{2}\Pi_{0;0}(\bm p)\\
\notag\Pi^{\tau\bar\tau}_{x;y}(\bm p)&=-\frac{1}{2}\Pi_{x;y}(\bm p)\\
\notag\Pi^{\tau\bar\tau}_{x;0}(\bm p)&=0\\
\notag\Pi^{\tau\bar\tau}_{y;0}(\bm p)&=0\\
\notag\Pi^{\tau\bar\tau}_{z;0}(\bm p)&=\frac{1}{2}\Pi_{zs;0}(\bm p)=N \frac{\eta}{8\pi}\\
\notag\Pi^{\tau\bar\tau}_{x;z}(\bm p)&=0\\
\Pi^{\tau\bar\tau}_{y;z}(\bm p)&=0.
\end{align}

\section{Interactions in Cooper channel}
\label{cooperchannel}

To obtain the interactions in band basis, we define the creation operator $\widetilde{\psi}_k^\dag$ which creates a fermion in the upper band, while $\psi^\dag_k$ creates a fermion of definite pseudospin. Changing to the band basis, we use
\begin{align}
 \psi_k&={\cal U}_{k,\tau,s} \widetilde{\psi}_k \nonumber \\
{\cal U}_{k,\tau,s}&=\begin{pmatrix} w^a_{\tau,s}(k) && \bar{w}^a_{\tau,s}(k) \\ w^b_{\tau,s}(k)e^{ i \tau \theta_k} && \bar{w}^b_{\tau,s}(k)e^{i \tau \theta_k} \end{pmatrix}
\end{align} 
with the wavefunction components given by, $w^a_{\tau,s}(k) \equiv v k/\sqrt{2\epsilon_k(\epsilon_k - \alpha \tau \eta)},  \ 
  w^b_{\tau,s}(k)\equiv (\tau \epsilon_k-\alpha\eta)/(v k) w^a_{\alpha, k, \tau}$, with functions $\bar{w}^a_{\tau,s}(k), \bar{w}^b_{\tau,s}(k)$ similarly defined for the lower band eigenstates, but not needed. 

We then keep only the upper band, e.g. we use the projections
\begin{align}
%{\cal U}_{[1]}^\dag \sigma_+ \ {\cal U}_{[2]}\Big|_{++}&=f_{[1]}g_{[2]}e^{ i \tau_2 \theta_{k_2}},\\
%{\cal U}_{[1]}^\dag \sigma_- \ {\cal U}_{[2]}\Big|_{++}&=f_{[2]}g_{[1]}e^{- i \tau_1 \theta_{k_1}},\\
\notag {\cal U}_{[1]}^\dag \sigma_x \ {\cal U}_{[2]}\Big|_{++}&=\left(w^a_{[1]}w^b_{[2]}e^{ i \tau_2 \theta_{k_2}}+w^a_{[2]}w^b_{[1]}e^{ -i \tau_1 \theta_{k_1}}\right),\\
\notag {\cal U}_{[1]}^\dag \sigma_y \ {\cal U}_{[2]}\Big|_{++}&=i\left(-w^a_{[1]}w^b_{[2]}e^{ i \tau_2 \theta_{k_2}}+w^a_{[2]}w^b_{[1]}e^{ -i \tau_1 \theta_{k_1}}\right),\\
\notag {\cal U}_{[1]}^\dag \sigma_z \ {\cal U}_{[2]}\Big|_{++}&=\left(w^a_{[1]}w^a_{[2]}- w^b_{[1]}w^b_{[2]}e^{ -i \tau_1 \theta_{k_1}+i \tau_2 \theta_{k_2}}\right),\\
{\cal U}_{[1]}^\dag \sigma_0 \ {\cal U}_{[2]}\Big|_{++}&=\left(w^a_{[1]}w^a_{[2]}+ w^b_{[1]}w^b_{[2]}e^{ -i \tau_1 \theta_{k_1}+i \tau_2 \theta_{k_2}}\right).
\end{align}
We compress notation such that indices are ${[1]}=\{k_1, \tau_1, s_1\}$. The notation ``$|_{++}$'' indicates that we consider just the upper-band contribution. The phase factors owe to the single particle Berry phase and play a central role in the pairing mechanism.

In the Cooper channel, $ \bm k_1=-\bm k_3$, $\bm k_2=-\bm k_4$, such that $\theta_{k_3}=\pi + \theta_{k_1}$, $\theta_{k_4}=\pi + \theta_{k_2}$. The matrix elements of the screened Coulomb interaction in the upper band, separated into intravalley and intervalley Cooper channels, are obtained as,
\begin{align}
\label{Vprojection}
%d^\dag_{[1]} d^\dag_{[3]}d_{[2]}d_{[4]} 
\notag {\cal V}_{intra}&=\sum_{\tau_i, s_i, \bm k_i} \left({\cal U}_{[1]}^\dag \otimes {\cal U}_{[3]}^\dag\hat{V}_I \ {\cal U}_{[2]} \otimes{\cal U}_{[4]} \right) \Big|_{++} \delta_{\tau_1,\tau_2,\tau_3,\tau_4} \delta_{s_1,s_2} \delta_{s_3,s_4}\delta_{\bm k_1, -\bm k_3} \delta_{\bm k_2, -\bm k_4},\\
\notag {\cal V}_{inter}&=\sum_{\tau_i, s_i, \bm k_i} \Big\{\left({\cal U}_{[1]}^\dag \otimes {\cal U}_{[3]}^\dag\hat{V}_I \ {\cal U}_{[2]} \otimes{\cal U}_{[4]} \right) \Big|_{++} \delta_{\tau_1,\tau_2,-\tau_3,-\tau_4} \delta_{s_1,s_2} \delta_{s_3,s_4}\delta_{\bm k_1, -\bm k_3} \delta_{\bm k_2, -\bm k_4}\\ 
 &+ \left({\cal U}_{[1]}^\dag \otimes {\cal U}_{[3]}^\dag\hat{V}_{II} \ {\cal U}_{[2]} \otimes{\cal U}_{[4]} \right) \Big|_{++} \delta_{\tau_1,-\tau_2,-\tau_3, \tau_4} \delta_{s_1,s_3} \delta_{s_2,s_4}\delta_{\bm k_1, -\bm k_3} \delta_{\bm k_2, -\bm k_4}\Big\}.
\end{align}
We compactly write this as a spin and valley tensor (pseudpspin has been removed since we work in band basis and  keep just the upper band), using the scattering angle  $\theta\equiv \theta_{\bm k_2}-\theta_{\bm k_1}$, $\hat{\cal V}(\theta)= g_{abcd} (\theta) s_a s_b \tau_c \tau_d+ j_{ab\alpha\beta} (\theta)s_a s_b \tau_\alpha \tau_\beta$, where $a,b,c,d\in \{0,z\}$ and $\alpha,\beta\in\pm$. We explicitly display the angular dependence of the interaction matrix elements.
Considering the angular momentum channel, $l$, 
\begin{align}
\notag \hat{\cal V}_{l}&= \int \frac{d\theta}{2\pi} e^{i l \theta} \left[g_{abcd}s_a s_b \tau_c \tau_d + j_{ab\alpha\beta} s_a s_b \tau_\alpha \tau_\beta \right]\\
&= g_{abcd}^ls_a s_b \tau_c \tau_d + j_{ab\alpha\beta}^l s_a s_b \tau_\alpha \tau_\beta
\end{align}
We find that $l=\pm 1$ ($p$-wave) and $l=0$ ($s$-wave) are the dominant channels. The resulting tensor is given in equations \eqref{Vcooperl0} and \eqref{Vcooperl1} in the main text.

\section{Competing Instabilities}
\label{competing_insta}

In this section we address the question of whether other instabilities may compete with superconductivity. As is expected in general, superconductivity dominates in the limit of weak interactions when the Fermi surface is not nested. We find a portion of the phase diagram at stronger couplings in which magnetic order dominates, as shown in Figure \ref{f:phase}. 

We consider three types of instabilities: ferromagnetism (FM), spin density wave (SDW), and charge density wave (CDW), where the SDW and CDW states are commensurate with the lattice with period $\bm K$. These phases are referred to as particle-hole instabilities, and are captured by the order parameters, respectively
\begin{align}
\notag \Phi_1&=\sum_{\bm k}  \psi^\dag_{\uparrow,\tau,\bm k}  \psi_{\downarrow,\tau,\bm k},\\
\notag \Phi_2&=\sum_{\bm k}  \psi^\dag_{\uparrow,\tau,\bm k}  \psi_{\downarrow,-\tau,\bm k},\\
\Phi_3&=\sum_{\bm k}  \psi^\dag_{\uparrow,\tau,\bm k}  \psi_{\uparrow,-\tau,\bm k}.
\end{align}
These orders take hold when their associated susceptibilities diverge, a condition which results in equations analogous to the BCS gap equation. Denoting the particle-hole susceptibility $\chi_{s_1 \tau_1,s_2 \tau_2}(p_0, \bm p)$, 
FM order corresponds to a divergence in $\chi_1=\chi_{s \tau,-s \tau}(0, \bm 0)$, SDW order to $\chi_2= \chi_{s \tau,-s -\tau}(0, \bm 0)$ and CDW to $\chi_3=\chi_{s \tau,s -\tau}(0, \bm 0)$.

In the previous section, we wrote the interactions in the basis of states in the upper band (ie at the Fermi surface) with $ \bm k_1=-\bm k_3$, $\bm k_2=-\bm k_4$, corresponding to the Cooper scattering channel. For the particle-hole instabilities, the relevant scattering amplitudes are the direct and exchange channels. We again restrict all momenta to lie on the Fermi surface $|\bm k_i|=k_F$. The scattering condition for the exchange channel is $ \bm k_1=\bm k_4$, $\bm k_2=\bm k_3$, with scattering angle $\theta\equiv \theta_{k_2}-\theta_{k_1}$. The corresponding couplings $\cal{V}$ are obtained analogously to the procedure in the Cooper channel (\ref{Vprojection}). We further define the $\ell=0$ component as $\int d\theta \ \hat{V}/(2\pi)\equiv \hat{\cal V}^X$. For the density channel, $ \bm k_1=\bm k_2$, $\bm k_3=\bm k_4$, with scattering angle $\theta\equiv \theta_{k_3}-\theta_{k_1}$, and we define the $\ell=0$ component as $\int d\theta \hat{V}/(2\pi)\equiv \hat{\cal V}^D$. The procedure of the previous section yields the tensor,
\begin{align}
\label{Xcoupling}
\notag \hat{\cal V}^X&= g_{0}^X  +  g_{1}^X \tau_z \otimes \tau_z  +g_{2}^X s_z \otimes  s_z + g_{3}^X s_z \tau_z \otimes  s_z \tau_z + g_4^X(1+s_z s_z)(\tau_z\tau_0+\tau_0\tau_z) + \left(j_{0}^X+ j_{1}^X s_z\otimes  s_z\right)  (\tau_+ \otimes  \tau_-+\tau_- \otimes  \tau_+)\\
\hat{\cal V}^D&= j_0^D (\tau_+\otimes  \tau_-+\tau_- \otimes \tau_+)
\end{align}
The ladder equations give the total susceptibilities in terms of the static susceptibilities and the direct/exchange scattering amplitudes, 
\begin{align}
\label{chiRPA}
%$\frac{ dq_{10} d^2\bm{q_1}}{(2\pi)^3} \frac{ dq_{20} d^2\bm{q_2}}{(2\pi)^3}
\notag {\chi}_{s_1 \tau_1,s_2 \tau_2}(0, \bm 0)&= \chi^0_{s_1 \tau_1,s_2 \tau_2}(0, \bm 0)+ \sum_{s_a,s_b,\tau_a,\tau_b}\int\frac{d^3q_1 d^3 q_2}{(2\pi)^6} G_{s_1 \tau_1}(q_{10}, \bm{q_1}) G_{s_2 \tau_2}(q_{10}, \bm{q_1}) 
 {V}_{s_1 \tau_1 s_a \tau_a,s_2 \tau_2 s_b \tau_b}(\theta_2-\theta_1) G_{s_a \tau_b}(q_{20}, \bm{q_2}) G_{s_b \tau_b}(q_{20}, \bm{q_2}) + ... \\
\notag  &= \chi^0_{s_1 \tau_1,s_2 \tau_2}(0, \bm 0) + \chi^0_{s_1 \tau_1,s_2 \tau_2}(0, \bm 0) \sum_{s_a,s_b,\tau_a,\tau_b}\left(\int \frac{d\theta}{2\pi} {V}_{s_1 \tau_1 s_a \tau_a,s_2 \tau_2 s_b \tau_b}(\theta_2-\theta_1)\right)  \chi_{s_a \tau_a,s_b \tau_b}(0, \bm 0) \\
\notag &= \chi^0_{s_1 \tau_1,s_2 \tau_2}(0, \bm 0)+ \chi^0_{s_1 \tau_1,s_2 \tau_2}(0, \bm 0) \sum_{s_a,s_b,\tau_a,\tau_b}{\cal V}^{X/D}_{s_1 \tau_1 s_a \tau_a,s_2 \tau_2 s_b \tau_b} {\chi}_{s_a \tau_a,s_b \tau_b}(0, \bm 0).
\end{align}
where the static susceptibilities are given by
\begin{align}
\chi^0_{s_1 \tau_1,s_2 \tau_2}(0, \bm 0)&\equiv -i \int\frac{d^3q }{(2\pi)^3} G_{s_1 \tau_1}(q_{0}, \bm{q}) G_{s_2 \tau_2}(q_{0}, \bm{q}).
\end{align}
For FM, SDW, and CDW order, we evaluate the relevant static susceptibilities as 
$ \chi^0_{s \tau,-s \tau}(q_0=0,\bm q=0)= \chi^0_{s \tau,-s -\tau}(q_0=0,\bm q=0)=\chi^0_{s \tau,s -\tau}(q_0=0,\bm q=0)=\mu/(2\pi)$. The resulting RPA equations for the susceptibilities reduce to
\begin{align}
\notag {\chi}_1&= \frac{\mu}{2\pi}+ \frac{\mu}{2\pi}\left(\hat{\cal V}^X_{s+s+,\bar{s}+\bar{s}+}+\hat{\cal V}^X_{s+s-,\bar{s}-\bar{s}+} \right) {\chi}_1,\\
\notag {\chi}_2&= \frac{\mu}{2\pi}+ \frac{\mu}{2\pi}\left(\hat{\cal V}^X_{s+s+,\bar{s}-\bar{s}-} \right) {\chi}_2,\\
{\chi}_3&= \frac{\mu}{2\pi}+ \frac{\mu}{2\pi}\left(\hat{\cal V}^D_{s+s-,s-s+}\right) {\chi}_3.
\end{align}
The condition for FM, SDW and CDW instabilities immediately follow, and can be written analytically in terms of the coupling constants found in (\ref{Xcoupling}),
\begin{align}
\label{pole_condition}
\notag \text{(FM):} \ \ 1&= \frac{\mu}{2\pi}\left(\hat{\cal V}^X_{s+s+,\bar{s}+\bar{s}+}+\hat{\cal V}^X_{s+s-,\bar{s}-\bar{s}+} \right)\\
\notag&= \frac{\mu}{2\pi}\left(g_0^X+g_1^X-g_2^X-g_3^X+j_0^X-j_1^X\right)\\
\notag \text{(SDW):} \ \ 1&= \frac{\mu}{2\pi} \hat{\cal V}^X_{s+s+,\bar{s}-\bar{s}-}  = \frac{\mu}{2\pi}\left(g_0^X-g_1^X-g_2^X+g_3\right)\\
\text{(CDW):} \ \ 1&= \frac{\mu}{2\pi} \hat{\cal V}^D_{s+s-,s-s+}= \frac{\mu}{2\pi} j_0^D.
\end{align}
From Eq. \eqref{pole_condition}, the system exhibits instabilities which compete with superconductivity when the dimensionless coupling constants are order unity -- as distinct from the superconducting instability which occurs for arbitrarily weak attractive coupling. Throughout most of the phase diagram we considered, this condition is not met, and so superconductivity is the sole instability of the system.

Increasing the chemical potential, the antiscreening mechanism causes the $g_0$ coupling to grow large. Since this coupling does not appear in the susceptibility for CDW order, antiscreening gives rise to only FM and SDW ordering. We do find one small region of phase diagram where $g_0$ grows large enough to give rise to FM and SDW order. The other couplings, which are much smaller than $g_0$, act to favor SDW over FM order. In the phase diagram we simply label this region as {\it magnetic instability} since we expect FM and SDW to be nearly degenerate. Moreover, this part of the phase diagram should not be taken too literally, since the large coupling constant means corrections to mean field theory are likely significant.

\newpage

 \section{Effective tight-binding model for the superlattice}

In order to derive the effective lattice model, we introduce a basis of Wannier orbitals $|\bm{R},\alpha\rangle$ localised at the sites $\bm{R}$ of the honeycomb lattice, with $\alpha$ being a spin index defined by the action of threefold rotations ($C_{3z}$)
\begin{gather}
u(C_{3z}) |\bm{R},\alpha\rangle = e^{\frac{2\pi i}{3} \alpha} |\Lambda\bm{R},\alpha\rangle \ \ ,
\end{gather}
where $\alpha = \{\pm \frac{3}{2},\pm \frac{1}{2}\}$. We consider only the four lowest-energy orbitals on each site.

The lattice Hamiltonian has the form
\begin{gather}
H = \sum{T_{\alpha,\alpha'}(\bm{R},\bm{R}') c^\dag_{\bm{R},\alpha} c_{\bm{R}',\alpha'}}
\end{gather}
where
\begin{gather}
T_{\alpha,\alpha'}(\bm{R},\bm{R}') = \langle \bm{R},\alpha|H_{2DHG}|\bm{R}',\alpha'\rangle \ \ ,
\end{gather}
with $H_{2DHG}$ defined in (\ref{HL}). There is a splitting between the on-site energies $T_{\alpha\alpha}(\bm{R},\bm{R}) = \varepsilon_\alpha = \varepsilon_{|\alpha|}$ for the $\alpha = \pm \frac{3}{2}$ and $\alpha = \pm\frac{1}{2}$ states, and we consider an effective model involving only the $\alpha = \pm \frac{3}{2}$ states, which are lowest in energy, and denote $\alpha = \frac{3}{2}s$ where $s$ is the spin index used throughout the main text, and $c_{\bm{R},\alpha}\rightarrow c_{\bm{R},s}$.

The topological mass term originates from nearest neighbour hopping terms which involve a spin transition $\alpha'-\alpha = \pm 2$. By symmetry we find, for hopping from a site $\bm{R}$ to a nearest neighbour $\bm{R}+\bm{d}$,
\begin{gather}
T_{\mp\frac{1}{2},\pm\frac{3}{2}}(\bm{R}+\bm{d},\bm{R}) = \lambda d_\pm^2 \ \ .
\end{gather}
An effective spin-conserving next nearest neighbour hopping term arises due to two consecutive hoppings with initial, intermediate and final sites $\bm{R},\bm{R}+\bm{d}$, and $\bm{R}+\bm{d}+\bm{d}'$ respectively
\begin{gather}
T^{\text{eff}}_{ss}(\bm{R}+\bm{d}+\bm{d}',\bm{R}) = \frac{\lambda^2}{\varepsilon_{\frac{3}{2}} - \varepsilon_{\frac{1}{2}}}|\bm{d}|^2 e^{2is\sigma (\theta'-\theta)}
\end{gather}
where $\theta,\theta'$ are the hopping directions in the first and second steps respectively and $\sigma = +1,-1$ when $\bm{R}\in A,B$ respectively.

Choosing lattice vectors $\bm{a}_1 = (a,0), \bm{a}_2 = (a/2, a\sqrt{3}/2)$, and denoting the three nearest neighbour bonds $\bm{d}_i =\bm{R}-\bm{R}'$ with $\bm{R}'$ in the A sublattice and $\bm{R}$ a neighbouring site, and the six next nearest neighbour bonds $\widetilde{\bm{d}}_n$ which are vectors of length $a$ directed along angles $\theta_n = \frac{n\pi}{3}$ for $n = \{0,1,2,3,4,5\}$, we obtain an effective Hamiltonian involving only the $|\alpha| = \frac{3}{2}$ states (after absorbing the on-site potential into the chemical potential)
\begin{gather}
H = -t\sum_{\langle \bm{R}+\bm{d}_i,\bm{R}\rangle}{ c^\dag_{\bm{R}+\bm{d}_i,s} c_{\bm{R},s}} - t'\sum_{\langle \langle \bm{R}+\widetilde{\bm{d}}_n,\bm{R}\rangle\rangle}{ e^{is\sigma \varphi_n} c^\dag_{\bm{R}+\widetilde{\bm{d}}_n,s} c_{\bm{R},s}}
\end{gather}
where $\varphi_n = \frac{2\pi}{3}$ for $n = 0, 2, 4$ and $\varphi_n = -\frac{2\pi}{3}$ for $n = 1, 3, 5$.

We make contact between the two forms of the normal state Hamiltonian by expanding the Hamiltonian near the $K$ points, and reproduce the effective Dirac Hamiltonian
\begin{gather}
\mathcal{H}(\tau\bm{K}+\bm{k}) \approx v(\tau k_x \sigma_x + k_y  \sigma_y) + \eta \tau \sigma_z s_z
\end{gather}
where we find the relation between the parameters in the Dirac theory and in the real space model
\begin{gather}
v = \frac{\sqrt{3}at}{2}, \ \  \ \eta = \frac{9}{2}t' 
\end{gather}
Near the $K$ points we have the upper band eigenstates
\begin{gather}
\widetilde\psi^\dag_{\bm{k} \tau s} = \sum_{\bm{R}}{\varphi_{\bm{k}\tau s}(\bm{R}) c^\dag_{\bm{R}s}}
\label{Dirac:lattice}
\end{gather}
with symmetry properties $\varphi_{\bm{k}\tau s}(-\bm{R}) = \varphi_{-\bm{k}\bar{\tau}s}(\bm{R})$ and $\varphi_{-\bm{k}\bar{\tau}\downarrow}(\bm{R}) = \varphi^*_{\bm{k}\tau\uparrow}(\bm{R}) $. Explicitly,
\begin{gather}
\varphi_{\bm{k}\tau s}(\bm{R}) = \frac{1}{\sqrt{2}} e^{i(\tau\bm{K}+\bm{k})\cdot\bm{R}}\left( w^a_{\tau,s}(k) a(\bm{R}) + e^{i\tau\theta_{\bm{k}}} w^b_{\tau,s}(k) b(\bm{R}) \right)
\end{gather}
with $a(\bm{R}) = \{1,0\}$, $b(\bm{R}) = \{0,1\}$, for $\bm{R}\in A,B$ respectively, and the functions $w^a_{\tau,s}(k)$ and $w^b_{\tau,s}(k)$ are defined in Section \ref{cooperchannel}. We shall use these wavefunctions to obtain a real space form for the  the superconducting gap functions presented in momentum space in Section \ref{gapeqsect}.

\section{The pairing term in the lattice representation}

The mean field BdG Hamiltonian is
\begin{gather}
H = \sum_{\bm{k},\tau, s} \varepsilon_{\bm{k}} \widetilde\psi^\dag_{\bm{k} \tau s} \widetilde\psi_{\bm{k} \tau s} + \frac{1}{2} \sum_{\bm{k},-\bm{k},\tau,\tau', s,s'}\widetilde\psi^\dag_{\bm{k} \tau s}({\Delta}_{\bm{k}})_{\tau s, \tau' s' } \widetilde\psi^\dag_{-\bm{k} \tau', s'} + \text{h.c.}
\end{gather}
where we have used $\psi_{\bm{k} \tau s}$ to refer to the upper band creation operator, as in the previous subsection. The three superconducting phases we study are given by
\begin{gather}
{\Delta}_{\bm{k}} =  \Delta_{k} \times \begin{cases}
\ d^z_s {s}_z \tau_0 \left( {\tau}_y s_y \right) \\
\ e^{\pm i \theta_{\bm{k}}} d_s^z {s}_z {\tau}_z  \left( {\tau}_y s_y\right)  \\
\ e^{i{\tau}_z(\phi-\theta_{\bm{k}})} (d^x_s{s}_x + d^y_s{s}_y) \tau_y \left( {\tau}_y s_y \right) 
\end{cases}
\end{gather}
for the $s_\tau$, $p+ip$ and $p+i\tau p$ phases respectively. We have added a factor $\Delta_k$ absent in the main text. This is a smooth function peaked at the Fermi momentum, encapsulating the fact that pairing should only occur near the Fermi surface, and should be retained in deriving the correct real space gap function. Writing
\begin{gather}
\Delta_{\bm{k};\tau\tau';ss'} = \Delta_{\tau\tau'}(\bm{k}) (d^\mu s_\mu i \hat{s}_y)_{ss'} \ \ ,
\end{gather}
to separate out the spin structure, we can use the upper band wavefunctions to go to the coordinate representation in terms of the full real space creation operator $c^\dag_{\bm r s}$,
\begin{gather}
H = \frac{1}{2}\sum{ (d^\mu s_\mu i \hat{s}_y)_{ss'}\varphi_{\bm{k}\tau s}(\bm{R}) \Delta_{\tau\tau'}(\bm{k}) \varphi_{-\bm{k}\tau's'}(\bm{R}') c^\dag_{\bm{R}s} c^\dag_{\bm{R}'s'}} = \frac{1}{2}\sum{\Delta_{ss'}(\bm{R},\bm{R}') c^\dag_{\bm{R}s} c^\dag_{\bm{R}'s'}}
\end{gather}
Note that under inversion, $\bm{R}\rightarrow -\bm{R},\bm{R}'\rightarrow -\bm{R}'$ we have
\begin{align}
\Delta_{ss'}(-\bm{R},-\bm{R}') &=\sum_{\bm{k}}{ (d^\mu \hat{s}_\mu i \hat{s}_y)_{ss'} \varphi_{-\bm{k}\bar{\tau} s}(\bm{R}) \Delta_{\tau\tau'}(\bm{k}) \varphi_{\bm{k}\bar{\tau}'s'}(\bm{R}')} \nonumber \\
&= \sum_{\bm{k}}{(d^\mu \hat{s}_\mu i \hat{s}_y)_{ss'} \varphi_{\bm{k}\tau s}(\bm{R}) \Delta_{\bar{\tau}\bar{\tau}'}(-\bm{k}) \varphi_{-\bm{k}\tau's}(\bm{R})}
\end{align}
and the valley structures are explicitly given by
\begin{gather}
\Delta_{\tau\tau'}(\bm{k}) =  \Delta_k \times \begin{cases}
\ (i\hat{\tau}_y)_{\tau\tau'}  \\
\ e^{\pm i \theta_{\bm{k}}}(\hat{\tau}_z i\hat{\tau}_y)_{\tau\tau'}  \\
\  e^{i\hat{\tau}_z(\phi -\theta_{\bm{k}})} 
\end{cases}
\end{gather}
We can now explicitly evaluate the functions $\Delta(\bm{R}, \bm{R}')$. We begin with the intervalley phases, which take the form
\begin{gather}
H_\Delta = \sum_{\bm{k};s}{\Delta_k e^{i\ell \theta_{\bm{k}}} \widetilde\psi^\dag_{\bm{k}+s} \widetilde\psi^\dag_{-\bm{k}-\bar{s}}}
\end{gather}
where $\ell = 0$ for the $s_\pm$ phase and $\ell = \pm 1$ for the $p\pm ip$ phases. Expanding $\widetilde\psi^\dag_{\bm{k}\tau s}$ in the position basis (\ref{Dirac:lattice}) we find
\begin{gather} 
H_\Delta = \sum{\Delta_k e^{i\ell \theta_{\bm{k}}} \varphi_{\bm{k}+s}(\bm{R}) \varphi_{-\bm{k}-\bar{s}}(\bm{R}') c^\dag_{\bm{R}s} c^\dag_{\bm{R}'\bar{s}}} = \sum{\Delta(\bm{R},\bm{R}') c^\dag_{\bm{R}\uparrow} c^\dag_{\bm{R}'\downarrow}}\end{gather}
where we may write
\begin{gather}
\Delta(\bm{R},\bm{R}') = 
\sum_{\bm{k}}{\Delta_k e^{i\ell \theta_{\bm{k}}}\left[
\varphi_{\bm{k}+\uparrow}(\bm{R}) \varphi^*_{\bm{k}+\uparrow}(\bm{R}') - \varphi_{\bm{k}+\uparrow}(-\bm{R}) \varphi^*_{\bm{k}+\uparrow}(-\bm{R}')\right]} \ .
\end{gather}

In order to perform the summation over $k$ we introduce the functions $f^{\sigma\sigma'}_m(l)$ defined by
\begin{gather}
f^{\sigma\sigma'}_m(l) = \int{\Delta_k w^\sigma_{++}(k) w^{\sigma'}_{++}(k) J_{m}(kl) \frac{kdk}{2\pi}}  \ \ , 
\end{gather}
with $m=0,1,2,\dots$, and the relation
\begin{gather}
\sum{w^\sigma_{++}(k) w^{\sigma'}_{++}(k)\Delta_k e^{i(\bm{k}\cdot(\bm{R}-\bm{R}') + \ell \theta_{\bm{k}})}} = 
i^{|\ell|} e^{i\ell \theta} f^{\sigma\sigma'}_{|\ell|}(|\bm{R}-\bm{R}'|) \ \ ,
\end{gather}
where $\theta = \theta_{\bm{R}} - \theta_{\bm{R}'}$.

The function $f^{\sigma\sigma'}_0(l)$ is peaked at $l=0$ and oscillates over length scales $\sim k_F^{-1}$ with decaying amplitude, while for $m>0$ the functions $f^{\sigma\sigma'}_m(l)$ vanish at $l=0$, increase to a global maximum at $l \approx k_F^{-1}$ and then decays for larger values of $l$.

In terms of the functions $f^{\sigma\sigma}_m(l)$ the gap $\Delta(\bm{R},\bm{R}')$ is given by
\begin{gather}
\Delta(\bm{R},\bm{R}')\nonumber \\
= \frac{1}{2}\times\begin{cases}
i^{|\ell|} e^{i\ell \theta} \left( f^{\sigma\sigma}_{|\ell|}(|\bm{R}-\bm{R}'|) e^{i\bm{K}\cdot(\bm{R}-\bm{R}')} - (-1)^{\ell} f^{\bar{\sigma}\bar{\sigma}}_{|\ell|}(|\bm{R}-\bm{R}'|) e^{-i\bm{K}\cdot(\bm{R}-\bm{R}')}\right) \ \ , \ &\bm{R},\bm{R}'\in \sigma \\
e^{i\ell \theta}\left(
i^{|\ell-1|} e^{i(\bm{K}\cdot(\bm{R}-\bm{R}')-\theta)} f^{AB}_{|\ell-1|}(|\bm{R}-\bm{R}'|) + (-1)^{\ell} i^{|\ell+1|} e^{-i(\bm{K}\cdot(\bm{R}-\bm{R}') - \theta)} f^{AB}_{|\ell+1|}(|\bm{R}-\bm{R}'|)\right) \ \ , \ &\bm{R}\in A, \bm{R}'\in B \\
e^{i\ell\theta}\left( i^{|\ell+1|} e^{i(\bm{K}\cdot(\bm{R}-\bm{R}') + \theta)} f^{AB}_{|\ell+1|}(|\bm{R}-\bm{R}'|) + (-1)^\ell i^{|\ell-1|} e^{-i(\bm{K}\cdot(\bm{R}-\bm{R}') + \theta)} f^{AB}_{|\ell-1|}(|\bm{R}-\bm{R}'|)\right) \ \ , \ &\bm{R}\in B, \bm{R}'\in A
\end{cases}
\label{Delta:lattice}
\end{gather}
Note that for $\bm{R},\bm{R}'\in \sigma$ we have
\begin{gather}
\Delta(\bm{R},\bm{R}')= -\frac{1}{2}i^{|\ell|}  e^{i\ell\theta} \left(
f^{\bar{\sigma}\bar{\sigma}}_{|\ell|}(|\bm{R}-\bm{R}'|) e^{i\bm{K}\cdot(\bm{R}-\bm{R}')}-(-1)^{\ell}f^{\sigma\sigma}_{|\ell|}(|\bm{R}-\bm{R}'|) e^{-i\bm{K}\cdot(\bm{R}-\bm{R}')} \right)
\end{gather}
while for $\bm{R}'\in A, \bm{R}\in B$ we have
\begin{gather}
\Delta(\bm{R},\bm{R}') = \frac{1}{2}\left(
i^{|\ell+1|} e^{i(\bm{K}\cdot(\bm{R}-\bm{R}') + (\ell+1)\theta_{|\bm{R}-\bm{R}'|})} f^{AB}_{|\ell+1|}(|\bm{R}-\bm{R}'|)
+(-1)^{\ell}  i^{|\ell-1|}e^{i(-\bm{K}\cdot(\bm{R}-\bm{R}') + (\ell-1)\theta)} f^{AB}_{|\ell-1|}(|\bm{R}-\bm{R}'|)
\right)
\end{gather}
For nearest neighbours, $\bm{R}-\bm{R}'=\bm{d}_i$, recall that we have $\bm{K}\cdot\bm{d}_i = \{0, -2\pi/3, 2\pi/3\}$ and $\theta_{\bm{R}} = \theta_i =  \{ \pi/2,\pi/2 + 2\pi/3, \pi/2 + 4\pi/3 \}$, giving us $\bm{K}\cdot\bm{R} + \theta_{\bm{R}} = \frac{\pi}{2}$. Thus for $\bm{R}'\in A$ we have
\begin{gather}
\Delta(\bm{R}'+\bm{d}_i,\bm{R}') = \frac{1}{2}e^{i\ell \theta_i}\left(
i^{|\ell+1|+1} f^{AB}_{|\ell+1|}(\tfrac{a}{\sqrt{3}}) + (-1)^\ell i^{|\ell-1|-1}  f^{AB}_{|\ell-1|}(\tfrac{a}{\sqrt{3}})\right) 
\label{Delta:NN}
\end{gather}
For next nearest neighbours, $\bm{R}-\bm{R}' = \widetilde{\bm{d}}'_n$ we have
\begin{gather}
e^{i\bm{K}\cdot(\bm{R}-\bm{R}')} = \begin{cases}
e^{-\frac{2\pi i}{3}} \ \ , \  &i = 1,3,5 \\
e^{\frac{2\pi i}{3}} \ \ , \ &i=2,4,6
\end{cases}
\label{Delta:NNN}
\end{gather}
which gives us, for $\bm{R}-\bm{R}'=\widetilde{\bm{d}}_n$
\begin{gather}
\Delta(\bm{R},\bm{R}') =\begin{cases} \frac{1}{2} i^{|\ell|}e^{i\ell \theta_i'}\left( f^{\sigma\sigma}_{|\ell|} (a) e^{-\frac{2\pi i}{3}} - (-1)^\ell f^{\bar{\sigma}\bar{\sigma}}_{|\ell|}(a) e^{\frac{2\pi i}{3}}\right) \ \ , \ & i = 1,3,5\\
\frac{1}{2}i^{|\ell|}e^{i\ell \theta_i'} \left(f^{\sigma\sigma}_{|\ell|}(a) e^{\frac{2\pi i}{3}} - (-1)^\ell f^{\bar{\sigma}\bar{\sigma}}_{|\ell|}(a) e^{-\frac{2\pi i}{3}}\right) \ \ , \ &i=2,4,6
\end{cases}
\end{gather}
with $\theta'_i$ being the angle between $\bm{d}'_i$ and the $x$ axis.

\subsubsection{$s_\tau$}

We obtain the gap in the $s_\tau$ phase by setting $\ell = 0$ in (\ref{Delta:lattice}). For neighbour pairing we find (\ref{Delta:NN})
\begin{gather}
\Delta(\bm{R}'+\bm{d}_i,\bm{R}') =\frac{1}{2} f^{AB}_1(\tfrac{a}{\sqrt{3}})\left( -1 + 1\right) = 0
\end{gather}
and therefore restrict the pairing to next nearest neighbours only. The gap depends on the function $f^{\sigma\sigma}_0(a)$. Writing $f^{AA}_0(a) = \alpha+\beta$, $f^{BB}_0(a) = \alpha-\beta$ we have from (\ref{Delta:NNN})
\begin{gather}
\Delta(\bm{R},\bm{R}') = \begin{cases}
\frac{1}{2}\left(
(\alpha+\beta) e^{-\frac{2\pi i}{3}} - (\alpha-\beta) e^{\frac{2\pi i}{3}} \right)\ \ , \ &i=1,3,5 \\
\frac{1}{2}\left(
(\alpha+\beta) e^{\frac{2\pi i}{3}} - (\alpha-\beta) e^{-\frac{2\pi i}{3}}\right) \ \ , \ &i=2,4,6
\end{cases} \nonumber \\
= \begin{cases}
-\frac{1}{2}(\beta+i\sqrt{3}\alpha) \ \ , \ &i = 1,3,5 \\
-\frac{1}{2}(\beta-i\sqrt{3}\alpha)\ \ , &i = 2,4,6
\end{cases}
\end{gather}
In the limit of spin-orbit interaction we have $\beta \rightarrow 0$. For the numerical diagonalisation we choose a gap in which $\beta = 0$ (since the spin-orbit interaction is weak) and $\alpha = i\Delta'$, so that $\Delta(\bm{R},\bm{R}')$ is purely real.

\subsubsection{$p+ip$}
For exact diagonalisation we take only the nearest neighbour pairing terms. For $\bm{R}'\in A$, $\bm{R}=\bm{R}'+\bm{d}_i$, the gap is given by setting $\ell = +1$ in (\ref{Delta:NN}),
\begin{gather}
\Delta(\bm{R},\bm{R}') = \frac{1}{2}e^{i\theta}\left[-i f^{AB}_2(R) +i f^{AB}_0(R) \right] \nonumber \\
= \Delta' e^{i\theta}
\end{gather}
where $\theta = \theta_{\bm{R}}-\theta_{\bm{R}'}$.

\subsubsection{$p+i\tau p$}
We now consider the $p+i\tau p$ phase. Since the $\bm d_s$ vector is pinned in-plane for this phase, pairing is between the same spin species, i.e. the gap is proportional to $s_z$. This way, BdG Hamiltonian can be decomposed into spin blocks, the Majoranas corner states associated to each of which are related by time-reversal symmetry. 

The derivation then proceeds through more or less the same manipulations as above (c.f. Appendix A4 of \cite{Li2021}). It is possible to decompose the pairing term into two identical spin blocks,
\begin{gather}
H_\Delta = \frac{1}{2}\sum_{k, \tau,s}{\Delta_k e^{i\tau \phi} e^{-i\tau \theta_{\bm{k}}} \widetilde\psi^\dag_{\bm{k},\tau,s} \widetilde\psi^\dag_{-\bm{k},\tau,s}} = \frac{1}{2}\sum_{s}{\Delta(\bm{R},\bm{R}') c^\dag_{\bm{R},s} c^\dag_{\bm{R}',s}} \ \ .
\end{gather}
Expanding $\widetilde{\psi}_{\bm{k},\tau,s}$ in the position basis (\ref{Dirac:lattice}) we find
\begin{gather}
\Delta(\bm{R},\bm{R}') = \sum_{\bm{k}}{}\Delta_k\{
\tfrac{1}{2}e^{i\{\bm{K}\cdot(\bm{R}+\bm{R}') + \bm{k}\cdot(\bm{R}-\bm{R}') + \phi -\theta_{\bm{k}}\}}\left[
\left(w^a_{+,s}(k) a(\bm{R}) + e^{i\theta_{\bm{k}}} w^b_{+,s}(k)  b(\bm{R})\right)
\left(w^a_{+,s}(k) a(\bm{R'}) - e^{i \theta_{\bm{k}}} w^b_{+,s}(k)  b(\bm{R'})\right)\right] \nonumber \\
+ \left.\tfrac{1}{2}e^{i\{ -\bm{K}\cdot(\bm{R}+\bm{R}')+\bm{k}\cdot(\bm{R}-\bm{R}')  -\phi +\theta_{\bm{k}}\}}\left[
\left(w^a_{-,s}(k) a(\bm{R}) + e^{-i\theta_{\bm{k}}} w^b_{-,s}(k)  b(\bm{R})\right)
\left(w^a_{-,s}(k) a(\bm{R'}) - e^{-i \theta_{\bm{k}}} w^b_{-,s}(k)  b(\bm{R'})\right)\right] \right\} \ \ .
\end{gather}
Performing the summation over $\bm{k}$ yields functions $f^{\sigma\sigma'}_m(|\bm{R}-\bm{R}'|)$ which all vanish at small separations $\bm{R}-\bm{R}' \ll k_F^{-1}$ except for $m = 0$. For purposes of exact diagonalisation, we keep only terms involving nearest neighbours, which correspond to those that cancel the winding factor $e^{i\theta_{\bm{k}}}$. This gives
\begin{gather}
\Delta(\bm{R},\bm{R}') = \sum_{\bm{k}}{}\Delta_k\left\{
\tfrac{1}{2}e^{i\{\bm{K}\cdot(\bm{R}+\bm{R}') + \bm{k}\cdot(\bm{R}-\bm{R}') + \phi -\theta_{\bm{k}}\}}
w^a_{+,s}w^b_{+,s}e^{i \theta_{\bm{k}}} \left(- a(\bm{R}) b(\bm{R'})+ b(\bm{R}) a(\bm{R'}) \right)\right. \nonumber \\
+ \left.\tfrac{1}{2}e^{i\{ -\bm{K}\cdot(\bm{R}+\bm{R}')+\bm{k}\cdot(\bm{R}-\bm{R}')  -\phi +\theta_{\bm{k}}\}}
w^a_{-,s}w^b_{-,s}e^{-i \theta_{\bm{k}}} \left(- a(\bm{R}) b(\bm{R'})+ b(\bm{r}) a(\bm{R'}) \right) \right\}
\end{gather}
We note that $w^a_{-,s}w^b_{-,s}=-w^a_{+,s}w^b_{+,s}$, and is independent of spin index $s$. Performing the summation over $\bm{k}$
\begin{gather}
\int{w^a_{+,s}w^b_{+,s} \Delta_k e^{i\bm{k}\cdot(\bm{R}-\bm{R}')} \frac{d^2\bm{k}}{(2\pi)^2}} = \int{\frac{v k}{\varepsilon_{\bm k}} \Delta_k J_0(k|\bm{R}-\bm{R}'|) \frac{kdk}{2\pi}} = f^{AB}_0(|\bm{R}-\bm{R}'|)
\end{gather}
we find
\begin{gather}
\Delta(\bm{R},\bm{R}') =\frac{1}{2} f^{AB}_0(|\bm{R}-\bm{R}'|)\left[
e^{i\{\bm{K}\cdot(\bm{R}+\bm{R}')  + \phi\}} -e^{i\{ -\bm{K}\cdot(\bm{R}+\bm{R}') -\phi\}} \right]
\left[- a(\bm{R}) b(\bm{R'})+ b(\bm{R}) a(\bm{R'}) \right]\nonumber \\
 =i f^{AB}_0(|\bm{R}-\bm{R}'|)\left[
\sin(\bm{K}\cdot(\bm{R}+\bm{R}')  + \phi) \right]
\left[- a(\bm{R}) b(\bm{R'})+ b(\bm{R}) a(\bm{R'}) \right],
\end{gather}
and therefore, with $\bm{R}\in A,\bm{R}'\in B$, 
\begin{gather}
\label{hotsgap}
H_\Delta = \sum_{\langle \bm{R},\bm{R}'\rangle}{
\Delta' \left[\sin(\bm{K}\cdot(\bm{R}+\bm{R}') + \phi) c^\dag_{\bm{R}} c^\dag_{\bm{R}'} + \text{h.c.}\right]}.
\end{gather}
where $\Delta' = i f^{AB}_0(|\bm{R}-\bm{R}'|)$.

\end{document}